\documentclass[preprint, resetfootnote]{aastex701}
\usepackage{amsmath,amssymb}
\usepackage{float}
\usepackage [autostyle, english = american]{csquotes} 
\MakeOuterQuote{"}
\usepackage{hyperref}
\usepackage{xspace}
\usepackage{multirow}
\DeclareRobustCommand{\LMCt}{\hyperlink{cite.lam18_freqopt}{LMC18}\xspace}
\DeclareRobustCommand{\LMCp}{\hyperlink{cite.lam18_freqopt}{(LMC18)}\xspace}
\DeclareRobustCommand{\NGXVt}{\hyperlink{cite.ng15_timing}{NG15}\xspace}

\DeclareRobustCommand{\NGXVgwbt}{\hyperlink{cite.ng15_gwb_evidence}{NG15gwb}\xspace}

\newcommand{\degrees}{{^\circ}}
\newcommand{\Dec}{{\mathrm{Dec.}}}
\newcommand{\MHz}{{\; \mathrm{MHz}}}

\newcommand{\logten}{{\log_{10}}}
\newcommand{\Npsrs}{{N_\mathrm{psr}}}
\newcommand{\tint}{{t_\mathrm{int}}}
\newcommand{\tswap}{{t_\mathrm{swap}}}
\newcommand{\tmax}{{t_\mathrm{max}}}
\newcommand{\tmaxi}{{\tmax_{,i}}}
\newcommand{\tmin}{{t_\mathrm{min}}}
\newcommand{\tmini}{{\tmin_{,i}}}
\newcommand{\sigmaTOA}{{\sigma_\mathrm{TOA}}}
\newcommand{\sigmaTOAsq}{{\sigma_\mathrm{TOA}^2}}

\newcommand{\sigmaDMnu}{\sigma_{\mathrm{DM}(\nu)}}
\newcommand{\sigmatel}{\sigma_\mathrm{tel}}
\newcommand{\sigmaTF}{\sigma_\mathrm{S/N}}
\newcommand{\sigmaJ}{\sigma_\mathrm{J}}
\newcommand{\sigmaJsing}{\sigma_\mathrm{J,1}}
\newcommand{\sigmaDISS}{\sigma_\mathrm{DISS}}
\newcommand{\sigmaDMhat}{\sigma_{\widehat{\mathrm{DM}}}}
\newcommand{\sigmadtC}{\sigma_{\delta t_{C}}}
\newcommand{\taud}{\tau_d}
\newcommand{\dtaud}{\Delta \tau_d}
\newcommand{\dnud}{\Delta \nu_d}
\newcommand{\etat}{\eta_t}
\newcommand{\etanu}{\eta_\nu}
\newcommand{\Tsys}{T_\mathrm{sys}}
\newcommand{\Trcvr}{T_\mathrm{rcvr}}
\newcommand{\Tgal}{T_\mathrm{gal}}
\newcommand{\Aeff}{A_\mathrm{eff}}

\newcommand{\Weff}{W_\mathrm{eff}}
\newcommand{\Uscale}{U_\mathrm{scale}}
\newcommand{\DM}{\mathrm{DM}}
\newcommand{\DMbar}{\overline{\mathrm{DM}}}
\newcommand{\nuh}{\nu_{\mathrm{j}}}
\newcommand{\nul}{\nu_{\mathrm{i}}}
\newcommand{\xvec}{\mathbf{x}}

\newcommand{\sigmaDMbar}{\sigma_{\overline{\mathrm{DM}}}}
\newcommand{\Cdmbarnn}{C_{\delta\overline{\mathrm{DM}}}(\nuh,\nuh)}
\newcommand{\Cdmbarnpnp}{C_{\delta\overline{\mathrm{DM}}}(\nul,\nul)}
\newcommand{\Cdmbarnnp}{C_{\delta\overline{\mathrm{DM}}}(\nuh,\nul)}

\newcommand{\Dnud}{\Delta \nu_d}

\newcommand{\diag}{\mathrm{diag}}
\newcommand{\Var}{\mathrm{Var}}
\newcommand{\Cov}{\mathrm{Cov}}
\newcommand{\fyr}{{f_\mathrm{yr}}}
\newcommand{\tvec}{{\vec{t}_\mathrm{int}}}
\newcommand{\topt}{{\tvec\mkern-2mu{}^\ast}}
\newcommand{\teq}{{\vec{t}_\mathrm{eq}}}
\newcommand{\tepoch}{{t_\mathrm{epoch}}}
\newcommand{\Pdot}{{\dot{P}}}
\newcommand{\Pdotint}{{\Pdot_\mathrm{int}}}
\newcommand{\AGWB}{{A_\mathrm{GWB}}}
\newcommand{\gammaGWB}{{\gamma_\mathrm{GWB}}}
\newcommand{\alphaGWB}{{\alpha_\mathrm{GWB}}}
\newcommand{\Ared}{{A_\mathrm{red}}}
\newcommand{\gammared}{{\gamma_\mathrm{red}}}
\newcommand{\Tobs}{{T_\mathrm{obs}}}
\newcommand{\NcalInv}{{\mathcal{N}^{-1}}}
\newcommand{\Tcal}{{\mathcal{T}}}
\newcommand{\Rcal}{{\mathcal{R}}}
\newcommand{\SNR}{{\mathrm{S/N}}}
\newcommand{\Qmat}{{\bf Q}}
\newcommand{\tildep}{{\tilde{p}}}
\newcommand{\tildeQmat}{{\bf \tilde{Q}}}



\shorttitle{Optimal Pulsar Timing Array Strategies with the Deep Synoptic Array}
\shortauthors{T.~Cohen et al.}

\begin{document}

\title{Optimal Pulsar Timing Array Strategies with the Deep Synoptic Array}
\correspondingauthor{Tyler Cohen}
\email{tyler.cohen@nanograv.org}

\author[0000-0001-7587-5483]{Tyler Cohen}
\affiliation{Department of Physics, New Mexico Institute of Mining and Technology, 801 Leroy Place, Socorro, NM 87801, USA}
\email{}
\author[0000-0002-6664-965X]{Paul B. Demorest}
\affiliation{National Radio Astronomy Observatory, 1003 Lopezville Rd., Socorro, NM 87801, USA}
\email{}
\author[0000-0002-4972-1525]{Jeremy G. Baier}
\affiliation{Department of Physics, Oregon State University, Corvallis, OR 97331, USA}
\email{}
\author[0000-0001-5229-7430]{Ryan S. Lynch}
\affiliation{Green Bank Observatory, P.O. Box 2, Green Bank, WV 24944, USA}
\email{}
\author[0000-0001-5799-9714]{Scott M. Ransom}
\affiliation{National Radio Astronomy Observatory, 520 Edgemont Road, Charlottesville, VA 22903, USA}
\email{}
\author[0000-0003-0721-651X]{Michael T. Lam}
\affiliation{SETI Institute, 339 N Bernardo Ave Suite 200, Mountain View, CA 94043, USA}
\email{}
\author[0000-0001-5134-3925]{Gabriella Agazie}
\affiliation{Center for Gravitation, Cosmology and Astrophysics, Department of Physics and Astronomy, University of Wisconsin-Milwaukee,\\ P.O. Box 413, Milwaukee, WI 53201, USA}
\email{}
\author[0000-0002-8935-9882]{Akash Anumarlapudi}
\affiliation{Department of Physics and Astronomy, University of North Carolina, Chapel Hill, NC 27599, USA}
\email{}
\author[0000-0003-0638-3340]{Anne M. Archibald}
\affiliation{Newcastle University, NE1 7RU, UK}
\email{}
\author[0009-0008-6187-8753]{Zaven Arzoumanian}
\affiliation{X-Ray Astrophysics Laboratory, NASA Goddard Space Flight Center, Code 662, Greenbelt, MD 20771, USA}
\email{}
\author[0000-0003-2745-753X]{Paul T. Baker}
\affiliation{Department of Physics and Astronomy, Widener University, One University Place, Chester, PA 19013, USA}
\email{}
\author[0009-0008-9349-1086]{Ava L. Battaglia}
\affiliation{Department of Physics, Lafayette College, Easton, PA 18042, USA}
\email{}
\author[0000-0003-3053-6538]{Paul R. Brook}
\affiliation{Institute for Gravitational Wave Astronomy and School of Physics and Astronomy, University of Birmingham, Edgbaston, Birmingham B15 2TT, UK}
\email{}
\author[0009-0008-4771-2255]{Anabelle L. Brueggemann}
\affiliation{Department of Physics, Lafayette College, Easton, PA 18042, USA}
\email{}
\author[0000-0002-6039-692X]{H. Thankful Cromartie}
\affiliation{Department of Physics and Astronomy, Vanderbilt University, 2301 Vanderbilt Place, Nashville, TN 37235, USA}
\email{}
\author[0000-0002-1529-5169]{Kathryn Crowter}
\affiliation{Department of Physics and Astronomy, University of British Columbia, 6224 Agricultural Road, Vancouver, BC V6T 1Z1, Canada}
\email{}
\author[0000-0002-2185-1790]{Megan E. DeCesar}
\altaffiliation{Resident at the Naval Research Laboratory}
\affiliation{Department of Physics and Astronomy, George Mason University, Fairfax, VA 22030}
\email{}
\author[0000-0001-8885-6388]{Timothy Dolch}
\affiliation{Department of Physics and Astronomy, University of New Mexico, Albuquerque, NM 87131, USA}
\affiliation{Department of Physics, Hillsdale College, 33 E. College Street, Hillsdale, MI 49242, USA}
\affiliation{Eureka Scientific, 2452 Delmer Street, Suite 100, Oakland, CA 94602-3017, USA}
\affiliation{SETI Institute, 339 N Bernardo Ave Suite 200, Mountain View, CA 94043, USA}
\email{}
\author[0000-0001-7828-7708]{Elizabeth C. Ferrara}
\affiliation{Department of Astronomy, University of Maryland, College Park, MD 20742, USA}
\affiliation{Center for Research and Exploration in Space Science and Technology, NASA/GSFC, Greenbelt, MD 20771}
\affiliation{NASA Goddard Space Flight Center, Greenbelt, MD 20771, USA}
\email{}
\author[0000-0001-5645-5336]{William Fiore}
\affiliation{Department of Physics and Astronomy, University of British Columbia, 6224 Agricultural Road, Vancouver, BC V6T 1Z1, Canada}
\email{}
\author[0000-0001-8384-5049]{Emmanuel Fonseca}
\affiliation{Department of Physics and Astronomy, West Virginia University, P.O. Box 6315, Morgantown, WV 26506, USA}
\affiliation{Center for Gravitational Waves and Cosmology, West Virginia University, Chestnut Ridge Research Building, Morgantown, WV 26505, USA}
\email{}
\author[0000-0001-7624-4616]{Gabriel E. Freedman}
\affiliation{NASA Goddard Space Flight Center, Greenbelt, MD 20771, USA}
\email{}
\author[0000-0001-6166-9646]{Nate Garver-Daniels}
\affiliation{Department of Physics and Astronomy, West Virginia University, P.O. Box 6315, Morgantown, WV 26506, USA}
\affiliation{Center for Gravitational Waves and Cosmology, West Virginia University, Chestnut Ridge Research Building, Morgantown, WV 26505, USA}
\email{}
\author[0000-0001-8158-683X]{Peter A. Gentile}
\affiliation{Department of Physics and Astronomy, West Virginia University, P.O. Box 6315, Morgantown, WV 26506, USA}
\affiliation{Center for Gravitational Waves and Cosmology, West Virginia University, Chestnut Ridge Research Building, Morgantown, WV 26505, USA}
\email{}
\author[0000-0003-4090-9780]{Joseph Glaser}
\affiliation{Department of Physics and Astronomy, West Virginia University, P.O. Box 6315, Morgantown, WV 26506, USA}
\affiliation{Center for Gravitational Waves and Cosmology, West Virginia University, Chestnut Ridge Research Building, Morgantown, WV 26505, USA}
\email{}
\author[0000-0003-1884-348X]{Deborah C. Good}
\affiliation{Department of Physics and Astronomy, University of Montana, 32 Campus Drive, Missoula, MT 59812}
\email{}
\author[0000-0003-2742-3321]{Jeffrey S. Hazboun}
\affiliation{Department of Physics, Oregon State University, Corvallis, OR 97331, USA}
\email{}
\author[0000-0003-1082-2342]{Ross J. Jennings}
\altaffiliation{NANOGrav Physics Frontiers Center Postdoctoral Fellow}
\affiliation{Department of Physics and Astronomy, West Virginia University, P.O. Box 6315, Morgantown, WV 26506, USA}
\affiliation{Center for Gravitational Waves and Cosmology, West Virginia University, Chestnut Ridge Research Building, Morgantown, WV 26505, USA}
\email{}
\author[0000-0001-6607-3710]{Megan L. Jones}
\affiliation{Center for Gravitation, Cosmology and Astrophysics, Department of Physics and Astronomy, University of Wisconsin-Milwaukee,\\ P.O. Box 413, Milwaukee, WI 53201, USA}
\email{}
\author[0000-0001-6295-2881]{David L. Kaplan}
\affiliation{Center for Gravitation, Cosmology and Astrophysics, Department of Physics and Astronomy, University of Wisconsin-Milwaukee,\\ P.O. Box 413, Milwaukee, WI 53201, USA}
\email{}
\author[0000-0002-0893-4073]{Matthew Kerr}
\affiliation{Space Science Division, Naval Research Laboratory, Washington, DC 20375-5352, USA}
\email{}
\author[0000-0003-1301-966X]{Duncan R. Lorimer}
\affiliation{Department of Physics and Astronomy, West Virginia University, P.O. Box 6315, Morgantown, WV 26506, USA}
\affiliation{Center for Gravitational Waves and Cosmology, West Virginia University, Chestnut Ridge Research Building, Morgantown, WV 26505, USA}
\email{}
\author[0000-0001-5373-5914]{Jing Luo}
\altaffiliation{Deceased}
\affiliation{Department of Astronomy \& Astrophysics, University of Toronto, 50 Saint George Street, Toronto, ON M5S 3H4, Canada}
\email{}
\author[0009-0004-3621-5422]{Hugo E. Marquis}
\affiliation{Department of Physics, Lafayette College, Easton, PA 18042, USA}
\email{}
\author[0000-0001-5481-7559]{Alexander McEwen}
\affiliation{Center for Gravitation, Cosmology and Astrophysics, Department of Physics and Astronomy, University of Wisconsin-Milwaukee,\\ P.O. Box 413, Milwaukee, WI 53201, USA}
\email{}
\author[0000-0001-7697-7422]{Maura A. McLaughlin}
\affiliation{Department of Physics and Astronomy, West Virginia University, P.O. Box 6315, Morgantown, WV 26506, USA}
\affiliation{Center for Gravitational Waves and Cosmology, West Virginia University, Chestnut Ridge Research Building, Morgantown, WV 26505, USA}
\email{}
\author[0000-0002-4642-1260]{Natasha McMann}
\affiliation{Department of Physics and Astronomy, Vanderbilt University, 2301 Vanderbilt Place, Nashville, TN 37235, USA}
\email{}
\author[0000-0001-8845-1225]{Bradley W. Meyers}
\affiliation{Australian SKA Regional Centre (AusSRC), Curtin University, Bentley, WA 6102, Australia}
\affiliation{International Centre for Radio Astronomy Research (ICRAR), Curtin University, Bentley, WA 6102, Australia}
\email{}
\author[0000-0002-3616-5160]{Cherry Ng}
\affiliation{Dunlap Institute for Astronomy and Astrophysics, University of Toronto, 50 St. George St., Toronto, ON M5S 3H4, Canada}
\email{}
\author[0000-0002-6709-2566]{David J. Nice}
\affiliation{Department of Physics, Lafayette College, Easton, PA 18042, USA}
\email{}
\author[0000-0001-5465-2889]{Timothy T. Pennucci}
\affiliation{Institute of Physics and Astronomy, E\"{o}tv\"{o}s Lor\'{a}nd University, P\'{a}zm\'{a}ny P. s. 1/A, 1117 Budapest, Hungary}
\email{}
\author[0000-0002-8509-5947]{Benetge B. P. Perera}
\affiliation{Arecibo Observatory, HC3 Box 53995, Arecibo, PR 00612, USA}
\email{}
\author[0000-0002-8826-1285]{Nihan S. Pol}
\affiliation{Department of Physics, Texas Tech University, Box 41051, Lubbock, TX 79409, USA}
\email{}
\author[0000-0002-2074-4360]{Henri A. Radovan}
\affiliation{Department of Physics, University of Puerto Rico, Mayag\"{u}ez, PR 00681, USA}
\email{}
\author[0000-0002-5297-5278]{Paul S. Ray}
\affiliation{Space Science Division, Naval Research Laboratory, Washington, DC 20375-5352, USA}
\email{}
\author[0000-0003-4391-936X]{Ann Schmiedekamp}
\affiliation{Department of Physics, Penn State Abington, Abington, PA 19001, USA}
\email{}
\author[0000-0002-1283-2184]{Carl Schmiedekamp}
\affiliation{Department of Physics, Penn State Abington, Abington, PA 19001, USA}
\email{}
\author[0000-0002-7283-1124]{Brent J. Shapiro-Albert}
\affiliation{Department of Physics and Astronomy, West Virginia University, P.O. Box 6315, Morgantown, WV 26506, USA}
\affiliation{Center for Gravitational Waves and Cosmology, West Virginia University, Chestnut Ridge Research Building, Morgantown, WV 26505, USA}
\affiliation{Giant Army, 915A 17th Ave, Seattle WA 98122}
\email{}
\author[0000-0001-9784-8670]{Ingrid H. Stairs}
\affiliation{Department of Physics and Astronomy, University of British Columbia, 6224 Agricultural Road, Vancouver, BC V6T 1Z1, Canada}
\email{}
\author[0000-0002-7261-594X]{Kevin Stovall}
\affiliation{National Radio Astronomy Observatory, 1003 Lopezville Rd., Socorro, NM 87801, USA}
\email{}
\author[0000-0002-2820-0931]{Abhimanyu Susobhanan}
\affiliation{Max-Planck-Institut f{\"u}r Gravitationsphysik (Albert-Einstein-Institut), Callinstra{\ss}e 38, D-30167 Hannover, Germany\\Leibniz Universit{\"a}t Hannover, D-30167 Hannover, Germany}
\email{}
\author[0000-0002-1075-3837]{Joseph K. Swiggum}
\altaffiliation{NANOGrav Physics Frontiers Center Postdoctoral Fellow}
\affiliation{Department of Physics, Lafayette College, Easton, PA 18042, USA}
\email{}
\author[0000-0001-9678-0299]{Haley M. Wahl}
\affiliation{Department of Physics and Astronomy, West Virginia University, P.O. Box 6315, Morgantown, WV 26506, USA}
\affiliation{Center for Gravitational Waves and Cosmology, West Virginia University, Chestnut Ridge Research Building, Morgantown, WV 26505, USA}
\email{}


\begin{abstract}
Pulsar timing array (PTA) experiments have seen evidence for a nanohertz-frequency gravitational wave background (GWB) through coordinated radio timing observations of millisecond pulsars.
The significance of this evidence is expected to grow and the GWB's progenitor(s)
uncovered with continued observations, 
especially as the next generation of radio telescopes comes online. We present a framework
for simulating timing observations of a set of pulsars with a set of telescopes to 
predict the per-pulsar optimal instrument, observing frequency, and integration time.
We apply this methodology to the NANOGrav source list of 85 pulsars observed with the
Green Bank Telescope, Very Large Array,
Canadian Hydrogen Intensity Mapping Experiment (CHIME), and the Deep Synoptic Array (DSA),
a radio dish array under construction in Nevada. We discuss the dominant
noise contributions to the timing precision of these pulsars and show that most
are not dominated by intrinsic pulse phase jitter noise. We also determine which pulsars 
should be timed with DSA and CHIME or only DSA and show that an ab initio DSA PTA
is more sensitive than one which uses current instruments. Finally, we solve for the optimal
integration time per pulsar, subject to a fixed time budget, that maximizes the GWB signal-to-noise.
We show that there is little benefit to optimizing over integration time for these pulsars 
observed with DSA. Through Monte Carlo realizations of the PTA, we show
that the GWB sensitivity improves by $\sim 5\%$ for time-optimized observations
compared to the equal-time-per-source scenario with an early-science, fixed source list.

\end{abstract}


\section{Introduction}
Millisecond pulsars (MSPs) are neutron stars which have been spun up to tremendous rotation
rates that are stable over cosmological timescales by accreting matter from a binary star companion.
MSPs send radio pulses toward Earth with such regularity that they can act as precise, naturally
occurring clocks. Nanohertz-frequency gravitational waves (GWs) passing through the Galaxy will cause
the pulsar-Earth light travel time to oscillate on periods of years. Pulsar timing array (PTA)
experiments measure these
oscillations by searching for correlations in high-precision timing observations of a set of MSPs 
carried out over decades. One possible source of nHz GWs is a stochastic gravitational wave background
(GWB). The prevailing theory is that such a background could be generated by a cosmological population
of supermassive black hole binaries (SMBHBs) in the centers of merging galaxies \citep{detweiler79}. 
Therefore, nHz GWs could act as a probe into the evolutionary history of galaxies and populations of SMBHBs
\citep{jaffe03, sesana04}.
Regional PTAs have recently seen evidence for the GWB 
\citep[][hereafter NG15gwb]{ppta_evidence, epta_inpta_evidence, ng15_gwb_evidence}, but determining
its source requires constraining the properties of the GWB spectrum through continued observations. 
Eventually we expect to be able to resolve signals from individual
SMBHBs amidst the background, potentially providing opportunities to study the mechanisms
that power active galactic nuclei \citep{Burke_Spolaor_2019}.

The North American Nanohertz Observatory for Gravitational Waves (NANOGrav) times 78
MSPs and has amassed 20 years of data using the Arecibo Observatory, Green Bank Telescope (GBT), Very Large
Array (VLA), and Canadian Hydrogen Intensity Mapping Experiment (CHIME). NANOGrav regularly 
releases datasets of its timing measurements, the latest being the 15-yr data set 
\citep[][hereafter NG15]{ng15_timing} consisting of 68 pulsars. 
For the last five years NANOGrav has conducted dual-band observations with the GBT and VLA, where
 59 pulsars are timed with the GBT only, 15 are
timed with the GBT and VLA, and one is timed with only the VLA.
In addition, daily cadenced, low-frequency (400--800 MHz)
observations are carried out by CHIME for 54 MSPs.
GBT and VLA observations have a monthly cadence, with observing time split equally among
the pulsars.
While an equal-time allocation is convenient from a scheduling perspective, it has not previously
been determined whether that is the optimal strategy for this set of pulsars.
NANOGrav is expected to transition most of its pulsar timing to the Deep Synoptic Array (DSA),
a dish array under construction in Eastern Nevada, by the late 2020s \citep{hallinan19_dsa2000}.
DSA is expected to have a significant increase in sensitivity and continuous bandwidth over
current instruments. It is critical to strategically plan how this instrument will be used now,
while there is still some flexibility in the design of the instrument and observing program.

There have been many efforts to optimize PTA observations for sensitivity to GWs,
from the optimal observing frequency per pulsar that minimizes timing noise \citep{lam18_freqopt} 
to the optimal integration time per pulsar that maximizes the signal-to-noise (S/N) ratio for the
GWB \citep{Lee2012, middleton25} and optimal integration time that maximizes single-SMBHB, continuous wave (CW)
sensitivity \citep{christy14, baier25}. Additionally, much work has been done to derive scaling laws that predict 
PTA time-to-detection of the GWB and its S/N \citep{siemens13, vigeland16_scalinglaws}. 
These works have shown that
PTA sensitivity to GWs is a function of radio telescope capabilities, the number of MSPs observed, 
the cadence with which they are observed, the total time baseline of observations, 
and uncertainties in their pulse times of arrival (TOA).
 Therefore, the time required to achieve PTA science objectives will be determined, in
 part, by how consistently new MSPs can be discovered and added to the PTA and how effectively
 next-generation telescopes like DSA, next-generation VLA (ngVLA)
 and Square Kilometer Array (SKA) are able to time these MSPs.
As new MSPs continue to be discovered and PTA experiment source lists continue to expand,
it becomes crucial to make the most of telescope resources in order to maximize scientific output.
We present a novel framework and software package, PTAOptimizer, for
simulating and predicting many sources of noise in pulsar timing data with radio telescopes
and predicting a simulated PTA's sensitivity to the GWB. 
We use this framework to optimize
the NANOGrav PTA for the simultaneous optimal instrument, observing frequency range, and integration time
per pulsar that maximizes GWB sensitivity. 

In Section \ref{sec:sigma_toa} we discuss the infinite-frequency TOA 
uncertainty figure-of-merit we use to evaluate the performance of a pulsar-receiver pair. 
In Section \ref{sec:gwb_snr} we discuss the GWB S/N objective function, which quantifies a PTA's
sensitivity to the GWB. 
In Section \ref{sec:psr-tele} we describe our models of the NANOGrav PTA and the telescopes/receivers
with which we simulate observations. 
In Section \ref{sec:opt_instr} we discuss the optimal instrument choice per pulsar and present 
our predictions for the improvement of a DSA PTA over current instruments.
In Section \ref{sec:jitter} we discuss the number of jitter-dominated pulsars and the efficacy
of subarrying and strategic beam placement. 
In Section \ref{sec:opt-tint} we discuss the geometry of the objective
function with respect to integration time and present our results for the optimal distribution
of integration time for a PTA using DSA. 
Finally, we summarize our conclusions and discuss
potential future work in Section \ref{sec:conclusion}.

\section{The Infinite-Frequency TOA Uncertainty}\label{sec:sigma_toa}
Pulsar timing requires one to sum many thousands of
pulses measured across time and observing frequency in order to accumulate
enough S/N to measure a TOA. "Folding" the pulses at the correct
rotational phase requires a timing model that can account for every
rotation of the pulsar. This timing model must account for all of the
effects along the line-of-sight to the pulsar that would result in
measuring an apparent change in the rotational frequency of the pulsar,
from spin-down due to magnetic dipole radiation, to parallactic motions,
to binary orbital accelerations. The model must also correct for dispersive delays as the pulses propagate through the 
interstellar medium (ISM), which are radio frequency-dependent and
spread out the pulse fluence in phase. The timing model
is fit to align the pulse phase over many rotations of the pulsar, producing
an average "pulse profile." This profile is then cross-correlated with a "template"
profile (often a smoothed profile from previously measured noisy data or a sum of 
Gaussian components) to compute the phase offset from the start of the observation
from which one can measure the TOA at the center of the observation.
Subtracting the TOAs predicted from the timing model from the measured TOAs produces residuals
that can then be cross-correlated between pulsars in the PTA
in order to search for low-frequency noise due to the GWB.

The ISM causes broadening of the pulse profile, frequency-dependent dispersive delays
in the TOA, and fluctuations in the observed flux, effects which all depend
on observing frequency, $\nu$. While TOAs are typically measured at multiple frequencies
to quantify and model these effects, one can compute a single TOA unperturbed
by these effects referenced to an "infinite frequency," $t_\infty$.
The uncertainty on $t_\infty$ depends on the 
sensitivity and configuration of the telescope-receiver system, the 
intrinsic properties of the pulsar, and the properties of the ISM along 
the line-of-sight. The $t_\infty$ uncertainty is a good figure of merit for a
given pulsar-telescope sensitivity to achromatic timing effects such as GW. We adopt
the epoch-averaged TOA uncertainty described in \citet[][hereafter LMC18]{lam18_freqopt}
with a few notable changes, described below. 

\LMCt derive the TOA uncertainty from the sum of variances
of difference noise contributions to the measurement of $t_\infty$. The noise component variances
are themselves computed from $N_\nu \times N_\nu$ covariance matrices which describe the 
correlation structure between frequency subbands of the TOA measurement errors and
propagation-induced perturbations within a single observing epoch,
where $N_\nu $ is the number of subbands. In Equation 4,
they define $\sigmaTOA$ as the quadrature sum of the achromatic (in radio frequency)
 noise uncertainty $\sigma_W$, standard error on $t_\infty$ from a fit for DM $\sigmaDMhat$,
uncertainty due to scattering delays $\sigmadtC$, frequency-dependent DM difference 
uncertainty $\sigmaDMnu$,
 and polarization calibration errors $\sigmatel$.
However, $\sigmaDMhat$ includes the error on the $t_\infty$ fit for DM, which 
is computed from the achromatic noise covariance matrix. The result is that the achromatic noise
contributes twice. Instead, we define the TOA uncertainty as 
\begin{equation}\label{eq:sigmaTOA}
	\sigmaTOAsq = \sigmaDMhat^2 + \sigmadtC^2 + \sigmaDMnu^2 + \sigmatel^2,
\end{equation}
where the achromatic noise uncertainties from fitting a profile to a template,
intrisic pulse phase jitter, and diffractive interstellar scintillation (DISS)
are propagated through $\sigmaDMhat$.

We relax the assumption in \LMCt that the receiver temperature, $\Trcvr$, is constant 
over the passband. The receiver temperature contains frequency-dependent brightness-temperature 
contributions from spillover, the receiver feed, and the low-noise amplifier (LNA). We treat the
brightness temperature of the Galaxy as sky position-dependent and compute it from a 
lookup table based on the pulsar Galactic latitude and longitude \citep{psrpoppy, haslam82},
and scaled to the observing frequency according to \citep{lawson87}
\begin{equation}
	\Tgal(l,b,\nu) = \Tgal(l,b,0.408\;\mathrm{GHz})\left(\frac{\mathrm{\nu}}{0.408\;\mathrm{GHz}} \right)^{-2.6}.
\end{equation}
We also allow the telescope gain (degrees per flux unit) to be a function of observing frequency,
\begin{equation}
	\mathrm{G}(\nu) = \frac{A_\mathrm{eff}(\nu)}{2k},
\end{equation}
where $k$ is Boltzmann's constant and $\Aeff$ is the effective illuminated area of the dish.
However, in the instruments and frequency ranges we consider in this paper, $G(\nu)$ is approximately
constant.
This results in an expression for the template-fitting uncertainty that differs slightly from
\LMCt,
\begin{equation}
	\sigmaTF = \frac{\Tsys(\nu) / \mathrm{G}(\nu) }{S(\nu_0) \left(\frac{\nu}{\nu_0}\right)^{\alpha}
	 \sqrt{N_{\rm pol} \Delta \nu \tint}}
	 \left(\frac{\Weff\left(\nu|{\taud}_{,0}\right)}{\Uscale\left(\nu|{\taud}_{,0}\right)} \right),
	\label{eq:sigmaw}
\end{equation}
where 
\begin{equation}
	\Tsys(\nu)  =  T_{\rm CMB} + \Trcvr(\nu) + \Tgal(l,b,\nu),
	\label{eq:Tsys}
\end{equation}
is the system temperature, $T_{\rm CMB}=2.73\ \mathrm{K}$ is the brightness temperature of the cosmic microwave background, $\Delta \nu$ is the bandwidth, $\tint$ is the integration time, $N_{\rm pol} $ is the number 
of polarizations, and $S(\nu_0)(\nu/\nu_0)^{\alpha}$ is the period-averaged flux density
at reference frequency $\nu_0$ scaled to the observing frequency by the pulsar spectral index $\alpha$
\footnote{We use the opposite sign convention to \LMCt for $\alpha$.}.
The factor $1/\Uscale$
converts the period-averaged flux to a peak flux and $\Weff$
accounts for the shape of the pulse profile, since a profile with sharper features 
will reduce the template-fitting uncertainty. 
The strongest frequency-dependence in $\sigmaTF$ is from the pulsar spectral index,
followed by the shape of the receiver passband,
 a small contribution from the Galactic background at the lowest observing frequencies,
 and scatter-broadening of the profile that increases $\Weff$ and $\Uscale$.
$\sigmaTF$ affects $\sigmaTOA$
via the achromatic noise covariance matrix that is propagated through $\sigmaDMhat$:
\begin{equation}\label{eq:WNcov}
	\bf{C}_{\nu_1\nu_2} = \bf{S}_{\nu_1\nu_2} + \bf{J}_{\nu_1\nu_2} + \bf{D}_{\nu_1\nu_2},
\end{equation}
where $\bf{S}$ is the template-fitting covariance matrix, $\bf{J}$ is the jitter covariance matrix,
and $\bf{D}$ is the DISS covariance matrix.

Pulsar timing observations are typically conducted over a wide bandwidth in order 
to obtain accurate measurements of DM, which is necessary to align the pulses in frequency.
However, widely-separated frequencies probe slightly different ISM paths with different DM,
leading to a systematic error on $t_\infty$ \citep{cordes16_freqDM}.
\LMCt estimate this contribution,
$\sigmaDMnu$, to the TOA uncertainty by epoch-averaging a covariance matrix
constructed from the root-mean-square DM difference for a 
two-frequency fit for DM from \citet{cordes16_freqDM}. But their expression (Equation 24)
does not form a valid covariance matrix since the DM difference drives the diagonal elements
to zero when the ratio between frequencies approaches unity. The proper expression comes from the variance on an
$N_\nu$-point fit for DM. We take an approach similar to the one that \LMCt 
take for the chromatic scattering delays by propagating the DM-difference covariances through the
achromatic noise fit for DM. Using the two-frequency structure function in Appendix B of
\citet{cordes16_freqDM}, we find that $\sigmaDMnu^2$ is the first diagonal element of 
the covariance matrix for the DM fit
\begin{equation}
	\langle \delta {\bf \theta}_{\delta \DMbar} \delta {\bf \theta}_{\delta \DMbar}^T\rangle =	
			\left({\bf X}^T {\bf C}^{-1}{\bf X}\right)^{-1}{\bf X}^T{\bf C}^{-1}{\bf C}_{T,\delta \DMbar}
			{\bf C}^{-1}{\bf X}\left({\bf X}^T {\bf C}^{-1}{\bf X}\right)^{-1}\\[1em]
\end{equation}
where ${\bf X} = \mathrm{matrix}(1\; \nu_k^{-2})$,  ${\bf \theta} = \mathrm{matrix}(t_\infty\; K\DM)$, 
the dispersion constant $K\approx 4.149 \mathrm{ms\; GHz^2 \; pc^{-1}\; cm^{3}}$ \citep{cordes16_freqDM}, 
${\bf C}$ is the achromatic noise covariance matrix, and ${\bf C}_{T,\delta \DMbar}$ is the 
covariance matrix of the timing perturbation induced by the DM difference. See Appendix 
\ref{appx:DMnu_covmat} for a full derivation of $\sigmaDMnu$.

$\sigmaTOA$ encapsulates the tradeoffs in the choice of observing parameters. 
For example, observing a pulsar at higher frequencies will reduce uncertainties due to 
scattering \citep[$\propto\nu^{-4.4}$ for a Kolmogorov medium,][]{cordes10} and dispersion ($\propto\nu^{-2}$),
but dim or steep spectrum pulsars will have increased
template-fitting noise at those frequencies. \LMCt search for the optimal 
observing frequencies by searching for the minimum $\sigmaTOA$ 
over a grid of center frequencies and bandwidths.
One can build increasingly more sensitive
telescopes with wider bandwidths and optimize over frequency but
will eventually hit the noise floor set by $\sigmatel$ and jitter noise,
\begin{equation}
	\label{eq:sigmaj}
	\sigmaJ = \frac{\sigmaJsing}{\sqrt{\tint/P}},
\end{equation}
where $\sigmaJsing$ is the single-pulse jitter uncertainty and $P$ is the
spin period of the pulsar. 
Template-fitting and jitter uncertainty depend strongly on integration time;
both are $\propto 1/\sqrt{\tint}$.
Jitter is essentially the standard deviation on the mean pulse and can only be
reduced by integrating for longer durations on the pulsar. Under the assumption
that jitter is constant with frequency, the jitter covariance matrix $\bf{J}_{\nu_1\nu_2}$
will have all values $\sigmaJ^2$.
DISS also depends on $\tint$. In the weak scattering regime \LMCp,
\begin{align}
	\label{eq:sigmaDISS}
    {\bf D_{ij}} &= \sigmaDISS(\nu_i)\sigmaDISS(\nu_j)\\
    			&= \frac{
                   \tau_\mathrm{d}(\nu_i)
                   }
                  {
                   \sqrt{
                         \left(1 + \eta_t\frac{\tint}{\Delta t_\mathrm{d}(\nu_i)}\right)
                         \left(1 + \eta_\nu\frac{\Delta \nu}{\Delta \nu_\mathrm{d}(\nu_i)}\right)
                        }
                  }
             \frac{
                   \tau_\mathrm{d}(\nu_j)
                   }
                  {
                   \sqrt{
                         \left(1 + \eta_t\frac{\tint}{\Delta t_\mathrm{d}(\nu_j)}\right)
                         \left(1 + \eta_\nu\frac{\Delta \nu}{\Delta \nu_\mathrm{d}(\nu_j)}\right)
                        }
                  }
\end{align}
where $\taud(\nu_i)$ is the scattering delay timescale in a subband with center frequency $\nu_i$,
$\dtaud\propto \nu^{6/5}$ and $\dnud \propto \nu^{22/5}$ are the scintillation bandwidth and 
timescale,
$\Delta \nu$ is the observing bandwidth and $\etat$ and $\etanu$ are filling factors based on the 
scattering geometry. In the strong scattering regime, when the denominator of Equation \eqref{eq:sigmaDISS}
$\gg 1$,
\begin{equation}
	\sigmaDISS(\nu_i) \propto \frac{\nu_i^{-8/5}}{\sqrt{\tint}}.
\end{equation}
The frequency-dependent DM difference uncertainty, variations in pulse broadening due
to scattering delays, and polarization calibration errors are all constant with $\tint$.
We refer the reader to \LMCt for a more thorough discussion of the frequency-dependent
timing effects and we summarize the integration time and frequency dependence
of all timing effects included in $\sigmaTOA$ in Table \ref{tab:tint_depdt}.
We use the companion software to \LMCt,
FrequencyOptimizer \citep{FrequencyOptimizer} to perform all frequency optimization and
calculations of $\sigmaTOA$. The current version includes all of the modifications 
described above.

\begin{center}
{
\renewcommand{\arraystretch}{1.3}
\begin{deluxetable}{lcccc}
\tabletypesize{\scriptsize}
\tablecolumns{5}
\tablecaption{Integration Time and Frequency Dependence of $\sigmaTOA$\label{tab:tint_depdt}}
\tablehead{
\colhead{Noise Term} & \colhead{Symbol} & \colhead{$\tint$ Dependence} & \colhead{$\nu$ Dependence}
														       & \colhead{Equation(s)}\\
}
\startdata
Template-fitting & $\sigmaTF$ & $\propto 1/\sqrt{\tint}$ & 
	${\propto}\frac{\Tsys(\nu)}{G(\nu)}\left(\frac{\nu}{\nu_0}\right)^{\alpha}
	\frac{\Weff\left(\nu|{\taud}_{,0}\right)}{\Uscale\left(\nu|{\taud}_{,0}\right)}$ & \eqref{eq:sigmaw}\\
Jitter & $\sigmaJ$ & $\propto 1/\sqrt{\tint}$ & $\sim$ constant & \eqref{eq:sigmaj}\\
DISS & $\sigmaDISS$ & $\underset{\sim}{\propto} 1/\sqrt{\tint}$ & $\propto \nu^{-8/5}$& \eqref{eq:sigmaDISS}\\
DM fit Standard Error & $\sigmaDMhat$ & $\underset{\sim}{\propto} 1/\sqrt{\tint}$ & 
	$\propto \frac{\sqrt{{\bf C}_{ii} + r^4{\bf C}_{jj} - 2r^2{\bf C}_{ij}}}{r^2 - 1}$ &\LMCt Sec. 2.3.1\\
DM fit Scattering Error & $\sigmadtC$ & $\sim$ constant & 
			$\propto \frac{r^{4.4} - r^2}{r^2 - 1}\nuh^{-4.4}$ & \LMCt Eq. (23)\\
DM Difference Systematic Error & $\sigmaDMnu$ & constant & 
			$\propto \left| \frac{r^2}{r^2-1} \right| F_{11/3}(r)\nuh^{-17/6}$ & 
									\eqref{eq:DMstrucfunc} \& \eqref{eq:phiF1}\\
Polarization Calibration & $\sigmatel$ & constant & $\propto \varepsilon (\nu) \pi_\mathrm{V}(\nu) W_{50}(\nu)$ &
									\LMCt Eq. (26)\\
\enddata
\tablecomments{\hspace{1em}$r\equiv \nuh / \nul$}
\end{deluxetable}
}
\end{center}

\section{The GWB Signal-to-Noise Objective Function}\label{sec:gwb_snr}
If all we seek to do is optimize for a single pulsar, then minimizing $\sigmaTOA$ is
sufficient. If however, we want to maximize sensitivity to the GWB, we must also consider
the spatial correlations across all of our pulsars and the effective noise spectrum of
the PTA as a whole which includes red (in GW frequency) noise from both the GWB and
intrinsic pulsar red noise. We adopt the optimal cross-correlation
statistic for a stochastic GWB from \citet{hazboun19} as our objective function,
\begin{equation} \label{eq:optcrosscorrstat}
	\rho^2\simeq
	2\Tobs \int_0^{f_{\rm Nyq}}{\rm d}f\>
	\frac{S_h^2(f)}{S_{\rm eff}^2(f)},
\end{equation}
where $\rho$ is the GWB S/N,
$f$ is the GW frequency, and $\Tobs$ is the timespan of PTA data.
\begin{equation}\label{eq:gwb_strain_pwrspec}
	S_h(f)=\AGWB^2\left(\frac{f}{\fyr}\right)^{2\alphaGWB}\frac{1}{f}
\end{equation}
is the GWB strain noise power spectrum, with dimensionless amplitude $\AGWB$ and spectral
index $\alphaGWB$ \citep{ng12p5_gwb}.
We assume the Bayesian posterior GWB amplitude from \NGXVgwbt ,
$\AGWB = 2.4 \times 10^{-15}$, and assume $\alphaGWB = -2/3$ for an ensemble
of circular binaries.
The effective strain noise power spectral density for the entire PTA is 
\begin{equation}
	S_\mathrm{eff}(f) =
	\left(\sum_I\sum_{J>I}
	\frac{T_{IJ}}{\Tobs}
	\frac{\chi^2_{IJ}}{S_I(f)S_J(f)}
	\right)^{-1/2}
\end{equation}
where $T_{IJ}$ is the overlapping timespan between pulsars $I$ and $J$ and $\chi_{IJ}$ are the Hellings and Downs coefficients for pulsars $I$ and $J$, which depend only on their on-sky angular separation.
We assume 15 years of data for every pulsar so that $\Tobs = T_{IJ} = 15\;\mathrm{yr}$.
We solve for the optimal
integration time allocation per epoch vector across pulsars, $\topt$,
at a single telescope which maximizes the S/N:
\begin{equation}
    \max_{\tvec \in \mathbb{R}^n} \quad \rho^2({\tvec})
\end{equation}
subject to the constraints
\begin{align}
    & \sum_{i=1}^\Npsrs t_{\mathrm{int},i} = B, \label{eq:budget_constraint} \\
    & t_{\mathrm{min},i}\le t_{\mathrm{int},i} \le t_{\mathrm{max},i} \quad \forall i, \label{eq:box_constraints}
\end{align}
where $B$ is the time budget, the total available telescope time in an observing epoch. We let $\tmin = 60\;{\rm s}$.
A pulsar which is optimally at $\tmin$ suggests that it will not provide GWB sensitivity and
is effectively discouraged from inclusion in the PTA. $\tmax$ is the pulsar time above the
horizon per observing epoch, up to the duration of the epoch, which we define to be 30 days in this case. 
The integration time enters the objective only through the pulsar strain noise power-spectral density (PSD)
\begin{equation}
	S_I(f,\tint_{,i})\equiv \frac{1}{\NcalInv_I(f,\tint_{,i})\Rcal(f)}
\end{equation}
where $\NcalInv_I(f,\tint_{,i})$ is the inverse-noise-weighted transmission function,
which accounts for the frequencies "absorbed" by the timing model fit.
\begin{equation}
	\Rcal(f) = \frac{1}{12\pi^2 f^2}
\end{equation} 
is the timing residual response of a pulsar to plane GWs averaged over source sky location, ($+$,$\times$) polarizations, and polarization angle, independent of the pulsar noise and direction \citep{hazboun19}.
Computing $\NcalInv_I(f,\tint_i)$ requires inverting a large $N_f \times N_f$ matrix (where $N_f$ is the number of GW frequency bins) for each pulsar
for each optimization iteration, so we use the approximation
\begin{align}
   \NcalInv(f,\tint_i)_I \approx \frac{\Tcal(f)}{P_I(f,\tint_i)}\\
   S_I(f,\tint_i)\approx \frac{P_I(f,\tint_i)}{\Tcal(f)\Rcal(f)}
\end{align}
where $\Tcal(f)$ depends only on the timing model and not the noise and $P_I(f,\tint_i)$
is the \textit{TOA residual power-spectral density} for pulsar $I$. Using this approximation
we need only compute the pulsar timing noise spectrum at each iteration however, whenever
we present the S/N for an observing configuration, $\rho(\tvec)$,
it is using the exact expression for $S_I$.
Our model of the pulsar residual PSD contains white noise, red
GWB self-noise, and spatially-uncorrelated, intrinsic pulsar red noise:
\begin{equation}\label{eq:psr_psd}
	    P_I(f,\tint_i) = 2\sigmaTOA_{,I}(\tint_{,i})^2\Delta t + \
	    \frac{\AGWB^2}{12\pi^2}\left(\frac{f}{\fyr} \right)^{-\gammaGWB}\fyr^{-3} + \
		\frac{\Ared^2}{12\pi^2}\left(\frac{f}{\fyr} \right)^{-\gammared}\fyr^{-3},
\end{equation}
where $\gammaGWB = 3 - 2\alphaGWB$ and $\Ared$ is the pulsar intrinsic red noise strain amplitude. 
We assume that the cadence of observations, 
$\Delta t = 365.26\ \mathrm{days} / 12\ \mathrm{epochs} = 30.44 \; \mathrm{days}$.

Brute force grid searches for $\topt$ with a PTA of only a few ($\leq 5$) pulsars
revealed that $\rho^2(\tvec)$ is generally a smooth, monotonically increasing i.e.
non-convex function. We use the trust-region algorithm \citep{conn00_trustregion}
offered in \texttt{scipy.optimize} package to minimize the scaled (by the equal-time-allocation
$\SNR^2$) objective, $F(\tvec)=-\rho^2(\tvec)/\rho^2(\teq)$. This is a quasi-Newton method that uses
curvature information to construct an approximate
model of the objective around the current point that is valid within a dynamic trust-region 
radius. At iteration $k$, a candidate step $s$ is found by solving the 
subproblem \citep{nocedalwright}
\begin{equation}
	\min_s m_k(t_k + s), \quad \text{where\ } t_k+s \text{\ lies inside the trust region}
\end{equation}
where the local model of the objective is approximated as
\begin{equation}
	m_k(t_k + s) = F_k + s^T\nabla F_k + \frac{1}{2}s^T{\bf B}_k s
\end{equation}
and ${\bf B}$ is the approximation to the Hessian matrix.
To make derivatives of $F(\tvec)$ analytically tractable and fast to compute, we write the objective
in the pseudo-quadratic form
\begin{equation}\label{eq:pseudoquadratic}
	F(\tvec) = \tildep(\tvec)^T\tildeQmat \tildep(\tvec),
\end{equation}
where $\tildeQmat$ is an $\Npsrs N_f \times\Npsrs N_f$, block-diagonal matrix that can be pre-computed for a PTA and
which contains all of the constants that don't depend on $\tint$.
Equation \eqref{eq:pseudoquadratic} is pseudo-quadratic because $F(\tvec)$
is not quadratic in $\tint$, but is quadratic in the frequency-discretized inverse pulsar PSDs,
\begin{equation}
	\tildep_{(k-1)\Npsrs + I} \equiv a_I(f_k,\tint_{,i}) = \frac{1}{S_I(f_k,\tint_{,i})}\; \mathrm{for}\ k=1,...,N_f,
\end{equation}
where $\tildep$ is an $\Npsrs N_f$-length vector.
Therefore, the gradient is
\begin{equation}
       (\nabla F(\tvec))_i = -8\Delta t \sigmaTOA_{,i}(\tint_{,i})\sigmaTOA_{,i}'(\tint_{,i})
                                    \sum_{k=1}^{N_f} \left[(Q_k a_k)_i \right]\frac{a_i(f_k;\tint_{,i})}
                                                {P_i(f_k;\tint_{,i})}.
\end{equation}
In general, $\sigmaTOA(\tint)$ is smooth and monotonically decreasing, so we pre-compute a lookup
table of $\sigmaTOA(\tint_{,k})$ and use PCHIP (Piecewise Cubic Hermite Interpolating Polynomial) interpolation to compute 
$\sigmaTOA_{,i}(\tint_{,i})$
and $\sigmaTOA_{,i}'(\tint_{,i})$ for further computational efficiency. See Appendix \ref{appx:GWB_SN_QP}
for a full derivation of the quadratic form and derivatives. We use the package \texttt{hasasia} 
\citep{hasasia} for modeling the PTA and computations of Equation \eqref{eq:optcrosscorrstat} and pulsar
PSDs. 

The trust region optimizer will terminate when the norm of the Lagrangian gradient, the change in $\tvec$,
or the constraint violation are below their respective tolerances, but this does not guarantee
an optimal solution. While we cannot prove conclusively that $\topt$ is a global optimum,
we perform three tests to verify the stationarity of the solution:
\begin{enumerate}
	\item $\topt$ is numerically consistent with the Karush-Kuhn-Tucker (KKT) conditions 
	\citep{karush39, kuhn_tucker51} necessary for a local optimum. See Appendix \ref{appx:kktresid}.
	\item Many multi-start optimizers where the intial vectors, $\tvec\mkern-2mu{}^0$,
	are uniformly sampled from the budget surface
	produce highly constrained distributions of  $\tint_{,i}^*$ and $\rho(\topt)$.
	\item Random, feasible time-swaps between two pulsars away from $\topt$ always result in 
	$\rho(\vec{t}_\mathrm{swap}) < \rho(\topt)$.
\end{enumerate}

While unlikely, it is possible that an extraordinarily bright pulsar with little intrinsic red noise could
have $\topt \leq 60\ \mathrm{s}$ and still contribute sensitivity to the PTA. Setting $\tmin$ much below 60 s can
potentially create pathological geometries for the optimizer when $\Npsrs$ is large, so we perform an additional
sanity check: we compute the relative S/N loss from dropping each pulsar from the PTA,
\begin{equation}
	\Delta \rho (\topt)_{\rm i} = \frac{\rho_\mathrm{full} - \rho_{-i}}{\rho_\mathrm{full}}
\end{equation}
and verify that pulsars with $\tint^*_{,i} = \tmin$ have the lowest $\Delta \rho (\topt)_{\rm i}$.
The optimization framework described above is available as a python package at \url{https://github.com/tycohen/PTAOptimizer}.
\section{Modeling the NANOGrav PTA}\label{sec:psr-tele}
\subsection{Pulsar Models}\label{ssec:psrmodels}
We consider a PTA consisting of 85 pulsars. While a DSA PTA is expected to be
much larger, this represents a plausible source list during commissioning and early science.
In addition to the aforementioned 78 pulsars
that are currently being timed by NANOGrav, we also include five pulsars that were dropped 
due to poor timing performance after the loss of the 300-m Arecibo telescope in
2020: J0621+2514, J1327+3423, J1630+3550, J1312+0051, and J0557+1551. We also 
include J0218+4232, which has not been timed since the NANOGrav 9-year dataset
and J0102+4839, a new Fermi Large Area Telescope unassociated gamma-ray source 
confirmed with the GBT \citep{gbt350_fermi}. We include these pulsars so that we
may test whether they become capable of contributing sensitivity to the PTA with 
optimized observing parameters and a more powerful next-generation telescope.

Data for these pulsars are aggregated from a variety of sources. Spin period ($P$),
period derivative ($\dot{P}$), DM, and sky coordinates for 68 pulsars are taken from the 
\NGXVt timing models or from the most recent NANOGrav timing model for the rest. 
$\dtaud$, $\dnud$, $\taud$, and distance $d$, were all estimated from the DM using
the NE2001 electron density map \citep{ne2001}. $\Weff$, $\Uscale$, and $W_{50}$
were computed according to \LMCt from \NGXVt narrow-band, L-band profile templates (or the 
most recent NANOGrav template), except for 
J1102+0249, J1852--1310, and J0102+4839 for which profile templates were constructed from archival GBT data.
To estimate the flux densities at 1 GHz, $S_{1000}$, for each pulsar 
the median flux at each frequency band was computed from all \NGXVt observations
that produced a TOA where the S/N of the profile was $\geq 8 $. 
Spectral indices, $\alpha$, were computed as $\logten(\text{flux ratio}) / \logten(\nu\; \text{ratio})$
using either (in order of precedence)
$800 \MHz/ 1400 \MHz$ for GBT observations, $430 \MHz / 1400 \MHz$ for Arecibo
observations where available, or $1400 \MHz / 2000 \MHz$ with the GBT or VLA.
Then the L-band flux (at Arecibo or the GBT), $S_{1400}$, was scaled to 1 GHz
\begin{equation}\label{eq:1GHzflux}
	S_{1000} = S_{1400} \left(\frac{1000 \MHz}{1400 \MHz} \right)^{\alpha}.
\end{equation}
We are somewhat limited by modeling pulsar spectral energy densities as a power-law, extrapolated over
a wide frequency range. While most MSP spectra appear to follow a power law, \citet{kuniyoshi15} 
found evidence for low-frequency spectral turn-over in a small number of MSPs and \citet{al22} 
found strong evidence in favour of the MSP population having 
a two-component power-law spectral model with a positive spectral index below 300 MHz
and a negative spectral index above. Additionally, excluding flux measurements that produced
low S/N profiles can potentially bias the distributions from which we estimate flux and spectral index.
For these reasons we set an upper limit on our spectral index computation of $\alpha = -1$.
We attempt to alleviate these assumptions' impact on the optimal integration time through Monte Carlo draws
of fluxes and spectral indices (and other pulsar attributes) from plausible distributions, described in
Section \ref{sec:opt-tint}.
A few pulsars did not have sufficient observations to perform
this flux computation, so we used published fluxes and spectra indices:
J0218+4232 \citep{kramer98, kuniyoshi15},
J0621+2514 \citep[2-frequency fit using 820 MHz and 1.4 GHz from][]{sanpaarsa16, spiewak22_meertime},
J0102+4839 \citep[2-frequency fit using 320 MHz and 820 MHz from][]{bangale24},
J2205+6012 \citep{nancay22_J2205}, J1911--1114 \citep{psrcat}, and J0437--4715 \citep{psrcat}.
Spectral indices for J1102+0249 and J1852--1310 were assumed to be the high-frequency population mean, --1.47,
from model D in \citet{al22} and their fluxes were scaled to 1 GHz from 820 MHz and 1.5 GHz GBT 
observations respectively.
Single-pulse jitter estimates for pulsars in the NANOGrav 12.5-year dataset
were taken from the constant-in-frequency models in \citet{lam19_jitter}.
For pulsars not in that dataset, we use the scaling relationship between
jitter and pulse duty cycle, $W_{50}/P$, shown in Figure 7 of the same paper:
\begin{equation}\label{eq:jitterscaling}
	\sigmaJsing = 10^{-0.218}\left(\frac{W_{50}}{P}\right)^{1.216}P.
\end{equation}
Finally, we chose to assume intrinsic red timing noise is present in all pulsars.
NANOGrav has measured significant red noise in only 12 pulsars so far, after accounting 
for a red noise process that is common to every pulsar, consistent with a GWB \citep{ng15_noisebudget}.
As longer pulsar data time spans will increase low-frequency GW sensitivity
and more sensitive telescopes like DSA beat down the pulsar white noise,
we anticipate measuring intrinsic red noise in an increasing number of pulsars. 
For each pulsar we assumed $\Ared$ from the mean of the distribution
implied by \citet{sc10_rednoise},
with updated parameters from an excess noise (EN) analysis performed by
\citet{ng9_excessnoise}, parameterized by $P$ and 
the intrinsic period derivative corrected for transverse motion
of the pulsar \citep{shklovskii, nicetaylor95}, $\Pdotint$:
\begin{equation}\label{eq:sc10RN}
		\Ared = \frac{\sqrt{12}\pi}{3.154\times 10^{13}}
				(\gammared-1)^{1/2} C_2 P^{-\alpha_\mathrm{EN}}
					\left(\frac{|\Pdotint|}{P^2}\right)^{\beta_\mathrm{EN}},
\end{equation}
where
$\ln C_2 = -1.2$, 
$\alpha_\mathrm{EN} = -0.9$, and $\beta_\mathrm{EN} = 0.8$.
This model assumes a fixed spectral index for all pulsars, $\gammared = 5.6$.
While this approach attempts to avoid overcounting GWB contributions to red noise
(which we include separately through $\AGWB$ and $\gammaGWB$), the red noise measurements
used to constrain the distribution of $\Ared$ were collected before a common red process
was measured and therefore may still be contaminated by the GWB.
See Appendix \ref{appx:rednoise} for the derivation of the implied $\Ared$ distribution. 
All assumed parameters are shown in Table \ref{tab:NG20_optinstr_pulsar_params}.

\newcommand{\InsertPsrtabCaption}{Spin periods and DMs were taken from NANOGrav timing data.
$\taud$, $\dtaud$, $\dnud$, and DM-distance $d$ were estimated from \citet{ne2001}. Fluxes
and spectral indices were computed from NANOGrav data as described in Section \ref{ssec:psrmodels}
except where noted otherwise. Profile shape parameters $\Weff$, $\Uscale$, and $W_{50}$ were 
computed from NANOGrav profile templates or smoothed profile data. Single-pulse jitter was taken
from \citet{lam19_jitter}, except where noted it was computed using Equation \eqref{eq:jitterscaling}.
Red noise amplitude is compute from Equation \eqref{eq:sc10RN} and a single red noise spectral index is
assumed for all pulsars.}
\begin{deluxetable}{lcccccccccccccc}
\tabletypesize{\scriptsize}
\tablecaption{Adopted Pulsar Parameters\label{tab:NG20_optinstr_pulsar_params}}
\tablehead{
\colhead{Pulsar} & \colhead{$P$} & \colhead{DM} & \colhead{$\tau_d$} & \colhead{$\Delta \tau_d$} & \colhead{$\Delta \nu_d$} & \colhead{$d$} & \colhead{$S_{1000}$} & \colhead{$\alpha$} & \colhead{$W_{50}$} & \colhead{$W_\mathrm{eff}$} & \colhead{$U_\mathrm{scale}$} & \colhead{$\sigma_\mathrm{J,1}$} & \colhead{$A_\mathrm{red}$} & \colhead{$\gamma_\mathrm{red}$}\\
\colhead{} & \colhead{(ms)} & \colhead{($\mathrm{pc\,cm^{-3}}$)} & \colhead{($\mathrm{\mu s}$)} & \colhead{(s)} & \colhead{(GHz)} & \colhead{(kpc)} & \colhead{(mJy)} & \colhead{} & \colhead{($\mathrm{\mu s}$)} & \colhead{($\mathrm{\mu s}$)} & \colhead{} & \colhead{($\mathrm{\mu s}$)} & \colhead{} & \colhead{}\\
}
\startdata
J0023+0923 & 3.05 & 14.33 & $1.5 \times 10^{-2}$ & 746.9 & $1.3 \times 10^{-2}$ & 0.69 & 0.61 & -1.67 & 201.68 & 431.69 & 11.63 & 8.16 & $1.33 \times 10^{-15}$ & 5.60\\
J0030+0451 & 4.87 & 4.33 & $3.4 \times 10^{-4}$ & 3343.0 & $5.4 \times 10^{-1}$ & 0.32 & 1.98 & -1.37 & 507.27 & 543.89 & 7.80 & 131.68 & $9.29 \times 10^{-16}$ & 5.60\\
J0102+4839 & 2.96 & 53.50 & $2.9 \times 10^{-1}$ & 306.0 & $6.3 \times 10^{-4}$ & 2.29 & 0.60$^{\rm a}$ & -1.47$^{\rm b}$ & 301.66 & 621.32 & 7.99 & 111.47 & $1.46 \times 10^{-15}$ & 5.60\\
J0125--2327 & 3.68 & 9.57 & $6.2 \times 10^{-3}$ & 913.6 & $3.0 \times 10^{-2}$ & 0.43 & 2.81 & -1.00 & 115.21 & 467.30 & 15.20 & 33.01 & $1.36 \times 10^{-15}$ & 5.60\\
J0154+1833 & 2.36 & 19.80 & $3.3 \times 10^{-2}$ & 557.2 & $5.6 \times 10^{-3}$ & 0.86 & 0.31 & -2.38 & 42.13 & 256.06 & 25.05 & 10.68 & $3.38 \times 10^{-16}$ & 5.60\\
J0218+4232 & 2.32 & 61.24 & $4.1 \times 10^{-1}$ & 279.3 & $4.5 \times 10^{-4}$ & 2.67 & 2.00$^{\rm c}$ & -2.41$^{\rm d}$ & 947.54 & 479.51 & 2.69 & 472.57 & $7.95 \times 10^{-15}$ & 5.60\\
J0340+4130 & 3.30 & 49.58 & $2.9 \times 10^{-1}$ & 265.5 & $6.3 \times 10^{-4}$ & 1.73 & 1.01 & -1.86 & 211.14 & 499.75 & 10.56 & 265.04 & $9.02 \times 10^{-16}$ & 5.60\\
J0406+3039 & 2.61 & 49.37 & $2.7 \times 10^{-1}$ & 280.0 & $7.0 \times 10^{-4}$ & 1.74 & 0.93 & -1.45 & 72.21 & 358.17 & 12.85 & 20.14 & $1.22 \times 10^{-15}$ & 5.60\\
J0437--4715 & 5.76 & 2.64 & $1.1 \times 10^{-3}$ & 1258.0 & $1.8 \times 10^{-1}$ & 0.14 & 210.00$^{\rm e}$ & -1.00$^{\rm f}$ & 138.92 & 662.08 & 16.26 & 37.62 & $1.35 \times 10^{-15}$ & 5.60\\
J0509+0856 & 4.06 & 38.33 & $1.5 \times 10^{-1}$ & 343.8 & $1.3 \times 10^{-3}$ & 1.45 & 2.43 & -1.43 & 639.56 & 885.89 & 3.82 & 259.78 & $5.02 \times 10^{-16}$ & 5.60\\
J0557+1551 & 2.56 & 102.58 & 4.9 & 84.9 & $3.8 \times 10^{-5}$ & 2.92 & 0.06 & -1.00 & 90.02 & 361.08 & 19.66 & 26.45 & $1.09 \times 10^{-15}$ & 5.60\\
J0605+3757 & 2.73 & 20.95 & $6.0 \times 10^{-2}$ & 372.0 & $3.1 \times 10^{-3}$ & 0.70 & 0.83 & -1.76 & 150.97 & 452.04 & 11.34 & 48.91 & $7.09 \times 10^{-16}$ & 5.60\\
J0610--2100 & 3.86 & 60.65 & $2.9 \times 10^{-1}$ & 379.9 & $6.3 \times 10^{-4}$ & 3.54 & 1.32 & -1.46 & 185.33 & 576.87 & 13.15 & 58.22 & $1.29 \times 10^{-15}$ & 5.60\\
J0613--0200 & 3.06 & 38.78 & $9.7 \times 10^{-2}$ & 458.8 & $1.9 \times 10^{-3}$ & 1.71 & 4.28 & -2.02 & 272.91 & 333.20 & 7.17 & 47.01 & $1.09 \times 10^{-15}$ & 5.60\\
J0614--3329 & 3.15 & 37.05 & $7.0 \times 10^{-2}$ & 568.5 & $2.6 \times 10^{-3}$ & 1.90 & 1.49 & -1.43 & 73.08 & 296.32 & 8.31 & 19.62 & $1.95 \times 10^{-15}$ & 5.60\\
J0621+2514 & 2.72 & 83.63 & 1.9 & 121.9 & $9.8 \times 10^{-5}$ & 2.33 & 0.17$^{\rm g}$ & -2.35$^{\rm h}$ & 155.36 & 472.64 & 9.72 & 50.67 & $2.85 \times 10^{-15}$ & 5.60\\
J0636+5128 & 2.87 & 11.10 & $1.1 \times 10^{-2}$ & 719.8 & $1.6 \times 10^{-2}$ & 0.49 & 1.44 & -1.70 & 128.41 & 458.99 & 18.53 & 58.32 & $5.67 \times 10^{-16}$ & 5.60\\
J0645+5158 & 8.85 & 18.25 & $3.6 \times 10^{-2}$ & 480.5 & $5.1 \times 10^{-3}$ & 0.70 & 0.58 & -1.33 & 118.11 & 621.93 & 42.90 & 10.82 & $2.96 \times 10^{-16}$ & 5.60\\
J0709+0458 & 34.43 & 44.28 & $1.7 \times 10^{-1}$ & 351.9 & $1.1 \times 10^{-3}$ & 1.79 & 0.42 & -1.00 & 884.30 & 2751.21 & 30.81 & 242.71 & $4.33 \times 10^{-15}$ & 5.60\\
J0732+2314 & 4.09 & 44.67 & $2.1 \times 10^{-1}$ & 309.1 & $8.9 \times 10^{-4}$ & 1.66 & 1.31 & -1.42 & 946.21 & 648.25 & 3.72 & 417.51 & $7.56 \times 10^{-16}$ & 5.60\\
J0740+6620 & 2.89 & 14.96 & $1.9 \times 10^{-2}$ & 658.8 & $9.8 \times 10^{-3}$ & 0.68 & 1.62 & -1.37 & 132.91 & 271.36 & 8.71 & 30.69 & $1.02 \times 10^{-15}$ & 5.60\\
J0751+1807 & 3.48 & 30.25 & $9.0 \times 10^{-2}$ & 391.5 & $2.1 \times 10^{-3}$ & 1.15 & 1.55 & -1.00 & 174.66 & 503.38 & 10.24 & 55.41 & $7.77 \times 10^{-16}$ & 5.60\\
J0931--1902 & 4.64 & 41.49 & $1.5 \times 10^{-1}$ & 393.7 & $1.3 \times 10^{-3}$ & 1.88 & 1.43 & -1.58 & 352.42 & 412.07 & 6.73 & 93.95 & $3.77 \times 10^{-16}$ & 5.60\\
J1012--4235 & 3.10 & 71.65 & $5.8 \times 10^{-1}$ & 228.5 & $3.2 \times 10^{-4}$ & 2.51 & 0.60 & -1.00 & 143.35 & 444.53 & 15.27 & 44.67 & $8.02 \times 10^{-16}$ & 5.60\\
J1012+5307 & 5.26 & 9.02 & $7.1 \times 10^{-3}$ & 830.8 & $2.6 \times 10^{-2}$ & 0.41 & 5.15 & -1.53 & 586.56 & 629.73 & 6.68 & 26.04 & $1.13 \times 10^{-15}$ & 5.60\\
\enddata
\tablecomments{\InsertPsrtabCaption}
\tablenotetext{a}{\vspace{-2ex}scaled from 820 MHz \citep{bangale24}}
\tablenotetext{b}{\vspace{-2ex}2-$\nu$ fit from \citet{bangale24}}
\tablenotetext{c}{\vspace{-2ex}scaled from 1.4 GHz \citep{kramer98}}
\tablenotetext{d}{\vspace{-2ex}\citet{kuniyoshi15}}
\tablenotetext{e}{\vspace{-2ex}\citet{psrcat}}
\tablenotetext{f}{\vspace{-2ex}\citet{psrcat}}
\tablenotetext{g}{\vspace{-2ex}scaled from 1.44 GHz \citep{spiewak22_meertime}}
\tablenotetext{h}{\vspace{-2ex}2-$\nu$ fit from \citet{sanpaarsa16, spiewak22_meertime}}
\end{deluxetable}
\addtocounter{table}{-1}
\begin{deluxetable}{lcccccccccccccc}
\tabletypesize{\scriptsize}
\tablecaption{continued}
\tablehead{
\colhead{Pulsar} & \colhead{$P$} & \colhead{DM} & \colhead{$\tau_d$} & \colhead{$\Delta \tau_d$} & \colhead{$\Delta \nu_d$} & \colhead{$d$} & \colhead{$S_{1000}$} & \colhead{$\alpha$} & \colhead{$W_{50}$} & \colhead{$W_\mathrm{eff}$} & \colhead{$U_\mathrm{scale}$} & \colhead{$\sigma_\mathrm{J,1}$} & \colhead{$A_\mathrm{red}$} & \colhead{$\gamma_\mathrm{red}$}\\
\colhead{} & \colhead{(ms)} & \colhead{($\mathrm{pc\,cm^{-3}}$)} & \colhead{($\mathrm{\mu s}$)} & \colhead{(s)} & \colhead{(GHz)} & \colhead{(kpc)} & \colhead{(mJy)} & \colhead{} & \colhead{($\mathrm{\mu s}$)} & \colhead{($\mathrm{\mu s}$)} & \colhead{} & \colhead{($\mathrm{\mu s}$)} & \colhead{} & \colhead{}\\
}
\startdata
J1022+1001 & 16.45 & 10.27 & $8.3 \times 10^{-3}$ & 807.8 & $2.2 \times 10^{-2}$ & 0.45 & 6.12 & -1.08 & 715.59 & 1395.24 & 16.42 & 126.33 & $1.28 \times 10^{-15}$ & 5.60\\
J1024--0719 & 5.16 & 6.49 & $1.0 \times 10^{-3}$ & 2140.0 & $1.8 \times 10^{-1}$ & 0.39 & 2.00 & -1.02 & 139.97 & 572.62 & 8.58 & 33.87 & $1.71 \times 10^{-16}$ & 5.60\\
J1102+0249 & 4.05 & 27.37 & $7.1 \times 10^{-2}$ & 466.8 & $2.6 \times 10^{-3}$ & 1.29 & 0.14 & -1.47$^{\rm i}$ & 159.49 & 529.94 & 14.94 & 48.00 & $9.60 \times 10^{-16}$ & 5.60\\
J1125+7819 & 4.20 & 11.22 & $7.3 \times 10^{-3}$ & 1018.0 & $2.5 \times 10^{-2}$ & 0.63 & 1.61 & -1.98 & 479.70 & 651.30 & 8.27 & 124.67 & $2.87 \times 10^{-16}$ & 5.60\\
J1312+0051 & 4.23 & 15.34 & $2.3 \times 10^{-2}$ & 657.2 & $8.0 \times 10^{-3}$ & 0.83 & 0.66 & -1.34 & 257.74 & 650.04 & 9.92 & 85.26 & $1.21 \times 10^{-15}$ & 5.60\\
J1327+3423 & 41.51 & 4.18 & $4.6 \times 10^{-4}$ & 3512.0 & $4.0 \times 10^{-1}$ & 0.48 & 0.83 & -1.68 & 714.90 & 4151.87 & 38.01 & 179.99 & $1.65 \times 10^{-15}$ & 5.60\\
J1453+1902 & 5.79 & 14.06 & $1.8 \times 10^{-2}$ & 871.3 & $1.0 \times 10^{-2}$ & 1.15 & 0.58 & -1.44 & 296.15 & 862.38 & 11.49 & 265.45 & $8.08 \times 10^{-16}$ & 5.60\\
J1455--3330 & 7.99 & 13.57 & $5.6 \times 10^{-3}$ & 1062.0 & $3.3 \times 10^{-2}$ & 0.53 & 1.48 & -1.79 & 210.21 & 996.06 & 14.74 & 101.23 & $1.30 \times 10^{-15}$ & 5.60\\
J1600--3053 & 3.60 & 52.34 & $5.6 \times 10^{-1}$ & 187.5 & $3.3 \times 10^{-4}$ & 1.63 & 2.93 & -1.00 & 69.52 & 423.25 & 21.91 & 26.24 & $1.03 \times 10^{-15}$ & 5.60\\
J1614--2230 & 3.15 & 34.49 & $1.7 \times 10^{-1}$ & 295.5 & $1.1 \times 10^{-3}$ & 1.27 & 1.88 & -1.60 & 83.48 & 393.62 & 16.13 & 64.95 & $1.22 \times 10^{-15}$ & 5.60\\
J1630+3550 & 3.23 & 17.45 & $3.0 \times 10^{-2}$ & 669.7 & $6.2 \times 10^{-3}$ & 1.12 & 0.33 & -2.37 & 187.97 & 593.97 & 9.35 & 61.56 & $2.23 \times 10^{-15}$ & 5.60\\
J1630+3734 & 3.32 & 14.13 & $1.7 \times 10^{-2}$ & 801.9 & $1.1 \times 10^{-2}$ & 0.94 & 1.04 & -1.83 & 97.41 & 281.36 & 9.77 & 27.52 & $1.01 \times 10^{-15}$ & 5.60\\
J1640+2224 & 3.16 & 18.43 & $3.4 \times 10^{-2}$ & 634.9 & $5.4 \times 10^{-3}$ & 1.16 & 1.09 & -2.03 & 219.27 & 463.36 & 14.27 & 27.88 & $2.94 \times 10^{-16}$ & 5.60\\
J1643--1224 & 4.62 & 62.40 & $5.4 \times 10^{-1}$ & 231.5 & $3.4 \times 10^{-4}$ & 2.40 & 8.64 & -1.84 & 313.55 & 974.59 & 10.26 & 25.90 & $1.48 \times 10^{-15}$ & 5.60\\
J1705--1903 & 2.48 & 57.50 & $7.7 \times 10^{-1}$ & 160.5 & $2.4 \times 10^{-4}$ & 1.65 & 0.87 & -1.00 & 39.58 & 226.66 & 51.67 & 9.80 & $2.70 \times 10^{-15}$ & 5.60\\
J1713+0747 & 4.57 & 15.99 & $2.4 \times 10^{-2}$ & 663.7 & $7.6 \times 10^{-3}$ & 0.89 & 6.63 & -1.00 & 111.61 & 544.62 & 26.15 & 32.03 & $8.21 \times 10^{-16}$ & 5.60\\
J1719--1438 & 5.79 & 36.79 & $2.1 \times 10^{-1}$ & 262.9 & $8.8 \times 10^{-4}$ & 1.21 & 1.73 & -1.85 & 294.74 & 819.92 & 13.86 & 93.78 & $6.40 \times 10^{-16}$ & 5.60\\
J1730--2304 & 8.12 & 9.63 & $9.3 \times 10^{-3}$ & 824.8 & $2.0 \times 10^{-2}$ & 0.53 & 6.00 & -1.96 & 371.26 & 887.79 & 8.99 & 115.41 & $9.14 \times 10^{-16}$ & 5.60\\
J1738+0333 & 5.85 & 33.77 & $1.2 \times 10^{-1}$ & 371.2 & $1.5 \times 10^{-3}$ & 1.43 & 0.70 & -1.00 & 118.42 & 639.67 & 16.54 & 31.44 & $1.56 \times 10^{-15}$ & 5.60\\
J1741+1351 & 3.75 & 24.20 & $5.1 \times 10^{-3}$ & 1452.0 & $3.6 \times 10^{-2}$ & 0.90 & 0.56 & -2.08 & 83.90 & 386.37 & 28.12 & 41.52 & $2.62 \times 10^{-15}$ & 5.60\\
J1744--1134 & 4.07 & 3.14 & $1.2 \times 10^{-4}$ & 6504.0 & 1.6 & 0.41 & 3.57 & -1.80 & 137.44 & 511.35 & 27.15 & 42.03 & $8.00 \times 10^{-16}$ & 5.60\\
J1745+1017 & 2.65 & 23.97 & $4.9 \times 10^{-2}$ & 557.5 & $3.8 \times 10^{-3}$ & 1.26 & 1.08 & -1.20 & 266.43 & 389.86 & 8.41 & 98.18 & $3.68 \times 10^{-16}$ & 5.60\\
J1747--4036 & 1.65 & 152.97 & 13.6 & 54.7 & $1.4 \times 10^{-5}$ & 3.39 & 3.97 & -2.72 & 97.68 & 277.72 & 5.37 & 957.28 & $2.45 \times 10^{-15}$ & 5.60\\
J1751--2857 & 3.91 & 42.79 & $3.8 \times 10^{-1}$ & 186.0 & $4.8 \times 10^{-4}$ & 1.10 & 1.41 & -2.22 & 132.59 & 546.26 & 18.85 & 38.63 & $1.10 \times 10^{-15}$ & 5.60\\
J1802--2124 & 12.65 & 149.62 & 81.9 & 20.8 & $2.3 \times 10^{-6}$ & 2.94 & 2.24 & -2.40 & 178.64 & 1034.13 & 52.79 & 43.09 & $1.92 \times 10^{-15}$ & 5.60\\
J1803+1358 & 1.52 & 59.16 & $3.8 \times 10^{-1}$ & 290.5 & $4.8 \times 10^{-4}$ & 2.69 & 1.04 & -3.06 & 95.89 & 181.66 & 6.73 & 31.94 & $6.81 \times 10^{-17}$ & 5.60\\
J1811--2405 & 2.66 & 60.62 & 2.0 & 103.1 & $9.2 \times 10^{-5}$ & 1.77 & 2.92 & -1.88 & 109.17 & 313.05 & 11.78 & 33.15 & $1.57 \times 10^{-15}$ & 5.60\\
J1832--0836 & 2.72 & 28.20 & $1.1 \times 10^{-1}$ & 354.9 & $1.7 \times 10^{-3}$ & 1.11 & 2.21 & -2.23 & 47.69 & 189.53 & 13.53 & 57.31 & $7.25 \times 10^{-16}$ & 5.60\\
J1843--1113 & 1.85 & 59.96 & 1.0 & 140.4 & $1.8 \times 10^{-4}$ & 1.70 & 1.12 & -1.75 & 66.33 & 253.10 & 20.08 & 19.57 & $1.75 \times 10^{-15}$ & 5.60\\
J1852--1310 & 4.31 & 44.95 & $3.7 \times 10^{-1}$ & 216.4 & $5.1 \times 10^{-4}$ & 1.43 & 0.55 & -1.47$^{\rm j}$ & 130.67 & 495.63 & 12.65 & 37.16 & $1.03 \times 10^{-15}$ & 5.60\\
\enddata
\tablenotetext{i}{\vspace{-2ex}high-freq. population mean \citet{al22}}
\tablenotetext{j}{\vspace{-2ex}high-freq. population mean \citet{al22}}
\end{deluxetable}
\addtocounter{table}{-1}
\begin{deluxetable}{lcccccccccccccc}
\tabletypesize{\scriptsize}
\tablecaption{continued}
\tablehead{
\colhead{Pulsar} & \colhead{$P$} & \colhead{DM} & \colhead{$\tau_d$} & \colhead{$\Delta \tau_d$} & \colhead{$\Delta \nu_d$} & \colhead{$d$} & \colhead{$S_{1000}$} & \colhead{$\alpha$} & \colhead{$W_{50}$} & \colhead{$W_\mathrm{eff}$} & \colhead{$U_\mathrm{scale}$} & \colhead{$\sigma_\mathrm{J,1}$} & \colhead{$A_\mathrm{red}$} & \colhead{$\gamma_\mathrm{red}$}\\
\colhead{} & \colhead{(ms)} & \colhead{($\mathrm{pc\,cm^{-3}}$)} & \colhead{($\mathrm{\mu s}$)} & \colhead{(s)} & \colhead{(GHz)} & \colhead{(kpc)} & \colhead{(mJy)} & \colhead{} & \colhead{($\mathrm{\mu s}$)} & \colhead{($\mathrm{\mu s}$)} & \colhead{} & \colhead{($\mathrm{\mu s}$)} & \colhead{} & \colhead{}\\
}
\startdata
J1853+1303 & 4.09 & 30.57 & $3.1 \times 10^{-2}$ & 895.5 & $5.9 \times 10^{-3}$ & 2.09 & 0.88 & -1.99 & 129.27 & 344.52 & 9.96 & 62.78 & $9.18 \times 10^{-16}$ & 5.60\\
B1855+09 & 5.36 & 13.30 & $2.4 \times 10^{-2}$ & 766.5 & $7.8 \times 10^{-3}$ & 1.17 & 5.89 & -1.30 & 518.22 & 752.47 & 8.44 & 115.83 & $1.34 \times 10^{-15}$ & 5.60\\
J1903+0327 & 2.15 & 297.50 & $1.4 \times 10^{3}$ & 7.4 & $1.3 \times 10^{-7}$ & 6.37 & 0.85 & -1.00 & 197.08 & 401.64 & 9.08 & 105.76 & $2.56 \times 10^{-15}$ & 5.60\\
J1909--3744 & 2.95 & 10.39 & $9.1 \times 10^{-3}$ & 774.6 & $2.0 \times 10^{-2}$ & 0.46 & 1.96 & -1.92 & 40.33 & 258.73 & 62.40 & 12.57 & $1.27 \times 10^{-15}$ & 5.60\\
J1910+1256 & 4.98 & 38.07 & $5.0 \times 10^{-2}$ & 748.1 & $3.7 \times 10^{-3}$ & 2.33 & 0.95 & -1.57 & 132.70 & 635.09 & 25.09 & 84.31 & $7.82 \times 10^{-16}$ & 5.60\\
J1911+1347 & 4.63 & 30.98 & $3.0 \times 10^{-2}$ & 911.7 & $6.2 \times 10^{-3}$ & 2.07 & 1.17 & -1.35 & 88.85 & 463.64 & 25.33 & 38.27 & $1.43 \times 10^{-15}$ & 5.60\\
J1911--1114 & 3.63 & 30.98 & $1.1 \times 10^{-1}$ & 370.0 & $1.7 \times 10^{-3}$ & 1.22 & 1.30$^{\rm k}$ & -2.90$^{\rm l}$ & 389.03 & 623.17 & 6.04 & 145.41 & $1.50 \times 10^{-15}$ & 5.60\\
J1918--0642 & 7.65 & 26.59 & $6.2 \times 10^{-2}$ & 489.0 & $3.0 \times 10^{-3}$ & 1.24 & 2.85 & -1.95 & 149.26 & 874.30 & 17.80 & 36.21 & $1.35 \times 10^{-15}$ & 5.60\\
J1923+2515 & 3.79 & 18.86 & $1.0 \times 10^{-2}$ & 1387.0 & $1.8 \times 10^{-2}$ & 1.63 & 0.61 & -1.71 & 151.32 & 520.30 & 12.12 & 196.59 & $7.10 \times 10^{-16}$ & 5.60\\
B1937+21 & 1.56 & 71.02 & $7.8 \times 10^{-1}$ & 234.8 & $2.4 \times 10^{-4}$ & 3.56 & 39.58 & -3.13 & 37.14 & 145.27 & 21.80 & 16.99 & $1.35 \times 10^{-14}$ & 5.60\\
J1944+0907 & 5.19 & 24.36 & $1.7 \times 10^{-2}$ & 1109.0 & $1.1 \times 10^{-2}$ & 1.79 & 3.72 & -2.06 & 370.39 & 954.60 & 4.07 & 251.11 & $1.87 \times 10^{-16}$ & 5.60\\
J1946+3417 & 3.17 & 110.20 & 1.3 & 219.5 & $1.4 \times 10^{-4}$ & 5.14 & 1.41 & -1.33 & 221.54 & 481.95 & 9.14 & 197.56 & $4.37 \times 10^{-17}$ & 5.60\\
B1953+29 & 6.13 & 104.51 & 1.5 & 195.6 & $1.3 \times 10^{-4}$ & 4.64 & 1.73 & -2.05 & 224.07 & 832.27 & 8.06 & 330.38 & $1.82 \times 10^{-15}$ & 5.60\\
J2010--1323 & 5.22 & 22.16 & $4.0 \times 10^{-2}$ & 557.2 & $4.7 \times 10^{-3}$ & 1.02 & 1.22 & -1.53 & 246.13 & 523.92 & 20.45 & 60.16 & $4.49 \times 10^{-16}$ & 5.60\\
J2017+0603 & 2.90 & 23.92 & $2.3 \times 10^{-2}$ & 910.1 & $8.1 \times 10^{-3}$ & 1.57 & 0.28 & -1.00 & 64.42 & 240.80 & 11.48 & 24.91 & $1.11 \times 10^{-15}$ & 5.60\\
J2022+2534 & 2.65 & 53.66 & $1.4 \times 10^{-1}$ & 528.2 & $1.3 \times 10^{-3}$ & 3.31 & 1.62 & -1.91 & 168.42 & 385.28 & 9.19 & 56.24 & $6.06 \times 10^{-16}$ & 5.60\\
J2033+1734 & 5.95 & 25.08 & $1.8 \times 10^{-2}$ & 1168.0 & $1.1 \times 10^{-2}$ & 2.00 & 0.63 & -1.99 & 230.39 & 858.24 & 16.05 & 380.97 & $6.30 \times 10^{-16}$ & 5.60\\
J2039--3616 & 3.28 & 23.96 & $5.6 \times 10^{-2}$ & 442.8 & $3.3 \times 10^{-3}$ & 0.91 & 1.13 & -1.19 & 117.48 & 504.03 & 10.63 & 34.66 & $1.17 \times 10^{-15}$ & 5.60\\
J2043+1711 & 2.38 & 20.71 & $1.2 \times 10^{-2}$ & 1316.0 & $1.5 \times 10^{-2}$ & 1.76 & 0.42 & -1.78 & 20.49 & 177.88 & 16.61 & 15.65 & $6.93 \times 10^{-16}$ & 5.60\\
J2124--3358 & 4.93 & 4.60 & $7.6 \times 10^{-4}$ & 2075.0 & $2.4 \times 10^{-1}$ & 0.27 & 6.29 & -1.17 & 510.86 & 892.92 & 5.20 & 189.50 & $1.04 \times 10^{-15}$ & 5.60\\
J2145--0750 & 16.05 & 9.00 & $3.2 \times 10^{-3}$ & 1469.0 & $5.9 \times 10^{-2}$ & 0.57 & 8.80 & -2.04 & 338.29 & 1824.76 & 15.34 & 113.31 & $8.63 \times 10^{-16}$ & 5.60\\
J2150--0326 & 3.51 & 20.67 & $3.3 \times 10^{-2}$ & 615.3 & $5.5 \times 10^{-3}$ & 1.05 & 0.91 & -1.90 & 158.12 & 437.68 & 19.31 & 48.99 & $1.00 \times 10^{-15}$ & 5.60\\
J2205+6012 & 2.42 & 157.64 & 12.9 & 71.3 & $1.4 \times 10^{-5}$ & 5.48 & 1.03$^{\rm m}$ & -2.20$^{\rm n}$ & 81.15 & 295.72 & 20.31 & 23.60 & $2.52 \times 10^{-15}$ & 5.60\\
J2214+3000 & 3.12 & 22.56 & $1.9 \times 10^{-2}$ & 990.4 & $9.8 \times 10^{-3}$ & 1.54 & 0.39 & -1.00 & 180.83 & 562.84 & 10.78 & 103.36 & $1.22 \times 10^{-15}$ & 5.60\\
J2229+2643 & 2.98 & 22.72 & $2.5 \times 10^{-2}$ & 827.7 & $7.4 \times 10^{-3}$ & 1.43 & 0.86 & -2.14 & 436.65 & 580.81 & 7.69 & 113.38 & $2.28 \times 10^{-16}$ & 5.60\\
J2234+0944 & 3.63 & 17.83 & $2.0 \times 10^{-2}$ & 770.5 & $9.2 \times 10^{-3}$ & 1.00 & 1.27 & -1.00 & 233.56 & 549.95 & 9.38 & 40.15 & $1.18 \times 10^{-15}$ & 5.60\\
J2234+0611 & 3.58 & 10.77 & $4.5 \times 10^{-3}$ & 1340.0 & $4.1 \times 10^{-2}$ & 0.68 & 0.44 & -1.00 & 89.94 & 397.09 & 29.46 & 19.63 & $9.30 \times 10^{-16}$ & 5.60\\
J2302+4442 & 5.19 & 13.72 & $5.0 \times 10^{-3}$ & 1683.0 & $3.7 \times 10^{-2}$ & 1.18 & 2.45 & -1.50 & 343.65 & 644.21 & 4.86 & 270.74 & $1.12 \times 10^{-15}$ & 5.60\\
J2317+1439 & 3.45 & 21.90 & $1.1 \times 10^{-2}$ & 929.9 & $1.6 \times 10^{-2}$ & 0.83 & 0.96 & -2.32 & 152.69 & 379.11 & 8.67 & 25.24 & $3.69 \times 10^{-16}$ & 5.60\\
J2322+2057 & 4.81 & 13.39 & $6.2 \times 10^{-3}$ & 1243.0 & $3.0 \times 10^{-2}$ & 0.80 & 0.47 & -1.56 & 312.15 & 477.38 & 9.95 & 26.70 & $4.88 \times 10^{-16}$ & 5.60
\enddata
\tablenotetext{k}{\vspace{-2ex}\citet{psrcat}}
\tablenotetext{l}{\vspace{-2ex}\citet{psrcat}}
\tablenotetext{m}{\vspace{-2ex}\citet{nancay22_J2205}}
\tablenotetext{n}{\vspace{-2ex}\citet{nancay22_J2205}}
\end{deluxetable}

\subsection{Telescope Observation Simulations}
We simulate timing observations for each pulsar with all possible combinations of the 
current telescope/receiver pairs: 
\begin{itemize}
	\item CHIME 
	\item GBT 820 MHz
	\item GBT L-band\footnote{The recently-installed GBT Ultra-wideband Receiver covers 0.7--4 GHz simultaneously and 
	will replace most dual-band observations at the GBT but was not available at the time this analysis was conducted.
	It will be included in our simulations going forward.}
	\item VLA L-band
	\item VLA S-band.
\end{itemize}
Combinations that include CHIME are only valid for
pulsars visible to CHIME with declination, $\Dec, > -20\degrees$.
We also consider two future observing configurations: CHIME and DSA or DSA-only.
The simulated observing parameters for these instruments are summarized in Table \ref{tab:tel_rx}.
J0437--4715 is only visible with the VLA so it is
observed with VLA L-band and S-band and excluded from DSA optimization.
J1012--4235 is also below DSA's horizon, so it is excluded from DSA optimization as well.
For every pair of pulsar-instrument combination pair, we compute the frequency-optimized
$\sigmaTOA$ as long as the pulsar is visible to every instrument in the combination.
The optimal instrument and observing frequencies are chosen by the smallest $\sigmaTOA$.

A few caveats are necessary for simulations of multi-band observations. We treat 
the receiver combinations as a single, simultaneously-observed band in our computation of $\sigmaTOA$,
meaning $G(\nu)$, $\Trcvr(\nu)$, and $\Delta \nu$ cover the entire frequency range
from the bottom of the lowest band to the top of the highest band. For non-adjacent 
bands (CHIME and GBT L-band, GBT L-band and 820 MHz, GBT L-band and VLA S-band) 
we insert a single frequency channel between the bands with very low
$G$ ($\sim 10^{-5}\; \mathrm{K\; Jy^{-1}}$) to make the template-fitting noise covariances,
${\bf S}_{\nu_1 \nu_2} \propto G^{-2}$,
and therefore the achromatic noise covariances ${\bf C}_{\nu_1 \nu_2}$
that include this channel approximately infinte\footnote{This was done for convenience in software implementation.}.
$\sigmaDMhat$, $\sigmadtC$, and $\sigmaDMnu$ are all weighted by ${\bf C}^{-1}$, so
the contribution from this gap is effectively ignored in $\sigmaTOA$. However,
$\sigmaJ$, $\sigmaDISS$, and $\sigmatel$ don't depend on ${\bf S}_{\nu_1 \nu_2}$.
They will see 
a larger effective bandwidth and will be slightly overestimated from nonexistent 
frequency-frequency correlations when they are considered as individual noise components,
but the effect on $\sigmaTOA$ is suppressed by the achromatic noise covariances.
Whenever two bands overlap (e.g. CHIME and DSA),
we use the noise characteristics of the 
more sensitive instrument (DSA) in the overlapping bandwidth. Additionally, jitter
noise is not correlated between non-simultaneous observations, so $\sigmaJ$
will be slightly overestimated in all considered multi-instrument combinations. 
These effects can be addressed in the future
by masking the rows and columns of all of the noise covariance matrices for frequencies
in the gap between bands and between nonsimultaneous frequencies in the jitter covariance
matrix. 

\begin{figure}[t!]
	\centering
	\gridline{
		\fig{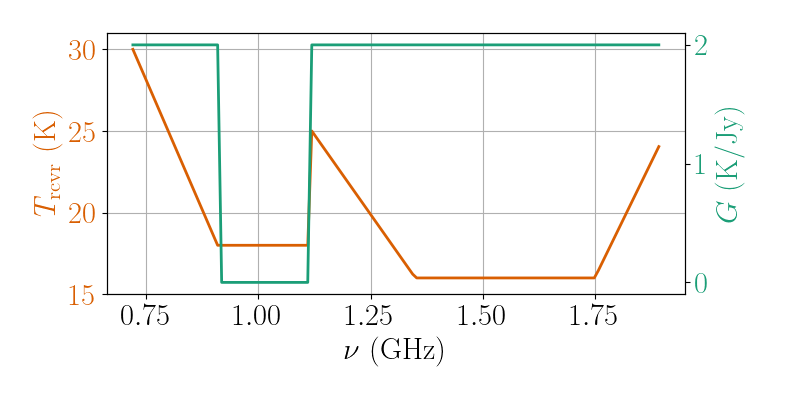}{0.5\textwidth}{(a)}
		\fig{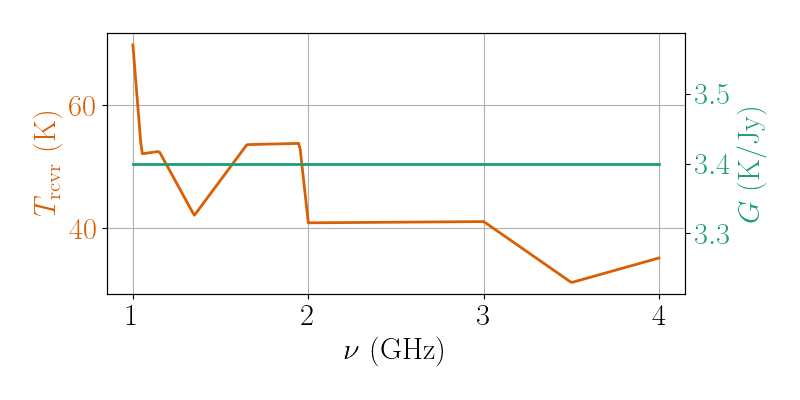}{0.5\textwidth}{(b)}
	}
	\vspace{-1em}
	\gridline{
		\fig{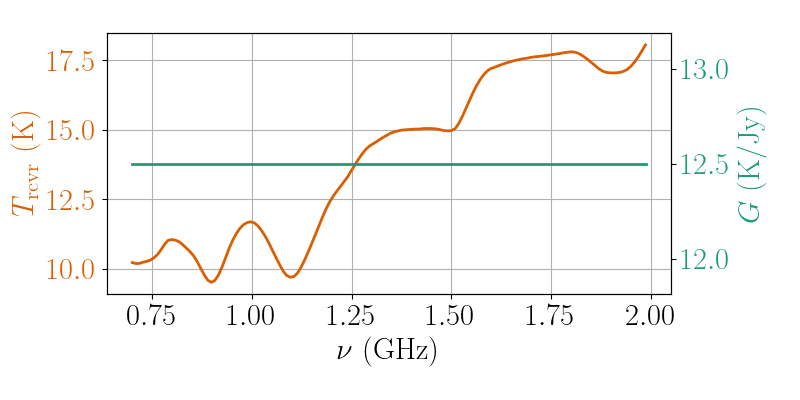}{0.5\textwidth}{(c)}
		\fig{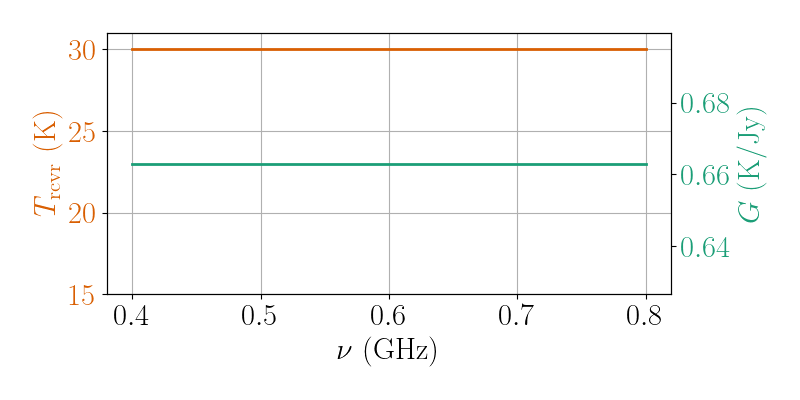}{0.5\textwidth}{(d)}
	}
	\caption{
	Receiver temperature and gain vs. observing frequency models for (a) GBT 820 MHz and L-band
	(b) VLA L-band and S-band
	(c) DSA including contributions from spillover, LNA, and feed \citep{flygare_dsa} and
	(d) CHIME for a $0\degrees$ Dec. pulsar.
	The dip in the GBT gain curve illustrates the low gain channel
	inserted between non-adjacent bands.
	}
	\label{fig:rcvr_trx}
\end{figure}

Models of the receiver passband shapes come from a few different sources. $\Trcvr(\nu)$
for the GBT receivers was measured from NANOGrav observations of the 
continuum flux calibrator source, quasar B1442+101, assuming a flat gain of 
$2\; \mathrm{K\; Jy^{-1}}$ \citep{gbtobsguide}.
For the VLA, we estimated $\Trcvr(\nu)$ using Figure 3.2.1
on the VLA Observational Status Summary webpage
\footnote{\url{https://science.nrao.edu/facilities/vla/docs/manuals/oss/performance/}},
to estimate $\Tsys(\nu)$ assuming a constant gain of $3.4\; \mathrm{K\; Jy^{-1}}$,
then computing $\Trcvr(\nu)$ according to Equation \eqref{eq:Tsys}. 
We assume that the CHIME $\Trcvr(\nu)=30\ \mathrm{K}$ is approximately a flat in the passband,
while the DSA $\Trcvr(\nu)$ is adopted from the feed, LNA,
and spillover brightness-temperature measurements at zenith shown in \citet{flygare_dsa}.
DSA gain is assumed to be $12.5\; \mathrm{K\; Jy^{-1}}$ to account for a recent 
antenna redesign, described in Section \ref{sssec:DSA}.
All four telescope's receiver models are shown in Figure \ref{fig:rcvr_trx}.
The current NANOGrav timing program observes pulsars with CHIME daily,
for as long as they are visible,
and approximately monthly with the GBT and/or VLA for 30 min/band for dual-band observations.
Since the dual-band configurations require separate observations to cover the full bandwidth,
the effective $\tint$ is 30 min/frequency channel, but the time allocation to observe the full
band is one hour.
Further details about DSA and CHIME observations are described in the following subsections.

In order to set the upper bounds of
the integration time optimization with DSA, we consider the total amount of time a pulsar is above the horizon, 
$\tmax$. The upper limit on $\tmax$ is the total length of an observing epoch, $\tepoch = 30.44 \; \mathrm{days}$. The arc-length above the minimum elevation is
\begin{equation}
	\ell = 2\cos^{-1}\left(\frac{\sin(\mathrm{alt.})}{\cos(\lambda)\cos(\delta)}
                        - \tan(\lambda)\tan(\delta)\right),
\end{equation}
where $\mathrm{alt.}$ is the minimum telescope elevation, $\delta$ is the Dec. of the pulsar,
and $\lambda$ is the telescope latitude. This translates to a maximum time of
\begin{equation}
	\tmax = \frac{86400 \times \ell \times \tepoch}{2\pi}\; \mathrm{s}.
\end{equation}

\subsubsection{DSA}\label{sssec:DSA}
The Deep Synoptic Array is a 1650-dish array under construction in Eastern Nevada, expected to make its first
observations in the late 2020s. 
We assume that pulsar observations
will be conducted down to a minimum elevation limit of $10\degrees$, yielding a Dec. range of
$-41\degrees < \Dec < 90\degrees$.
DSA is expected to cover 0.7-2 GHz simultaneously.
The simultaneous bandwidth offers at least a factor of $\sqrt{2}$ improvement in template-fitting uncertainty
by covering the full band with the same amount of observing time as current dual-band configurations,
assuming a fixed observing time.
The telescope was originally designed with 2000 dishes, but recently
underwent a redesign that reduced the number of dishes but increased their diameter from 5 m to 6.15 m
\citep{dsa_science_doc}.
Assuming the same 70\% aperture efficiency \citep{hallinan19_dsa2000} and $\Trcvr(\nu)$
\citep[Figure \ref{fig:rcvr_trx}c,][]{flygare_dsa}
as the original design, we account for the overall change in the
collecting area by increasing the gain from 10 to $12.5\; \mathrm{K\; Jy^{-1}}$.
Here, we consider an hypothetical, ambitious PTA observing program that utilizes 25\% of the total DSA observing time.
We assume 5\% of this time will be spent on slewing, calibration, and overhead, leaving
20\% of each month purely to integrate on PTA pulsars. If this time is fiducially distributed
equally amongst the pulsars, then the equal-time allocation vector is $\teq = 1.75\;\mathrm{hr\; /\; month}$. 
We expect this considerable increase in time allocation, simultaneous bandwidth,
and instrumental sensitivity to provide unparalleled TOA precision, even without optimization.

\subsubsection{CHIME}
The Canadian Hydrogen Intensity Mapping Experiment telescope is a transit instrument composed of four cylindrical reflectors located at the Dominion Radio Astrophysical Observatory (DRAO) at a latitude of $49.32^\circ$ with a declination range of $-20^\circ<\delta < 90^\circ$ \citep{ng18}.
The telescope's F-Engine covers 400-800 MHz,
but anywhere from 15\%-68\% of the band is masked on a given day due to RFI \citep{chime_overview}.
Because CHIME is a transit telescope, the (beam-center) gain depends strongly on declination,
which we model as (E. Fonseca, internal memo)
\begin{equation}
G_0(\delta) = \sqrt{0.7} \times 1.2\cos(\delta-49.32^\circ)\ \mathrm{K\ Jy^{-1}}.
\end{equation}
where $1.2\;\mathrm{K\ Jy^{-1}}$ is the nominal zenith gain \citep{andersen23}
and the $\sqrt{0.7}$ prefactor accounts for an assumed 70\% usable bandwidth due to RFI.
The cylinders of CHIME point N-S and so the field-of-view of the telescope in this direction is 120$^\circ$. 
In the E-W direction however, the field-of-view depends strongly (linearly) on frequency: 
2.5$^\circ$ at 400 MHz and 1.3$^\circ$ at 800 MHz. 
An order of magnitude estimate for the maximum $t_\mathrm{int}/\mathrm{day}$ for a source at $\delta = 0$ 
at 400 MHz is 10 minutes \citep{ng18, chime_frb1}. For patches of sky of equal area but at different declinations, 
it takes more time to cover the same area by an amount of $1/\cos(\delta)$. 
Therefore, a rough estimate for the number of seconds per 30 day epoch that a pulsar can be observed with CHIME as a function of frequency and declination is
\begin{equation}\label{eq:chime-uptime}
\tint = 1.8\times 10^4\ \mathrm{s\ /\ month}\left(\frac{400\ \mathrm{MHz}}{\nu}\right) \sec(\delta).
\end{equation}
For a circumpolar source which crosses the field-of-view twice per day the accessible time is approximately 
double the amount in Equation \eqref{eq:chime-uptime}.

CHIME observations are carried out with a daily cadence. We do not have the ability
in the current construction of our methodology to consider a pulsar PSD derived from
data with two different cadences, so we assume a single monthly epoch-averaged TOA
for CHIME (with the same total $\tint$/month) when combined with another instrument.
This assumption does not affect calculations of $\sigmaTOA$, only the estimate of
$\rho$ where white (in GW frequency) noise will be underestimated.

\begin{deluxetable}{llcccccc}
\tablecaption{Adopted Telescope and Receiver Parameters\label{tab:tel_rx}}
\tablehead{
\colhead{Telescope} &
\colhead{Receiver} &
\colhead{Latitude} &
\colhead{$\delta_\mathrm{min}$} &
\colhead{$\nu$ range} &
\colhead{$G_\mathrm{zenith}$} &
\colhead{nominal $\tint$}  &
\colhead{$\Delta t$} \\
\colhead{} &
\colhead{} &
\colhead{deg} &
\colhead{deg} &
\colhead{GHz} & 
\colhead{K Jy$^{-1}$} &
\colhead{hrs} &
\colhead{days}
}
\startdata
CHIME &  & 49.32 & $-20.0$ & 0.4--0.8 & 1.0 & $\delta,\, \nu$-dependent$^\mathrm{a}$ & 30.44 \\
\hline
\multirow{2}{*}{GBT} & Prime Focus 820 MHz & \multirow{2}{*}{38.43} & \multirow{2}{*}{$-46.0$} &
													0.72--0.92& 2.0 & 0.5 & \multirow{2}{*}{$30.44$}\\
                     & L-band &                         &                         & 
                     													 1.11--1.91 & 2.0 & 0.5 \\
\hline
\multirow{2}{*}{VLA} & L-band & \multirow{2}{*}{34.08} & \multirow{2}{*}{$-48.0$} & 
													1--2 & 3.4 & 0.5 & \multirow{2}{*}{$30.44$}\\
                     & S-band    &                         &                         & 2--4 & 3.4 & 0.5 \\
\hline
DSA &  & 39.0 & $-41.0$ & 0.7--2 & 12.5 & 1.75 & 30.44\\
\enddata
\tablenotetext{a}{See Equation \eqref{eq:chime-uptime}}
\end{deluxetable}

\section{Noise Estimates and Optimal Instruments}\label{sec:opt_instr}
Table \ref{tab:NG20_optinstr_optimal_instruments} lists the current instrument combination,
optimal instrument combination, $\sigmaTOA$ for current and
optimal instruments, and optimal frequency ranges with the current and optimal instrument ($\nu_\mathrm{min}^* 
,\;\nu_\mathrm{max}^*$) for each pulsar.
The left side of the table approximates the current NANOGrav timing program, 
but not necessarily the optimal instrument choices.
Four pulsars without post-Arecibo TOAs are chosen to have $\sigmaTOA$ from the optimal
combination of current instruments.
When the optimal bandwidth is smaller than the full bandwidth, it suggests that TOAs
should be constructed from only the optimal frequencies.
The right side of the table represents the optimal observing plan and projected 
TOA uncertainties at nominal $\tint$.
Among all considered observing configurations, all but four pulsars are optimally observed with
a combination that includes DSA. J0437--4715 and J1012--4235 are below the DSA horizon.
J1903+0327 and J1802--2145
have the highest $\taud$ of all pulsars 
in our dataset and are among the highest DM MSPs known \citep[ATNF Pulsar Catalog,][]{psrcat}.
Systematic scattering delay uncertainties, $\sigmadtC$, are overwhelmingly the largest component of $\sigmaTOA$
for these two pulsars. Since scattering effects are stronger at lower frequencies these two
pulsars are optimal with the VLA and only at the highest observing frequencies. In fact, 
J1903+0327 is optimal with only VLA S-band. In addition to the nominal equal-time allocation
of 1.75 hrs / epoch / pulsar with DSA we also considered 1 hr / epoch / pulsar, which is the equivalent time
allocation to current instruments. Even with this reduction in $\tint$, DSA or DSA + CHIME
are still the best performing configurations for all but five pulsars. 
In general, pulsars which are optimal with both CHIME and DSA are dominated by achromatic noise, $\sigmaDMhat$,
with relatively low $\sigmaDMnu$ and $\sigmadtC$.

\newcommand{\InsertOptInstrTabCaption}{"GBT8" = GBT 820 MHz, "GBTL" = GBT L-band, "VLAL" = VLA L-band,
"VLAS" = VLA S-band}
\startlongtable
\begin{deluxetable}{l|cccc|cccc}
\tabletypesize{\scriptsize}
\tablecaption{Optimal Instruments and Frequency Ranges\label{tab:NG20_optinstr_optimal_instruments}}
\tablehead{
\colhead{Pulsar} & \colhead{Current Instrument(s)} & \colhead{$\nu_\mathrm{min}^*$} & \colhead{$\nu_\mathrm{max}^*$} & \colhead{$\sigma_\mathrm{TOA}$} & \colhead{Optimal Instrument(s)} & \colhead{$\nu_\mathrm{min}^*$} & \colhead{$\nu_\mathrm{max}^*$} & \colhead{$\sigma_\mathrm{TOA}$}\\
\colhead{} & \colhead{} & \colhead{(GHz)} & \colhead{(GHz)} & \colhead{($\mathrm{\mu s}$)} & \colhead{} & \colhead{(GHz)} & \colhead{(GHz)} & \colhead{($\mathrm{\mu s}$)}
}
\startdata
J0023+0923 & CHIME+GBTL & 0.40 & 1.90 & 0.67 & CHIME+DSA & 0.41 & 1.99 & 0.07\\
J0030+0451 & CHIME+GBTL & 0.40 & 1.90 & 0.43 & CHIME+DSA & 0.45 & 1.97 & 0.12\\
J0102+4839 & CHIME & 0.48 & 0.79 & 2.88 & DSA & 0.80 & 1.98 & 0.23\\
J0125--2327 & GBT8+GBTL+VLAS & 0.72 & 4.00 & 0.13 & DSA & 0.70 & 2.00 & 0.03\\
J0154+1833 & CHIME+GBTL & 0.43 & 1.87 & 0.42 & DSA & 0.70 & 2.00 & 0.05\\
J0218+4232 & CHIME & 0.52 & 0.79 & 2.60 & CHIME+DSA & 0.46 & 1.70 & 0.35\\
J0340+4130 & CHIME+GBT8+GBTL+VLAS & 0.60 & 3.83 & 0.84 & DSA & 0.92 & 1.97 & 0.23\\
J0406+3039 & CHIME+GBTL & 0.61 & 1.90 & 0.49 & DSA & 0.99 & 1.98 & 0.09\\
J0437--4715 & VLAL+VLAS & 1.00 & 4.00 & 0.07 & VLAL+VLAS & 1.00 & 4.00 & 0.07\\
J0509+0856 & CHIME+GBTL & 0.42 & 1.90 & 1.24 & DSA & 0.78 & 1.99 & 0.25\\
J0557+1551 & CHIME+GBT8+GBTL+VLAS$^{\rm \dagger}$ & 0.69 & 4.00 & 3.53 & DSA & 1.11 & 1.98 & 0.96\\
J0605+3757 & CHIME+GBT8+GBTL & 0.43 & 1.88 & 0.57 & DSA & 0.72 & 1.99 & 0.08\\
J0610--2100 & GBT8+GBTL & 0.72 & 1.90 & 0.59 & DSA & 1.03 & 1.99 & 0.10\\
J0613--0200 & CHIME+GBT8+GBTL & 0.61 & 1.90 & 0.22 & DSA & 0.96 & 1.99 & 0.05\\
J0614--3329 & GBT8+GBTL & 0.72 & 1.90 & 0.41 & DSA & 0.84 & 1.99 & 0.05\\
J0621+2514 & CHIME+GBT8+GBTL+VLAS$^{\rm \dagger}$ & 0.58 & 3.94 & 5.28 & DSA & 0.93 & 1.98 & 0.92\\
J0636+5128 & CHIME+GBT8+GBTL & 0.40 & 1.65 & 0.18 & CHIME+DSA & 0.47 & 1.93 & 0.05\\
J0645+5158 & CHIME+GBT8+GBTL & 0.48 & 1.90 & 0.28 & DSA & 0.74 & 1.98 & 0.04\\
J0709+0458 & CHIME & 0.40 & 0.80 & 6.84 & DSA & 0.70 & 2.00 & 0.61\\
J0732+2314 & CHIME+GBTL & 0.43 & 1.88 & 1.73 & DSA & 0.78 & 1.99 & 0.39\\
J0740+6620 & CHIME+GBT8+GBTL & 0.43 & 1.88 & 0.19 & DSA & 0.70 & 2.00 & 0.03\\
J0751+1807 & CHIME+GBTL & 0.48 & 1.90 & 0.38 & DSA & 0.84 & 1.99 & 0.07\\
J0931--1902 & GBT8+GBTL & 0.72 & 1.90 & 0.78 & DSA & 0.84 & 1.99 & 0.12\\
J1012--4235 & GBT8+GBTL & 0.72 & 1.90 & 0.78 & GBT8+GBTL+VLAS & 0.72 & 4.00 & 0.56\\
J1012+5307 & CHIME+GBT8+GBTL & 0.40 & 1.90 & 0.15 & CHIME+DSA & 0.44 & 1.96 & 0.04\\
J1022+1001 & CHIME+GBTL & 0.40 & 1.90 & 0.34 & CHIME+DSA & 0.57 & 1.97 & 0.20\\
J1024--0719 & CHIME+GBT8+GBTL & 0.40 & 1.90 & 0.33 & CHIME+DSA & 0.40 & 2.00 & 0.04\\
J1102+0249 & GBTL & 1.11 & 1.90 & 10.35 & CHIME+DSA & 0.41 & 1.99 & 0.26\\
J1125+7819 & CHIME+GBT8+GBTL & 0.40 & 1.90 & 0.27 & CHIME+DSA & 0.45 & 1.97 & 0.11\\
J1312+0051 & CHIME+GBT8+GBTL+VLAS$^{\rm \dagger}$ & 0.40 & 4.00 & 0.91 & CHIME+DSA & 0.40 & 2.00 & 0.12\\
J1327+3423 & CHIME+GBTL+VLAS$^{\rm \dagger}$ & 0.40 & 4.00 & 1.35 & CHIME+DSA & 0.46 & 1.95 & 0.48\\
J1453+1902 & CHIME & 0.40 & 0.80 & 2.85 & CHIME+DSA & 0.40 & 2.00 & 0.29\\
J1455--3330 & GBT8+GBTL & 0.72 & 1.90 & 0.93 & DSA & 0.70 & 2.00 & 0.13\\
J1600--3053 & GBTL+VLAS & 1.11 & 4.00 & 0.12 & DSA & 1.40 & 1.99 & 0.06\\
J1614--2230 & GBT8+GBTL & 0.72 & 1.90 & 0.28 & DSA & 1.03 & 1.99 & 0.07\\
J1630+3550 & CHIME+GBTL & 0.40 & 1.90 & 1.60 & CHIME+DSA & 0.40 & 2.00 & 0.20\\
J1630+3734 & CHIME+GBT8+GBTL & 0.40 & 1.90 & 0.28 & DSA & 0.70 & 2.00 & 0.04\\
J1640+2224 & CHIME+GBTL & 0.43 & 1.87 & 0.39 & DSA & 0.70 & 2.00 & 0.05\\
J1643--1224 & GBT8+GBTL+VLAS & 0.82 & 3.96 & 0.28 & DSA & 1.21 & 1.99 & 0.08\\
J1705--1903 & GBTL+VLAS & 1.02 & 4.00 & 0.10 & DSA & 1.46 & 1.99 & 0.06\\
J1713+0747 & CHIME+GBT8+GBTL+VLAS & 0.60 & 4.00 & 0.06 & DSA & 0.96 & 1.99 & 0.03\\
J1719--1438 & CHIME+GBT8+GBTL & 0.52 & 1.86 & 0.66 & DSA & 0.90 & 1.99 & 0.13\\
J1730--2304 & GBT8+GBTL & 0.72 & 1.90 & 0.46 & DSA & 0.70 & 2.00 & 0.13\\
J1738+0333 & CHIME+GBTL+VLAS & 0.42 & 4.00 & 0.53 & DSA & 0.78 & 1.99 & 0.09\\
J1741+1351 & CHIME+GBTL & 0.40 & 1.90 & 0.29 & CHIME+DSA & 0.40 & 2.00 & 0.04\\
J1744--1134 & CHIME+GBT8+GBTL & 0.40 & 1.65 & 0.11 & CHIME+DSA & 0.49 & 1.98 & 0.04\\
J1745+1017 & CHIME+GBTL & 0.40 & 1.90 & 0.53 & DSA & 0.70 & 2.00 & 0.09\\
J1747--4036 & GBTL+VLAS & 1.29 & 4.00 & 1.70 & DSA & 1.58 & 1.99 & 0.94\\
J1751--2857 & GBT8+GBTL & 0.72 & 1.90 & 0.79 & DSA & 0.96 & 1.99 & 0.13\\
J1802--2124 & GBTL+VLAS & 1.84 & 4.00 & 1.74 & VLAL+VLAS & 1.83 & 4.00 & 1.64\\
J1803+1358 & CHIME+GBTL & 0.65 & 1.86 & 0.77 & DSA & 0.99 & 1.98 & 0.15\\
J1811--2405 & GBT8+GBTL & 0.91 & 1.90 & 0.65 & DSA & 1.40 & 1.99 & 0.17\\
J1832--0836 & CHIME+GBT8+GBTL & 0.61 & 1.90 & 0.24 & DSA & 0.96 & 1.99 & 0.05\\
J1843--1113 & GBT8+GBTL & 0.83 & 1.88 & 0.49 & DSA & 1.30 & 1.98 & 0.12\\
J1852--1310 & GBT8+GBTL & 0.72 & 1.90 & 1.44 & DSA & 0.90 & 1.99 & 0.18\\
J1853+1303 & CHIME+GBTL & 0.40 & 1.90 & 0.57 & DSA & 0.70 & 2.00 & 0.08\\
B1855+09 & CHIME+GBTL & 0.40 & 1.90 & 0.30 & CHIME+DSA & 0.55 & 1.98 & 0.11\\
J1903+0327 & VLAS & 2.64 & 3.99 & 6.67 & VLAS & 2.67 & 4.00 & 6.41\\
J1909--3744 & GBT8+GBTL+VLAS & 0.72 & 4.00 & 0.04 & DSA & 0.84 & 1.99 & 0.01\\
J1910+1256 & CHIME+GBTL & 0.40 & 1.90 & 0.43 & DSA & 0.72 & 1.99 & 0.09\\
J1911+1347 & CHIME+GBTL & 0.40 & 1.90 & 0.24 & DSA & 0.70 & 2.00 & 0.04\\
J1911--1114 & GBT8+GBTL & 0.72 & 1.90 & 2.28 & DSA & 0.70 & 2.00 & 0.19\\
J1918--0642 & CHIME+GBT8+GBTL & 0.48 & 1.90 & 0.30 & DSA & 0.84 & 1.99 & 0.06\\
J1923+2515 & CHIME+GBTL & 0.40 & 1.90 & 0.85 & CHIME+DSA & 0.40 & 2.00 & 0.18\\
B1937+21 & GBTL+VLAS & 1.51 & 3.98 & 0.05 & DSA & 1.61 & 1.98 & 0.04\\
J1944+0907 & CHIME+GBTL & 0.40 & 1.90 & 0.92 & CHIME+DSA & 0.43 & 1.98 & 0.24\\
J1946+3417 & GBTL+VLAS & 1.11 & 4.00 & 0.88 & DSA & 1.21 & 1.99 & 0.24\\
B1953+29 & CHIME & 0.59 & 0.79 & 6.63 & DSA & 1.09 & 1.99 & 0.46\\
J2010--1323 & GBT8+GBTL & 0.72 & 1.90 & 0.39 & DSA & 0.78 & 1.99 & 0.07\\
J2017+0603 & CHIME+GBTL+VLAS & 0.40 & 4.00 & 0.64 & CHIME+DSA & 0.41 & 1.99 & 0.08\\
J2022+2534 & CHIME+GBTL & 0.55 & 1.90 & 0.46 & DSA & 0.90 & 1.99 & 0.08\\
J2033+1734 & CHIME+GBTL & 0.40 & 1.90 & 1.06 & CHIME+DSA & 0.49 & 1.91 & 0.38\\
J2039--3616 & GBT8+GBTL & 0.72 & 1.90 & 0.67 & DSA & 0.72 & 1.99 & 0.06\\
J2043+1711 & CHIME+GBTL & 0.40 & 1.90 & 0.29 & DSA & 0.70 & 2.00 & 0.03\\
J2124--3358 & GBT8+GBTL & 0.72 & 1.90 & 0.53 & DSA & 0.70 & 2.00 & 0.17\\
J2145--0750 & CHIME+GBT8+GBTL & 0.40 & 1.65 & 0.33 & CHIME+DSA & 0.52 & 1.88 & 0.17\\
J2150--0326 & CHIME+GBT8+GBTL & 0.42 & 1.90 & 0.32 & DSA & 0.72 & 1.99 & 0.05\\
J2205+6012 & GBTL & 1.27 & 1.90 & 1.87 & DSA & 1.61 & 1.98 & 0.73\\
J2214+3000 & CHIME+GBTL+VLAS & 0.40 & 4.00 & 1.08 & CHIME+DSA & 0.40 & 2.00 & 0.15\\
J2229+2643 & CHIME+GBTL & 0.40 & 1.90 & 0.87 & CHIME+DSA & 0.42 & 1.93 & 0.13\\
J2234+0944 & CHIME+GBTL & 0.40 & 1.90 & 0.47 & DSA & 0.70 & 2.00 & 0.06\\
J2234+0611 & CHIME+GBTL & 0.40 & 1.90 & 0.31 & CHIME+DSA & 0.41 & 1.99 & 0.03\\
J2302+4442 & CHIME+GBT8+GBTL & 0.40 & 1.65 & 0.57 & CHIME+DSA & 0.49 & 1.98 & 0.24\\
J2317+1439 & CHIME+GBTL & 0.40 & 1.90 & 0.47 & CHIME+DSA & 0.41 & 1.99 & 0.06\\
J2322+2057 & CHIME+GBTL & 0.40 & 1.90 & 1.05 & CHIME+DSA & 0.40 & 2.00 & 0.10
\enddata
\tablecomments{\InsertOptInstrTabCaption}
\tablenotetext{\dagger}{\vspace{-2ex}best possible current combination}
\end{deluxetable}

\begin{figure}[t!]
	\centering
	\includegraphics[width=.8\textwidth]{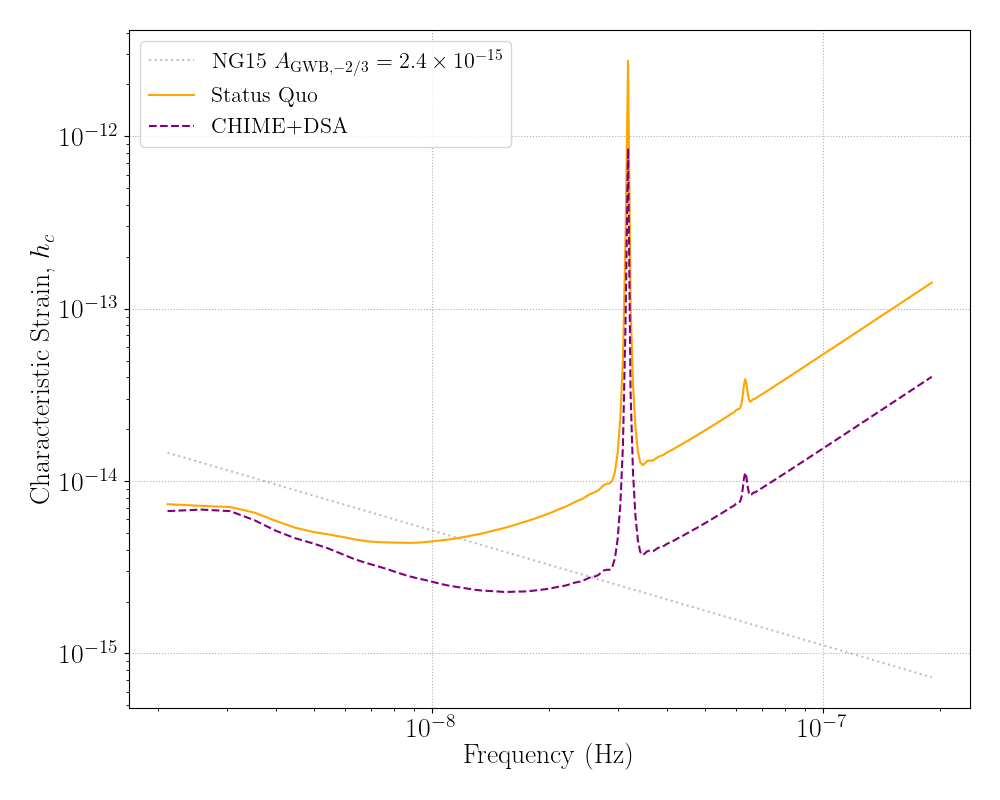}
	\caption{Strain-amplitude sensitivity curves for a status quo NANOGrav 15-year PTA 
	using current instruments (solid, orange curve) and a 15-year PTA using the optimal instruments
	(dashed, purple curve), primarily CHIME and DSA.
	The dotted line is a GWB power-law spectrum using the \NGXVgwbt Bayesian posterior amplitude.}
	\label{fig:nowfuture_sencurves}
\end{figure}

As a comparison between DSA and current instrumental sensitivity, we computed the GWB S/N for both a status quo PTA using the current instruments on the left side of Table 
\ref{tab:NG20_optinstr_optimal_instruments} and a PTA which effectively starts over with the option of including
DSA using the right side of the table,
assuming 15 years of data and monthly observations. The status quo PTA has $\rho = 6.85$,
while the optimal PTA using DSA has $\rho =14.44$, a considerable improvement.
In essence, if one were to start a 15-year timing program using any of the telescopes in Table \ref{tab:tel_rx},
these results show that observing with the DSA is by far the optimal choice.
We also compute the characteristic strain sensitivity curves,
\begin{equation}
	h_c(f) \equiv \sqrt{f S_\mathrm{eff}(f)},
\end{equation}
for both PTAs, shown in Figure \ref{fig:nowfuture_sencurves}.
We note that these figures are not intended to be predictions of NANOGrav's actual GWB S/N.
$\sigmaTOA_{,I}$ for NANOGrav datasets
is a piecewise function with multiple instrument changes (including data from defunct instruments)
and $\Tobs_{,I}$ ranging from a few years to two decades.
This is a comparison between a PTA observing program with
the current telescope combinations and an observing program with DSA, with all other variables being equal.
\section{Jitter-Dominated Pulsars}\label{sec:jitter}
\begin{figure}[t!]
	\centering
	\begin{minipage}{0.35\textwidth}
	    \centering
		\includegraphics[
	    	width=\linewidth,
	    	clip]{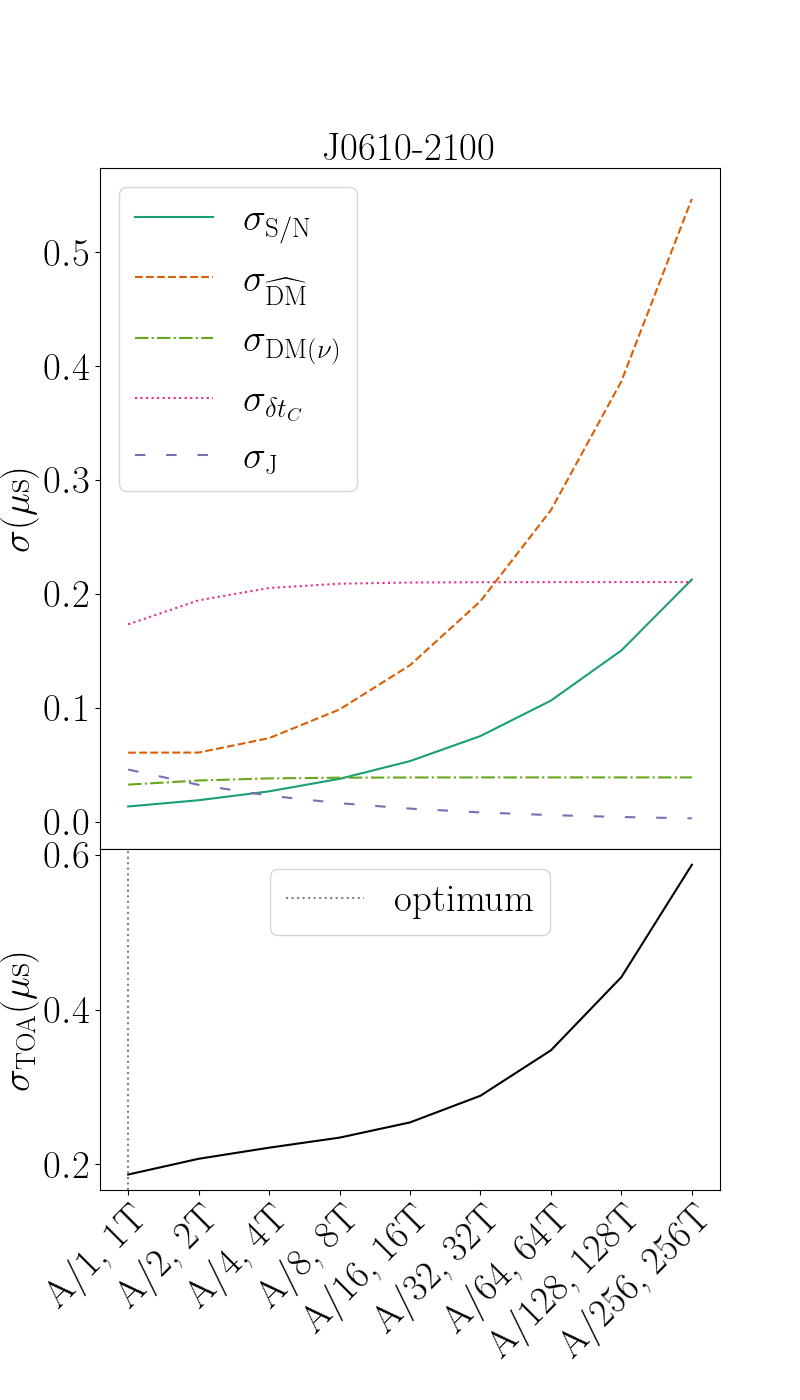}\\[-0.5ex]
	    	(a)
    	\end{minipage}
    	\hspace{-0.04\textwidth}
	\begin{minipage}{0.35\textwidth}
		\centering
		\includegraphics[
	    width=\linewidth,
	    clip]{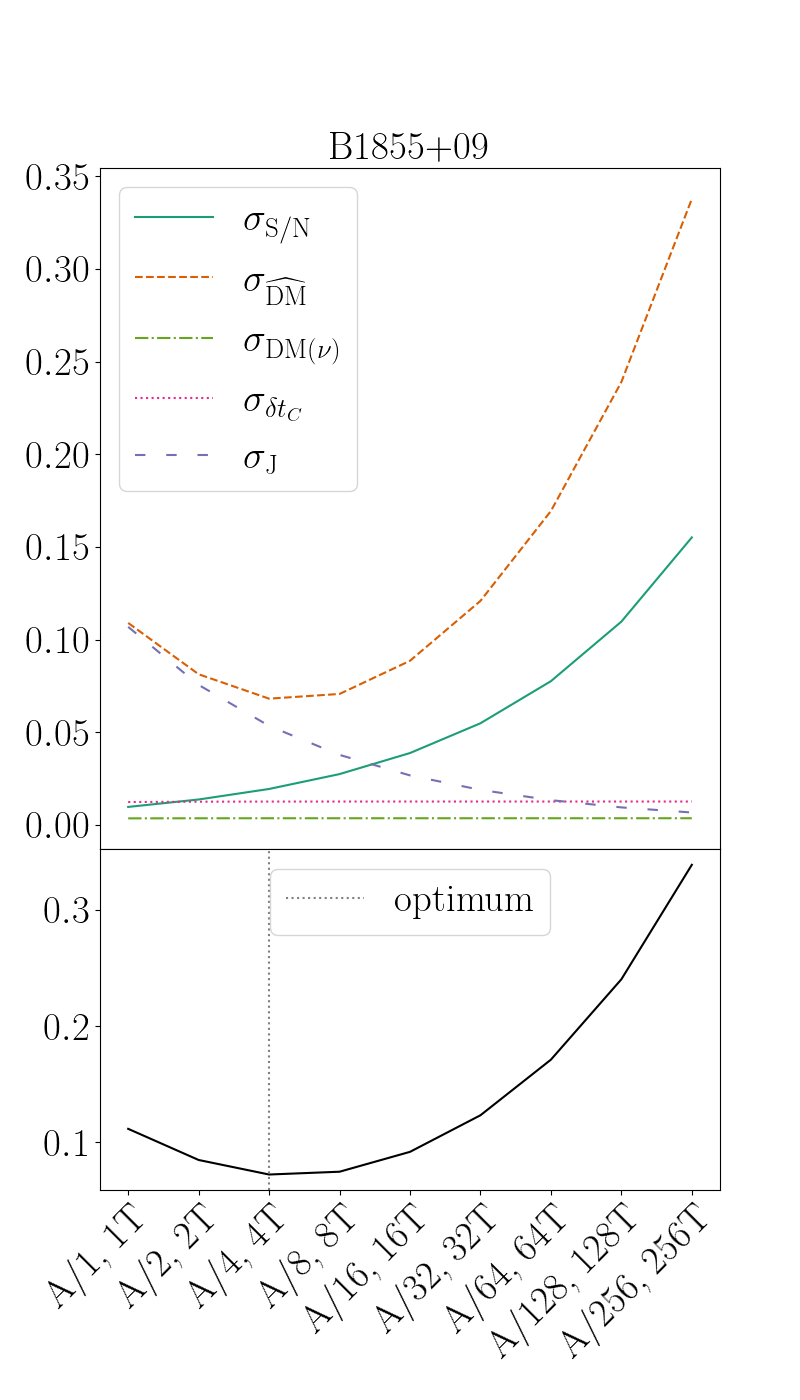}\\[-0.5ex]
	    (b)
    \end{minipage}
    \hspace{-0.04\textwidth}
    	\begin{minipage}{0.35\textwidth}
	    \centering
		\includegraphics[
	    width=\linewidth,
	    clip]{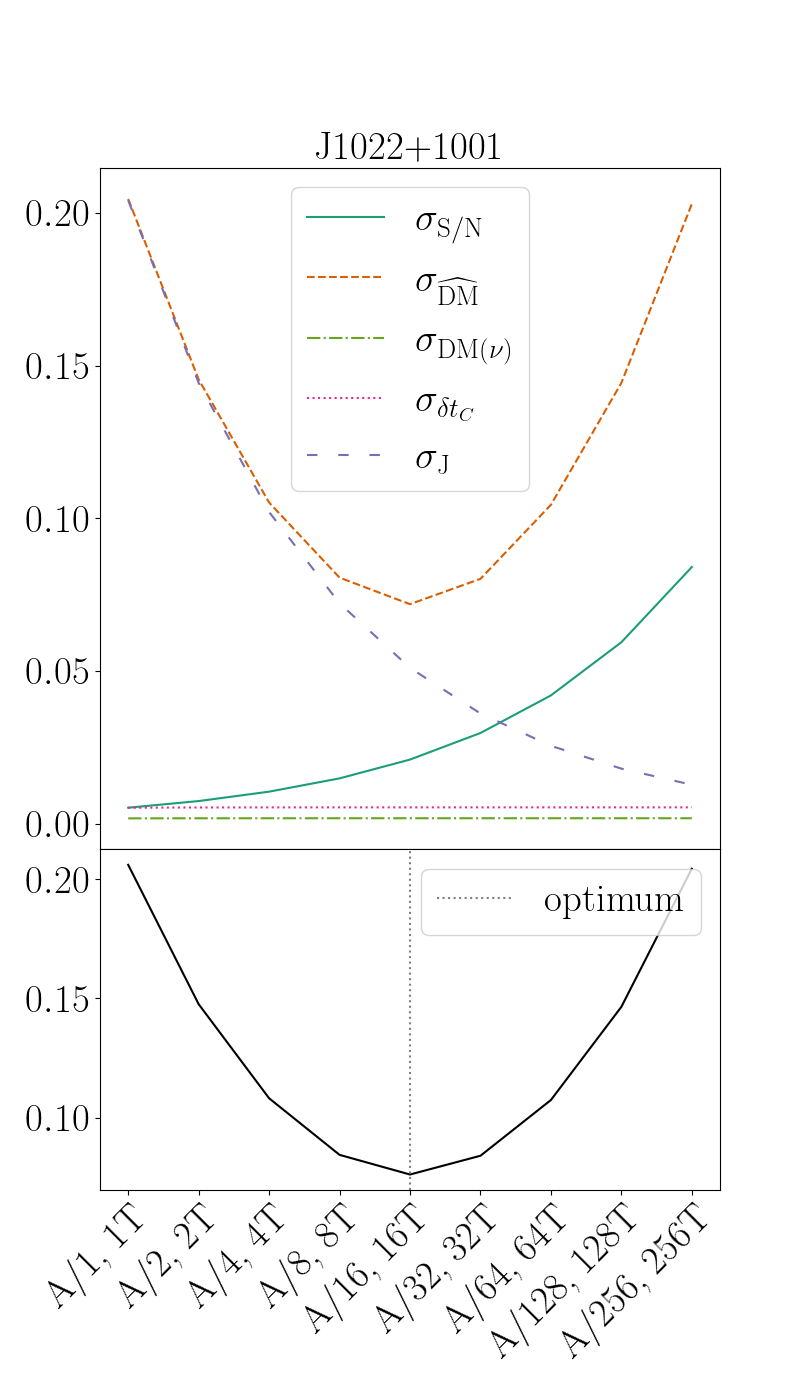}\\[-0.5ex]
	    (c)
    \end{minipage}
	\caption{
	Individual epoch-averaged noise components (top panels) and total TOA uncertainties 
	(bottom panels) vs. optimal integration factor for an (a) not jitter-dominated pulsar, J0610--2100, (b)
	a mildly jitter-dominated pulsar, B1855+09, and (c) a highly jitter-dominated
	pulsar, J1022+1001. Recall that $\sigmaTF$ and $\sigmaJ$ are weighted by the 
	frequency-frequency covariances and contained within 
	$\sigmaDMhat$, which is added in quadrature to $\sigmadtC$ and $\sigmaDMnu$ in 
	Eq. \eqref{eq:sigmaTOA}, so the curves in the top panels do not simply
	sum to the bottom panels.}

	\label{fig:jitterlim_examples}
\end{figure}

\begin{figure}[h!]
	\centering
	\includegraphics[width=0.7\textwidth]{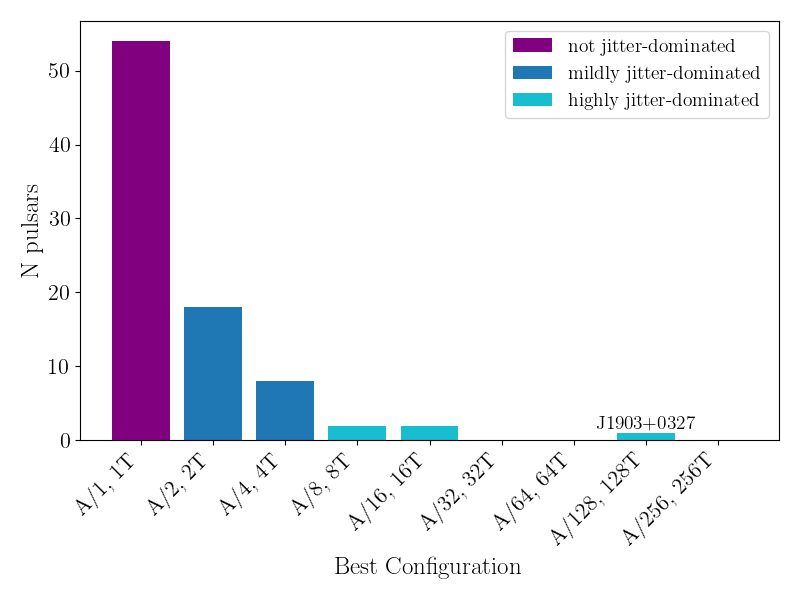}
	\caption{Histogram of optimal integration factor and jitter-dominated classification
	across the 83 pulsars in our dataset accessible to DSA. J1903+0327 is the sole pulsar with 
	$x^* > 16$ despite being dominated by $\sigmadtC$, not $\sigmaJ$.
	The majority of pulsars are not jitter-dominated.}
	\label{fig:jitterlim_hist}
\end{figure}

When designing a pulsar timing program,
it is often useful to ask whether a pulsar is dominated by jitter.
Some authors \citep[e.g.][]{cordes10, gitika25} define jitter-dominated as a pulsar which has $\sigmaTF< \sigmaJ$
and found a significant fraction of MSPs are jitter-dominated in timing observations.
We find it more useful to instead define jitter-dominated in the context of subarraying with
dish interferometers and the tradeoff between integration time and collecting area.
If multiple pulsars are jitter-dominated and at a sufficiently high elevation at the same time,
they will be timed to higher
precision by apportioning a fraction of the dishes of an interferometer to each pulsar
and increasing the integration time by the same factor, $x$.
We investigate which pulsars are dominated by jitter noise by iteratively halving
the DSA collecting area, doubling the integration time (starting from 1.75 hrs),
and computing $\sigmaTOA$
at each iteration until none of the pulsars improve.
While DSA is not currently 
expected to be capable of subarraying, we still find this method potentially
instructive for other dish arrays with finite time resources 
and a useful way to evaluate a PTA noise
budget's dependence on $\tint$ and telescope sensitivity. 
We define a pulsar to be highly jitter-dominated if the optimal integration factor that minimizes $\sigmaTOA$
is $x^* \geq 8$, mildly jitter-dominated for $x^* = 2,\;4$, and not jitter-dominated 
if $x^{*} = 1$. While these definitions are somewhat arbitrary they do have some
quantitative support. If the epoch-averaged jitter noise, $\sigmaJ$, 
is the overwhelmingly dominant component of the epoch-averaged achromatic noise, $\sigmaDMhat$,
which itself is the dominant component of $\sigmaTOA$, then it requires a larger optimal
integration factor for the template-fitting noise, $\sigmaTF$, to worsen more than $\sigmaJ$ improves.
Such a pulsar is highly dominated by jitter. If the jitter noise is larger, but not by many orders of 
magnitude, than the other sources of achromatic noise or if scattering or dispersion errors are significant,
then the cross-over between jitter and template-fitting uncertainties occurs at a lower optimal integration
factor and the pulsar is mildly jitter-dominated. If jitter never improves significantly and/or 
scattering and dispersive effects are the dominant components of noise, the pulsar is not jitter-dominated.
See Figure \ref{fig:jitterlim_examples} for an example from each classification. Figure \ref{fig:jitterlim_hist}
shows the distribution of jitter-dominated classification in our dataset and that most pulsars are not dominated by
jitter. We remind the reader that this is a fundamentally different quantity than the ratio of the jitter uncertainty
to template-fitting uncertainty.

\begin{figure}[t!]
	\centering
	\includegraphics[width=.9\textwidth]{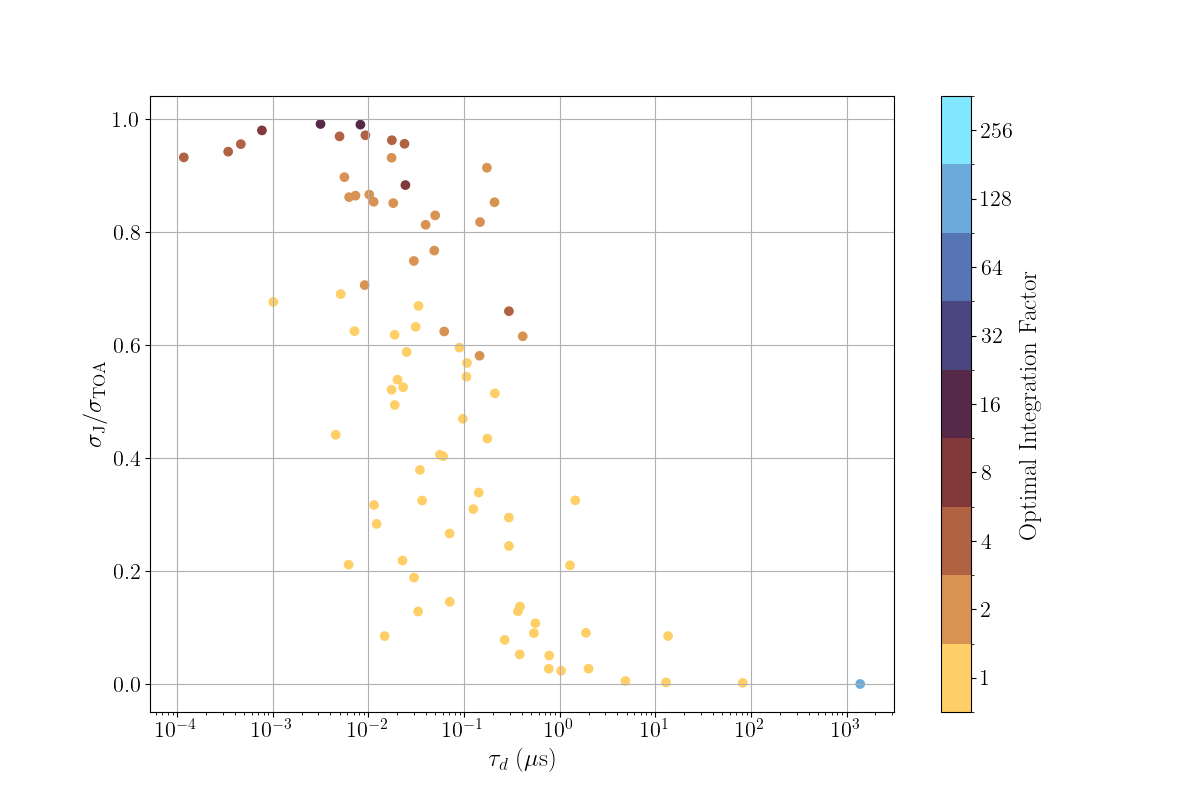}
	\caption{Scatter plot of jitter fraction, $\sigmaJ / \sigmaTOA$, 
	(at $\tint = 1.75\;\mathrm{hr}$ and full collecting area)
	 vs. scattering timescale for all DSA-timed pulsars.
	The color scale represents the optimal integration factor, with the light-blue point 
	representing the outlier J1903+0327, which is dominated by $\sigmadtC$.}
	\label{fig:jitterfrac_v_scatter}
\end{figure}

J1903+0327 breaks this classification scheme, as it is not dominated by jitter but still has a large
optimal integration factor due to large, variable scattering delays.
With $\taud=1.384\;\mathrm{ms}$,
the scattering timescale is more than half the pulse period
(likely overestimated by NE2001 -- \citet{crl08} measured $\taud=0.55\;\mathrm{ms}$)
and $\sigmadtC$ is the dominant noise component by several orders of magnitude.
Recall from Equation \eqref{eq:WNcov} that the achromatic noise covariance contains contributions 
from template-fitting and DISS. Decreasing $\Aeff$ and increasing $\tint$ results in a net
increase in $\bf{S}_{\nu_1\nu_2}$ relative to $\bf{D}_{\nu_1\nu_2}$ so  
the achromatic noise covariance goes from DISS-dominated to template-fitting-dominated. 
This changes the frequency weighting that the fixed chromatic scattering delay gets projected onto
in the fit for DM and $t_\infty$ (\LMCt Equation 23).
Because the chromatic scattering amplitude is so large for this pulsar, even this subtle reweighting
produces a small but measurable improvement in a very large $\sigmaTOA$ -- from $284\;\mathrm{\mu s}$ at $x=1$ to
$272\;\mathrm{\mu s}$ at $x^*=128$.
In general though, scattering strongly limits how much a pulsar can be improved by improving the jitter
noise, even for pulsars with $\sigmaJ / \sigmaTOA > 0.5$. Figure \ref{fig:jitterfrac_v_scatter}
shows that the fractional noise contribution from jitter and the optimal integration factor are
both negatively correlated with $\taud$.

\begin{figure}[h!]
	\centering
	\includegraphics[width=.9\textwidth]{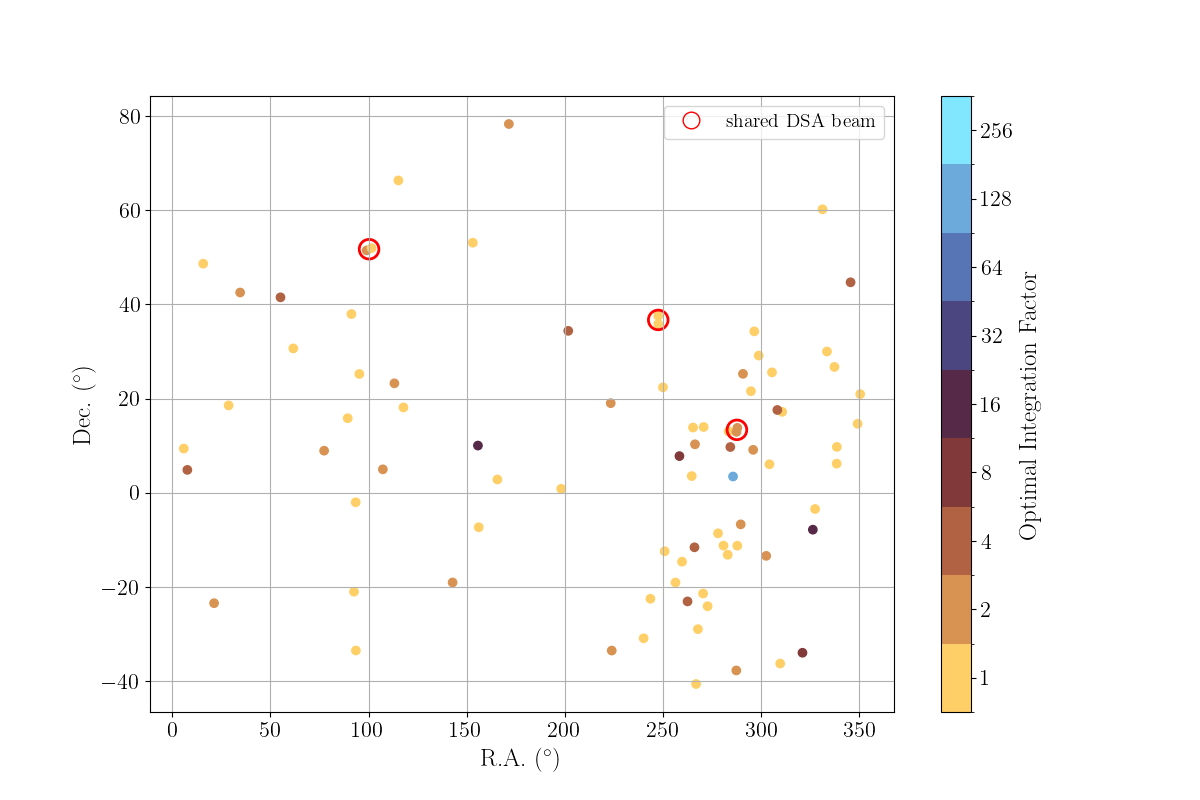}
	\caption{Sky locations for DSA-timed pulsars, colored by their optimal
	integration factor. Optimal DSA beams containing more than one pulsar
	are identified with a red circle. From left to right, the beams contain:
	J0636+5128 and J0645+5158, J1630+3550 and J1630+3734, J1910+1256 and J1911+1347.
	Beam size and shape not to scale.}
	\label{fig:sharedDSAbeams}
\end{figure}

Even without the ability to subarray, it may be possible to time multiple pulsars simultaneously 
if they fit within the same phased array beam. Positioning a pulsar off-axis will decrease
the gain and therefore the S/N, potentially more than can be made up by increasing $\tint$.
(E.g. if the gain is halved and $\tint$ is doubled the S/N still decreases by a factor $1/\sqrt{2}$.)
Assuming a Gaussian beam pattern, for $N$ pulsars to be placed off-axis while integrating for  
$N\tint$ without decreasing the S/N, all pulsars must be within
\begin{equation}
	\theta = \frac{\theta_\mathrm{FWHM}}{2} \sqrt{\frac{\ln N}{\ln 4}}
\end{equation}
of the beam center, where $\theta_\mathrm{FWHM}$ is the beam angular full-width at half-maximum.
We searched for groups of $N$ pulsars that satisfied this condition.
Accounting for the change in dish size, we assume DSA will have $\theta_\mathrm{FWHM} = 2.99\degrees$.
The result is only three beams, each containing two pulsars, none of which are highly jitter-dominated.
Figure \ref{fig:sharedDSAbeams} shows the location of these beams and the sky locations of the
DSA-accessible pulsars. 
Subarraying or strategic beam placement is of practical consideration only if there are enough 
jitter-dominated pulsars visible at the same time.
From this plot it is apparent that the field of jitter-dominated pulsars
is simply not dense enough to benefit greatly from shared beams or subarrays. This may potentially
change in the future as DSA makes new pulsar discoveries and the PTA grows, but newly discovered pulsars
are more likely to be found at larger distances with larger contributions from scattering \citep{meerkat_thousand_xv}.  
One possible remedy for increasing the number of jitter-dominated pulsars that could benefit from
re-distributing $\Aeff$ and $\tint$
is the use of cyclic spectroscopy, which can reduce or elimate scattering noise for sufficiently
high S/N pulsars \citep{dolch21_cyspec}.
\section{Optimal Integration Times}\label{sec:opt-tint}
\begin{figure}[t!]
	\centering
	\includegraphics[width=.85\textwidth]{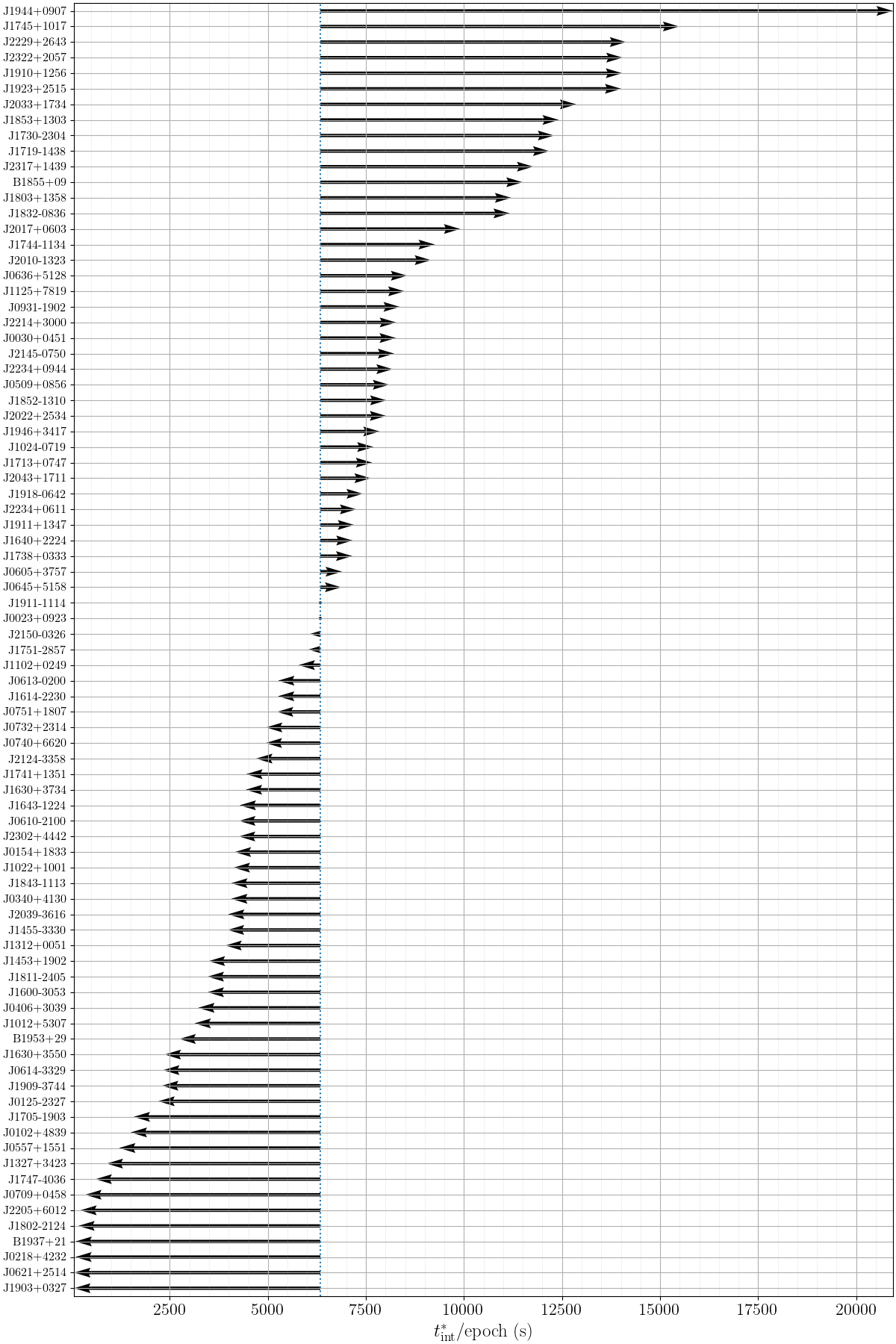}
	\caption{Optimal time shifts away from $\teq = 6300\ \mathrm{s}$ (dotted line). Pulsars
	which gain time in $\topt$ have a right-pointing arrow, while those that lose time have
	a left-pointing arrow.}
	\label{fig:timeshift}
\end{figure}

We searched for the optimal integration time vector, $\topt$, across the 83 DSA-accessible
pulsars in the PTA subject to the constraints in Equations \eqref{eq:budget_constraint} and
\eqref{eq:box_constraints}. Figure \ref{fig:timeshift} shows $\topt$ as the time shift for 
each pulsar away from the status-quo, equal-time allocation, $\teq$.
Thirty-nine pulsars lose time while fourty-four gain time. Two pulsars, J1903+0327 and J0621+2514,
get pushed to $\tmin$ and have the lowest $\Delta \rho (\topt)_{\rm i}$,
signifying that they do not contribute significant sensitivity to the GWB with DSA. 
B1937+21 is optimally observed for 93 s because while it is extremely bright, it has the most
predicted red noise.
\begin{figure}[t!]
	\centering
	\includegraphics[width=.85\textwidth]{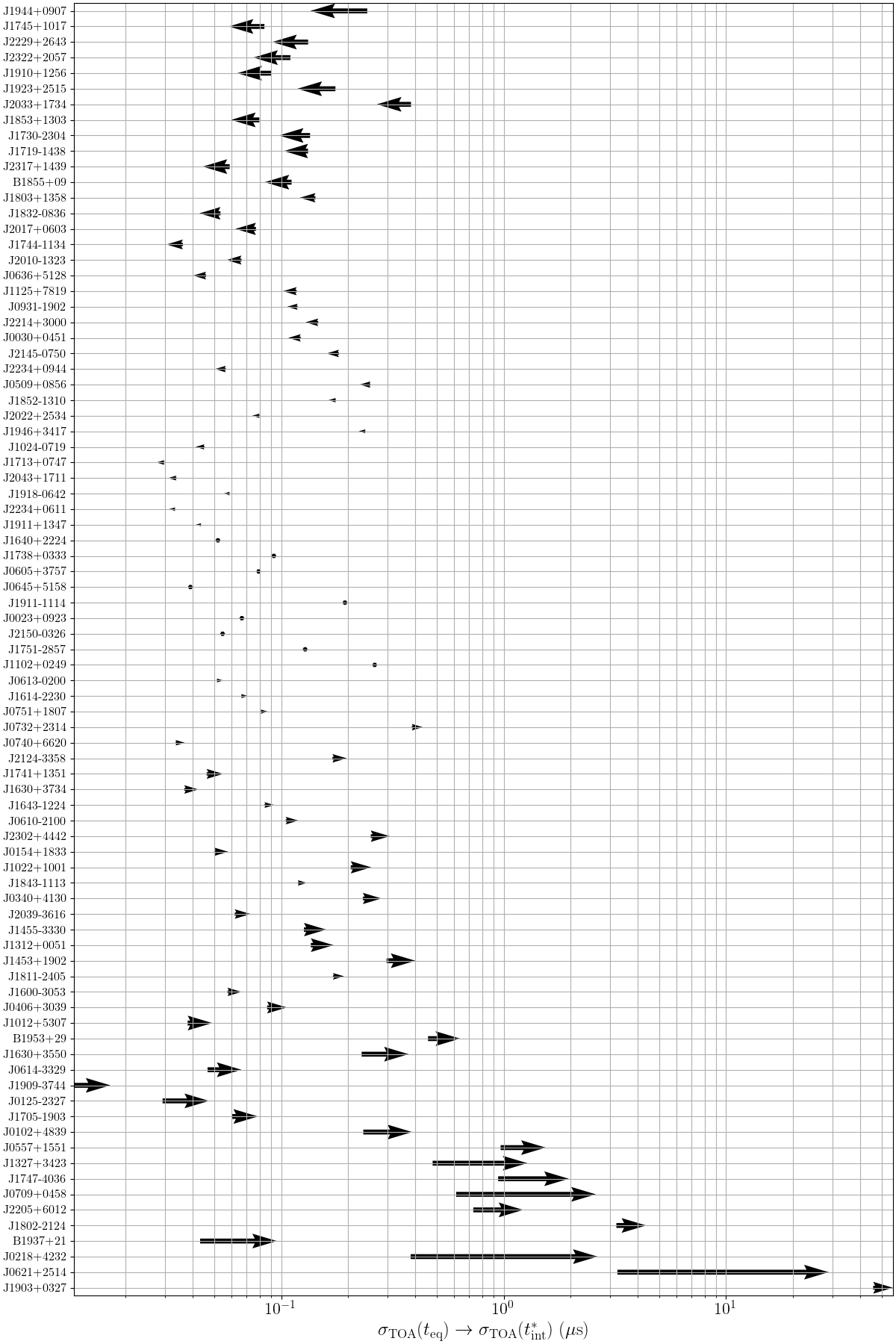}
	\caption{Changes in $\sigmaTOA$ from adding or subtracting time from a pulsar according to
	$\topt$. Pulsars are listed in the same order as Figure \ref{fig:timeshift}. Pulsars with
	right-pointing arrows are timed less precisely and pulsars with left-pointing arrows are timed
	more precisely.}
	\label{fig:sigmashift}
\end{figure}
Interestingly, four of five the dropped Arecibo pulsars are encouraged to be timed with DSA
because they are predicted to have relatively low red noise.
Figure \ref{fig:sigmashift}
shows how the change from $\teq$ to $\topt$ impacts the TOA uncertainty for each pulsar. 
This figure shows that the optimal solution sacrifices a relatively large amount of TOA
uncertainty in the $\sim$1/3 of the pulsars that are weakly influential for a relatively small
improvement in the TOA uncertainty in the $\sim$1/3 of pulsars that are strongly influential. 
Weakly influential pulsars are those which are dominated by red noise and/or scattering,
which set the limit for improvement by increasing $\tint$. We computed the power in the 
intrinsic red noise of each pulsar, 
\begin{equation}
	\sigma_{I,\mathrm{RN}}^2 = \int_{1/\Tobs}^{1/2\Delta t} P_{I,\mathrm{RN}}(f)df,
\end{equation}
where $P_{I,\mathrm{RN}}$ is the last term in Equation \eqref{eq:psr_psd}.
The four pulsars with $\tint^* < 100\ \mathrm{s}$ were among the largest intrinsic red noise power, which is 
strongly inversely correlated with $\topt$
(Figure \ref{fig:topt_vs_rnpwer}).
\begin{figure}[t!]
	\centering
	\includegraphics[width=.9\textwidth]{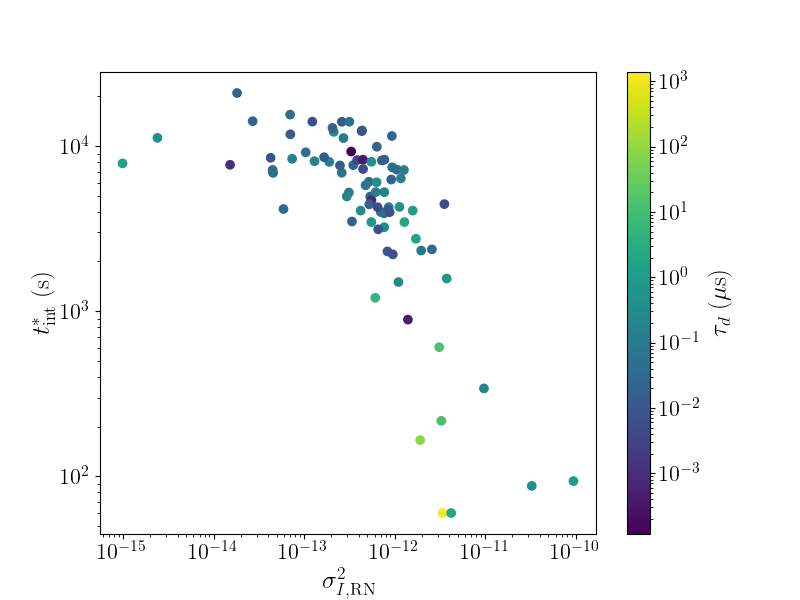}
	\caption{Optimal $\tint$ vs red noise power. Points are colored by their scattering
	timescale.}
	\label{fig:topt_vs_rnpwer}
\end{figure}
Unless a pulsar is very strongly dominated by $\sigmaDMhat$, 
having relatively large $\sigma_{I,\mathrm{RN}}^2$ is a good indicator that a pulsar will
not contribute significant GWB sensitivity. 
While we only simulate intrinsic pulsar spin red noise here, in practice, measured achromatic 
red noise can be contaminated by unmodeled or mis-modeled scattering delays \citep{larsen24}.
The influential pulsars are not necessarily the brightest, but 
do not have large predicted red noise and strong scattering and are dominated by
the components of $\sigmaTOA$ that improve by increasing $\tint$. In fact, many of the 
pulsars which would normally be considered the best timed
(e.g. see J1909--3744 in Figure \ref{fig:sigmashift}) actually sacrifice time to even out the
distribution of $\sigmaTOA_{,I}$.

\begin{figure}[t!]
	\centering
	\includegraphics[width=.8\textwidth]{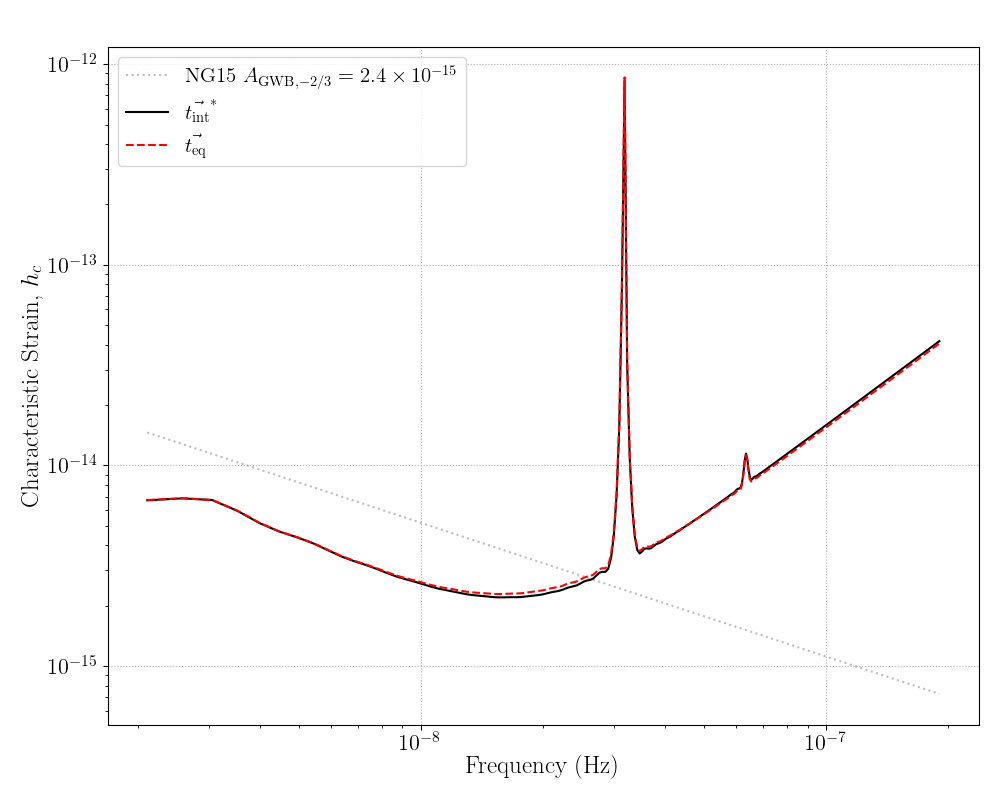}
	\caption{Strain-amplitude sensitivity curves for a NANOGrav 15-year PTA with DSA observing
	pulsars for the optimal time (solid, black curve) and equal time (dashed, red curve) per epoch.
	The dotted line is a GWB power-law spectrum using the \NGXVgwbt Bayesian posterior amplitude.}
	\label{fig:optveq_sencurves}
\end{figure}

In general, $\sigmaTOA_I(\tint_{,i})$ decreases smoothly and monotonically toward $\tmax_{,i}$
and $F(\tvec)$ increases smoothly toward the intersection of the per-pulsar hypercube constraints
and the ($\Npsrs - 1$) dimensional budget boundary hyperplane where $\topt$ lies.
Time swaps between random pairs of pulsars reveal that this optimum is stable but shallow.
In 700 feasible swaps, the largest decrease in $F(\tswap)$ away from $F(\topt)$ was only 0.04\%.
The improvement in the GWB S/N from equal time to optimal time is small:
from $\rho(\teq) = 14.28$ to $\rho(\topt) = 14.85$.
This can be explained by looking at the optimal- and equal-time solution strain-amplitude
sensitivity curves (Figure \ref{fig:optveq_sencurves}). 
The lowest GW frequencies are dominated by the GWB, so $\tint$ doesn't matter and the curves overlap.
At $\sim 10^{-8}\;\mathrm{Hz}$, white noise becomes significant but the noise floor is still set by the 
GWB and individual pulsar red noise so the improvement from optimizing $\tint$ is small, but
noticeable. At the highest GW frequencies, the white noise dominates and its noise floor is set
by the few "best" pulsars. But $\topt$ sacrifices time on the best pulsars effectively raising
the white noise floor at high frequencies where $h_c(f;\teq)$ appears slightly better. 
When we ran the optimization while holding the best five pulsars (by $\sigmaTOA(t_\mathrm{eq})$)
fixed at $t_\mathrm{eq}$, this effect went away.
Below $1.5\times 10^{-8}\;\mathrm{Hz}$, the GWB strain spectrum is above the PTA curve.
In this intermediate GWB signal limit, $\tint$ does not strongly impact $\rho$, leading to the shallow
optimum.

One could reasonably ask: what if the pulsar models are incorrect? Scintillation can lead to inaccurate
flux measurements and spectral index estimates and NANOGrav TOA selection biases flux measurements high.
NE2001 tends to underestimate pulsar distances at high Galactic latitude and systematically underestimates scattering for sight-lines towards the inner Galaxy \citep{ne2025}.
Intrinsic red noise measurements can often contain unmodeled effects like orbital perturbations
or chromatic noise from ISM dispersion and scattering \citep{larsen26},
and most \NGXVt pulsars do not have significant red noise measurements \citep{ng15_noisebudget}.
We address these issues by generating Monte Carlo realizations of our pulsar list with the most 
uncertain properties drawn from plausible distributions. We draw fluxes at two observing frequencies
from log-normal distributions
\begin{equation}
	\ln(S_{\nu}) \sim \mathcal{N}\left(\ln(q_{50}), 
						\left(\frac{\ln\left(q_{84}/ q_{16}\right)}{2z_{0.84}}\right)^2 \right)
\end{equation}
for the majority of pulsars where we have 16th, 50th, and 84th quantile flux measurements,
$q_{16}$, $q_{50}$, and $q_{84}$ respectively. Then we compute $S_{1000}$ according to Section \ref{ssec:psrmodels}.
For the pulsars with insufficient observations to draw from a measured distribution we drew pseudoluminosities
at 1.4 GHz from the MSP population distribution \citep{ridley_lorimer_10, psrpoppy}
\begin{equation}
	\log_{10} L_{1400} = \log_{10} \gamma_L + \alpha_L \log_{10} P + 
					\beta_L \log_{10}(\Pdotint/10^{-15}) + \delta_L
\end{equation}
where $\gamma_L=0.009$, $\alpha_L = -1.4$, $\beta_L=0.5$,
and $\delta_L \sim \mathcal{N}(0, 0.8^2)$ \citep{rajwade16}. For these pulsars we
computed the 1.4 GHz flux as $S_{1400} = L_{1400} / d^2$, drew a spectral index from
$\alpha \sim \mathcal{N}(-1.47, 0.7^2)$ \citep{al22}, and scaled the flux to 1 GHz according
to Equation \eqref{eq:1GHzflux}.
We sampled single pulse jitter from a regression model we fit to the 1.5 GHz jitter measurements in 
\citet{lam19_jitter}:
\begin{equation}
	\log_{10}(k_{\mathrm{J}, 1500,i}) \sim \mathcal{N}\left(a + b\log_{10}(W_{50;1500,i}/P_i),
											\sigma_{\delta \mathrm{J}}^2 \right)
\end{equation}
where $\sigma_\mathrm{J,1} = P k_\mathrm{J}$, $a = -0.54$, $b=0.93$, and 
$\sigma_{\delta \mathrm{J}}=0.29$. We re-sampled the histogram from Figure 10 of 
\citet{ne2001} for the multiplicative factor $r_\mathrm{scatt}$, which when multiplied
by $\taud$, creates scatter around the predicted value intrinsic to the uncertainties in the NE2001 model.
Finally, for the pulsars included in \citet{ng15_noisebudget}, we drew $\Ared$ and $\gammared$ from the posterior
distributions of these parameters, regardless of whether they had significant detection of red noise.
For all others, we drew $\Ared$ from the 
\citet{sc10_rednoise, ng9_excessnoise} population distribution described in Appendix \ref{appx:rednoise},
assuming a fixed spectral index, $\gammared = 5.6$.
We generated 100 realizations of the NANOGrav
PTA pulsars, searched for $\topt$, and computed $\rho^*$ and $h_c(f;\topt)$ for each.

\begin{figure}[t!]
	\centering
	\includegraphics[width=.8\textwidth]{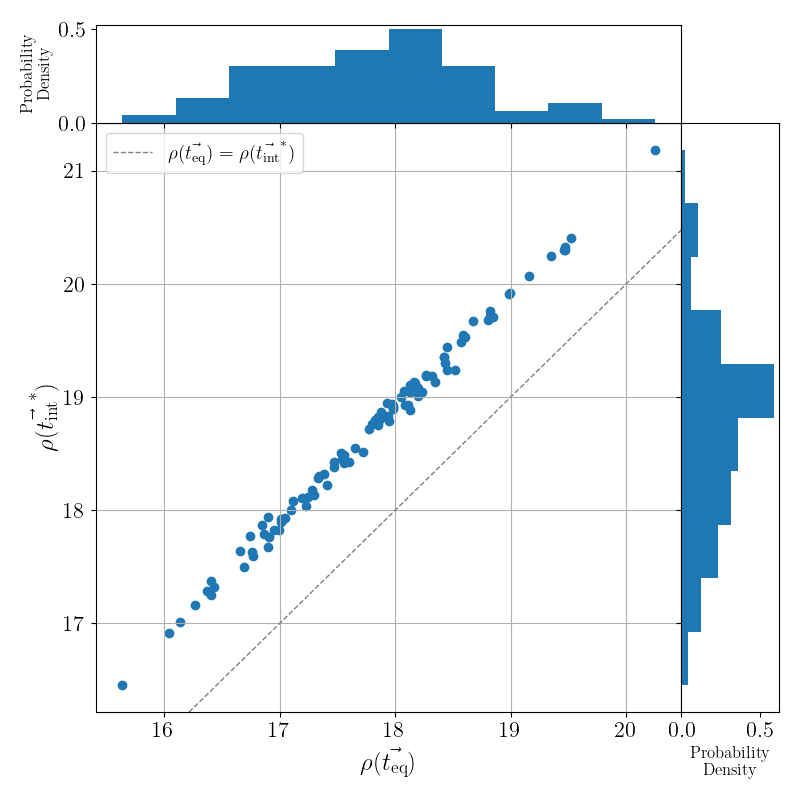}
	\caption{Optimal vs equal-time allocation GWB S/N for 100 Monte Carlo realizations of the PTA and
	marginal histograms for $\rho(\teq)$ and $\rho(\topt)$.
	The dashed line identifies no difference between the optimal and equal-time S/N. The correlation between
	$\rho(\teq)$ and $\rho(\topt)$ is $r = 0.997$ and $\rho(\topt)$ is consistently higher by 0.9.}
	\label{fig:snrscatterhist}
\end{figure}
\begin{figure}[t!]
	\centering
	\includegraphics[width=.9\textwidth]{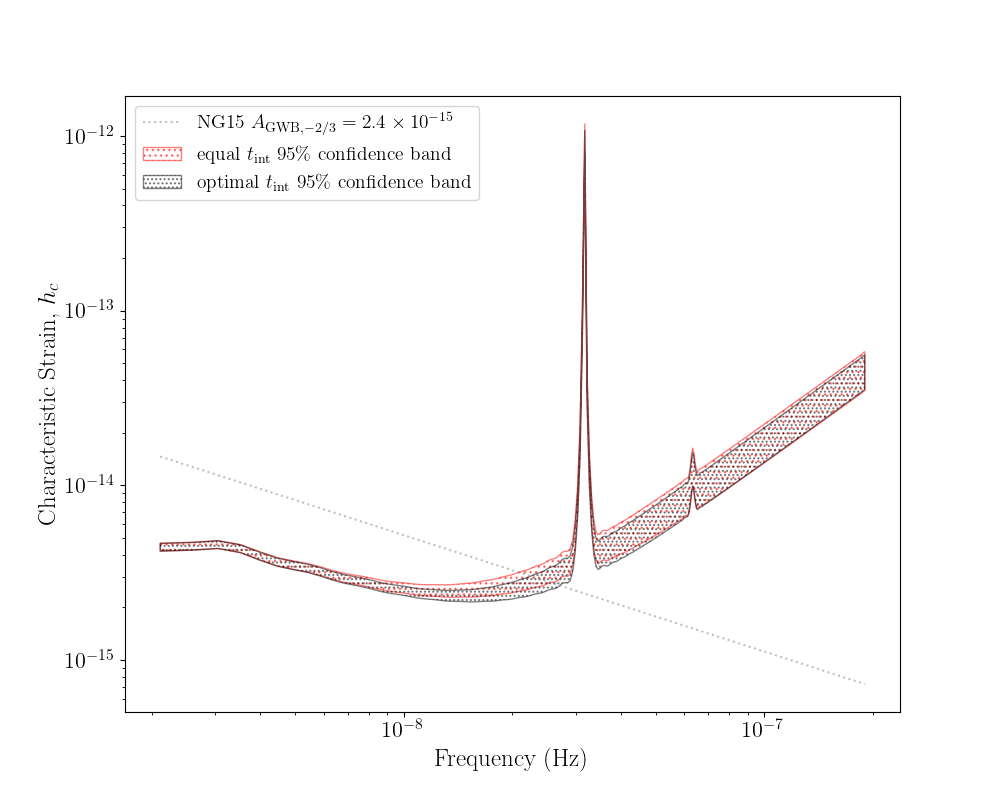}
	\caption{95\% confidence bands for $h_c(f;\teq)$ (red) and $h_c(f;\topt)$ (black) along with
	the assumed GWB power-law spectrum (dotted line).}
	\label{fig:strainspec_conintervals}
\end{figure}

Figure \ref{fig:snrscatterhist} shows the GWB S/N of these realizations and the marginal histograms
of $\rho(\teq)$ and $\rho(\topt)$. 
These simulations show that $\rho(\teq)$ and $\rho(\topt)$ are tightly correlated and that the relative
improvement by optimizing over $\tint$ does not depend strongly on the pulsar noise. A least-squares
regression gives the relationship a slope of 1.0 and an intercept of 0.9.
Figure \ref{fig:strainspec_conintervals} shows the 95\% confidence bands for the GW strain sensitivity
curves $h_c(f;\teq)$ and $h_c(f;\topt)$ over the Monte Carlo realizations. These bands exibit the same
trends as the assumed PTA model: no difference between optimal and equal-time at low frequencies and
a small difference above $10^{-8}$ Hz. However, the high-frequency cross-over is not prominent in the
Monte Carlo realizations, likely because independent sampling of flux, jitter, scattering, and red
noise changes the hierarchy of "best" pulsars from realization to realization leaving the high-frequency
sensitivity less concentrated in a small set of pulsars.
\begin{figure}[t!]
    \makeatletter
    \let\@makecaption\@gobbletwo
    \makeatother
    \caption[]{}
\end{figure}

These findings are in line with previous works exploring the scaling laws governing the
GWB sensitivity of PTA experiments. \cite{cordesshannon12} assessed that red timing noise and ISM
effects will be the limiting factors in optimizing detection significance. \citet{siemens13} and
\citet{vigeland16_scalinglaws} showed that in the intermediate-signal regime, when the GWB power is
much greater than the PTA white noise at low frequencies and much lower at high frequencies, reducing
the white noise by minimizing $\sigmaTOA$ does not significantly improve the GWB S/N. They showed that 
in this regime, the cross-correlations, $\chi_{IJ}$, are far more imporant. This is why $\topt$ still 
allocates time to pulsars with $\sigmaTOA(\teq) > 1\ \mathrm{\mu s}$; the number of pulsars in the PTA
is the most effective way to increase the S/N in this regime. 

\section{Conclusions}\label{sec:conclusion}
We have presented a novel framework for optimizing the observing frequency and integration
time of a PTA to maximize GWB sensitivity. We applied that framework to an expanded NANOGrav
source list observed with the DSA, but this methodology will be broadly applicable to other PTA
experiments and radio telescopes, expecially dish arrays like ngVLA and SKA. Our main results are
summarized below:
\begin{enumerate}
	\item An ab initio, 15-year DSA PTA with a 25\% total time allocation would have significantly improved
	GWB sensitivity over a 15-year PTA with current feasible instrument combinations, owing to
	an overall increase in time allocation for the observing program, instrumental sensitivity, 
	and simultaneous bandwidth. Even if the DSA time allocated to NANOGrav is equal to our current time allocation,
	pulsar timing observations will still be significantly more precise than with current instruments.
	\item Most pulsars in our source list will not be highly jitter-dominated with DSA, in the
	sense that their timing precision does not improve significantly with reduced instrumental
	sensitivity and increased integration time. The benefits of subarraying or strategic beam
	placement are highly limited due to the scarcity and distribution of highly jitter-dominated pulsars
	on the sky. This could change if new discoveries and/or techniques like cyclic spectroscopy,
	which can enable the removal of time-variable scattering delays, increase
	the number density of jitter-dominated pulsars accessible to DSA.
	\item In the DSA era, the GWB S/N will be dominated by red noise from the GWB itself.
	This limits the improvement in the GWB S/N achievable by optimizing observing frequency and integration
	time over the PTA subject to a fixed time budget to only a few percent. The opportunity cost of an
	early science DSA observing program with optimal per-pulsar integration time over equal-time allocation is low.
	This could change as new pulsar discoveries are made -- optimizing integration time could help cull the
	MSP population to only those which contribute GWB sensitivity.
	\item The overall GWB S/N is sensitive to the pulsar noise characteristics but the relative improvement
	from optimization is not. Monte Carlo simulations reinforced the conclusion that even if the pulsar noise
	characteristics are considerably different than what we've assumed, the improvement in GWB S/N is a
	small, constant offset. 
	\item In the intermediate GWB signal limit, the most important attributes for determining whether a pulsar will 
	contribute GWB sensitivity are intrinsic pulsar red noise and, to a lesser degree, the amount of 
	multi-path scattering by the ISM. 
\end{enumerate}

We give the caveat that our results are highly dependent on our assumed pulsar noise models. 
We sought to address this through our Monte Carlo analysis, but some of the predictive distributions are 
inconsistent with current measurements, especially for red noise.
Using custom noise modeling to mitigate unmodeled chromatic noise,
\citet{ng15_custnoise} found that only six pulsars in the \NGXVt dataset have
significant red noise measurements when accounting for a common red noise process.
As subsequent NANOGrav datasets will have increased timespans and tighter constraints on
spatially-correlated red noise from the GWB, constraints on intrinsic pulsar red noise will improve and
so too will the accuracy of our predictions.
There are many other opportunities for improving the robustness of our 
pulsar models.
Future optimal strategies analyses will use newer Galactic electron density models \citep[e.g.][]{ne2025}.
\citet{lam19_jitter} reported that a majority of NANOGrav pulsars have observing
frequency-dependent, single-pulse jitter noise.
Since some of the frequency-dependent models did not cover the full DSA band,
we assumed their constant-in-frequency model for all pulsars rather than extrapolate. It will be relatively
simple to include a frequency-dependent jitter model into our optimization framework in the future.
Our pulsar noise model accounts for stochastic, frequency-dependent variations in pulse profile shapes 
due to DISS, but some pulsars also exhibit frequency-evolution of their profile shape intrinsic to 
their emission \citep{kramer99, pennucci19}. It is difficult to disentangle intrinsic profile evolution
from the impulse response of the ISM, so we have assumed that $\Weff$ and $\Uscale$ are constant with
observing frequency. We are currently undertaking an analysis of MSP profile shapes and may be able to
use $\Weff(\nu)$ and $\Uscale(\nu)$ in future analyses. Both frequency-dependent jitter and profile shape
evolution will re-weight the pulsar achromatic noise covariances in the computation of $\sigmaTOA$.

We have only scratched the surface of possible GW sources and signals to which we can optimize our PTA.
In our analysis we have assumed a fixed GWB amplitude and spectral index, but we could explore whether
our optimal $\tint$ results hold for different GWB spectral indices or sample $A_\mathrm{GWB}$ from the
\NGXVt GWB amplitude posterior.
Additionally, in the post-GWB detection era,
the sensitivity to the GWB becomes less of a target for optimization than our ability to characterize its spectrum.
In the future, we aim to instead minimize an objective that is more akin to the measurement uncertainty 
on the GWB amplitude and spectral index. Such an objective will depend on the integration time per pulsar in a more nuanced way as the integration time also affects our ability to characterize the chromatic noise in our pulsars.
\citet{hazboun19} also presented a statistic for the S/N of a PTA to a single, deterministic GW source.
We could perform the same decomposition (as in Appendix \ref{appx:GWB_SN_QP}) of this statistic 
into its quadtratic form and write an objective function for CW sensitivity
across the whole sky or in targeted directions where we suspect strong CW sources to exist
without having to modify the rest of the simulation machinery.
GW spectrum parameter estimation is even more important for CWs, so we could modify the Fisher matrix
approach we took in \citet{liu23} to minimize amplitude and spectral index measurement uncertainties
for CWs with DSA to include the GWB and intrinsic pulsar red noise.
When pulsar and GWB red noise dominate, the cadence does not appreciably 
affect the GWB sensitivity \citep{cordesshannon12}. On the other hand, the cadence
increases the high-(GW) frequency
sensitivity and the detector volume to CW sources \citep{baier25}, thus we would seek to optimize over 
$\tint$ and $\Delta t$ simultaneously and explore the geometry of this $2\Npsrs$-dimensional
objective function. Finally, it is an open question as to how many total MSPs can or should be
included in a PTA in the DSA era. We plan to explore this through optimization of synthesized populations
of MSPs and simulations of next-generation surveys in a future paper.

\begin{acknowledgments}
	Thanks to Sarah Vigeland for helpful discussions about the gravitational wave background signal-to-noise ratio.
Thanks to Jonas Flygare at Owens Valley Radio Observatory for discussions about the DSA passband and for sharing data from his paper.
T.C.\ received support through the Grote Reber Fellowship Program administered by Associated Universities, Inc./National Radio Astronomy Observatory and from the NANOGrav National Science Foundation Physics Frontiers Center award no. 2020265.
P.R.B.\ is supported by the Science and Technology Facilities Council, grant number ST/W000946/1.
Pulsar research at UBC is supported by an NSERC Discovery Grant and by CIFAR.
K.C.\ is supported by a UBC Four Year Fellowship.
M.E.D.\ acknowledges support from cooperative agreement N00173-23-2-C603, between the Naval Research Laboratory and George Mason University.
T.D.\ and M.T.L.\ received support from an NSF Astronomy and Astrophysics Grant (AAG) award number 2511107 during this work.
E.C.F.\ is supported by NASA under award number 80GSFC24M0006.
D.C.G.\ is supported by NSF Astronomy and Astrophysics Grant (AAG) award \#2406919.
D.R.L.\ and M.A.M.\ are supported by NSF \#1458952.
M.A.M.\ is supported by NSF \#2009425.
The Dunlap Institute is funded by an endowment established by the David Dunlap family and the University of Toronto.
T.T.P.\ acknowledges support from the Extragalactic Astrophysics Research Group at E\"{o}tv\"{o}s Lor\'{a}nd University, funded by the E\"{o}tv\"{o}s Lor\'{a}nd Research Network (ELKH), which was used during the development of this research.
H.A.R.\ is supported by NSF Partnerships for Research and Education in Physics (PREP) award No.\ 2216793.
S.M.R.\ and I.H.S.\ are CIFAR Fellows.
Portions of this work performed at NRL were supported by ONR 6.1 basic research funding.
\end{acknowledgments}

\begin{contribution}
T.C. undertook all of the data analysis, wrote the PTAOptimizer software package
and analysis scripts, made the corrections and feature additions to the FrequencyOptimizer
software package described in the paper, and wrote the paper. 
P.B.D. provided mentorship, suggestions, and guidance regarding the analysis and writing of the paper. 
J.G.B. provided guidance on generating and interpreting PTA sensitivity curves with
hasasia and added capabilities to hasasia to enable rapid computation.
R.S.L. contributed frequency-dependent receiver passband capabilities to FrequencyOptimizer. 
M.T.L. provided useful discussions on FrequencyOptimizer and covariance matrix approaches
to TOA uncertainty estimation.
J.H.S provided helpful discussions about the gravitational wave background signal-to-noise ratio.
D.J.N., A.L.B, and H.E.M. provided NG15 flux-density estimates.
S.M.R. provided GBT profile data for J1102+0249, J1852-1310, and J0102+4839.
G.A., A.A., A.M.A., Z.A., P.T.B., P.R.B., H.T.C., K.C., M.E.D., P.B.D., T.D., E.C.F., W.F., E.F., G.E.F., N.G., P.A.G., J.G., D.C.G., J.S.H., R.J.J., M.L.J., D.L.K., M.K., M.T.L., D.R.L., J.L., R.S.L., A.M., M.A.M., N.M., B.W.M., C.N., D.J.N., T.T.P., B.B.P.P., N.S.P., H.A.R., S.M.R., P.S.R., A.S., C.S., B.J.S., I.H.S., K.S., A.S., J.K.S., and H.M.W. developed the NANOGrav 15-year dataset.
\end{contribution}

%
\facilities{GBT (NRAO), VLA(NRAO), CHIME (DRAO), DSA (Caltech)}



\appendix
\section{The Frequency-Dependent DM Error}\label{appx:DMnu_covmat}
The frequency-dependent DM error, $\sigmaDMnu$, is the RMS timing error
induced by measuring the DM at multiple frequencies.
Begin equating the two-frequency structure function in Appendix B of \citet{cordes16_freqDM},
	\begin{align}
		 \sigmaDMbar^2(\nuh, \nul) &= 
			\left\langle
				\left[
				\Delta \DMbar(\nuh, \nul, \,\xvec)
				\right]^2	 \nonumber	  
			\right\rangle\\
		&=
		 \Cdmbarnn +  \Cdmbarnpnp -2\Cdmbarnnp		\label{eq:DMstrucfunc}
	\end{align}
and Equation 12 of \citet{cordes16_freqDM},
	\begin{align}
		\sigmaDMbar^2(\nuh, \nul) = \left( 4.42 \times 10^{-5}\; \mathrm{pc\; cm^{-3}} g_\beta q_\beta  
			F_\beta(\nuh/\nul) \frac{\nuh \phi^2_\mathrm{F}(\nuh)}{1000} \right)^2	
	\end{align} 
where $\nuh > \nul$; $g_{11/3}=0.46$ and $q_{11/3}=1.15$ for a uniform Kolmogorov scattering medium ($\beta = 11/3$);
\begin{equation}
	\phi_F(\nu) \approx 
			9.6~{\rm rad} 
			\left(\frac{\nu/\Dnud}{100} \right)^{5/12},
	\label{eq:phiF1}
\end{equation}
$\dnud \propto \nu^{22/5}$,
and
\begin{equation}
	F_\beta(r\equiv\nuh/\nul) = \left\{ 2^{(4-\beta)/2} \left[ 1 + r^{2\beta/(\beta-2)} \right]^{(\beta-2)/2} - r^{\beta} -1 \right\}^{1/2}.
	\label{eq:Fbeta}
\end{equation}
Let 
	\begin{equation}
		D_{ij} =  \sigmaDMbar^2(\nuh, \nul)
		\qquad\mathrm{and}\qquad
		C_{ij} = \Cdmbarnnp.
	\end{equation}
Then (\eqref{eq:DMstrucfunc}) becomes
\begin{align}
	D_{ij} &= C_{ii} + C_{jj} - 2C_{ij}\\
	{\bf D} &= \mathrm{diag}({\bf C}){\bf 1}^T + {\bf 1}\mathrm{diag}({\bf C})^T -2{\bf C}.
\end{align}
Because the structure function determines the covariance only up to an unresolved common mode,
require that ${\bf C}$ be a centered covariance matrix and apply the centering matrix 
\begin{equation}
	{\bf J} = {\bf I} - \frac{1}{N}{\bf 1}{\bf 1}^T
\end{equation}
to both sides so the centered cross-frequency covaraiance matrix of $\delta \DMbar$ is
\begin{equation}
	{\bf C}_{\delta \DMbar} = -\frac{1}{2}{\bf J}{\bf D}{\bf J}.
\end{equation}
Let
\begin{equation}
	{\bf T} = {\bf T}_0 + {\bf T}_{\delta \DMbar},
\end{equation}
the data vector of TOAs across the frequency channels within an epoch
where ${\bf T}_0$ are the components of the TOAs modeled by the DM fit and 
\begin{equation}
	({\bf T}_{\delta \DMbar})_k = K\nu_k^{-2}\delta \DMbar(\nu_k).
\end{equation}
with covariance
\begin{equation}
	{\bf C}_{T,\delta \DMbar} = K^2\diag(\nu_k^{-2}){\bf C}_{\delta \DMbar}\diag(\nu_k^{-2}).
\end{equation}
Propagate this through
the fit for $t_\infty$ and DM, where the systematic error is the difference between the fit when accounting for DM
only and when accounting for the DM difference
\begin{equation}
	\delta {\bf \theta}_{\delta \DMbar} = \left({\bf X}^T {\bf C}^{-1}{\bf X}\right)^{-1}
											{\bf X}^T{\bf C}^{-1}{\bf T}_{\delta \DMbar}.
\end{equation}
where ${\bf X} = \mathrm{matrix}(1\; \nu_k^{-2})$,  ${\bf \theta} = \mathrm{matrix}(t_\infty\; K\DM)$, and ${\bf C}$ is the achromatic noise covariance matrix.
The covariance of this perturbation is 
\begin{align}
	\langle \delta {\bf \theta}_{\delta \DMbar} \delta {\bf \theta}_{\delta \DMbar}^T\rangle &=	
			\left({\bf X}^T {\bf C}^{-1}{\bf X}\right)^{-1}{\bf X}^T{\bf C}^{-1}{\bf C}_{T,\delta \DMbar}
			{\bf C}^{-1}{\bf X}\left({\bf X}^T {\bf C}^{-1}{\bf X}\right)^{-1}\\[1em]
			&= \begin{bmatrix}
				\Var(t_\infty) & \Cov(t_\infty,\,K\DM)\\
				\Cov(K\DM,\,t_\infty) & \Var(K\DM).
			\end{bmatrix}
\end{align}
$\sigmaDMnu^2 = \Var(t_\infty)$, the first diagonal element of this matrix.
\section{The Quadratic Form of the GWB S/N}\label{appx:GWB_SN_QP}
\subsection{White Noise-only}

For illustrative purposes, we begin assuming the pulsar residual PSD
consists only of white noise.
In order to write the GWB S/N in a quadratic structure,
\begin{equation}\label{eq:quadratic}
	F(\tvec) = p(\tvec)^T\Qmat p(\tvec),
\end{equation}
when 
\begin{equation}
	    P_I(f,\tint_i) = 2\sigmaTOA_{,I}(\tint_{,i})^2\Delta t,
\end{equation}
substitute the PTA PSD expression over pulsar pairs into $\rho$:
\begin{equation}
    \rho^2\simeq
    2T_{\rm obs}\int_0^{f_{\rm Nyq}}{\rm d}f\>
    S_h^2(f)
    \sum_{I<J}
    \frac{T_{IJ}}{T_{\rm obs}}
    \frac{\chi^2_{IJ}}{S_I(f)S_J(f)}.
\end{equation}
Define the per-pulsar weight
\begin{equation}
    a_I(f) = \frac{1}{S_I(f)}.
\end{equation}
Then 
\begin{equation}
    \rho^2\simeq
    2T_{\rm obs}\int_0^{f_{\rm Nyq}}{\rm d}f\>
    S_h^2(f)
    \sum_{I<J}
    \frac{T_{IJ}}{T_{\rm obs}}
    \chi^2_{IJ}a_I(f)a_J(f).
\end{equation}
Define the $N_\mathrm{psr} \times N_\mathrm{psr}$ symmetric matrix $\bf W$ such that
\begin{equation}
    \sum_{I<J} w_{IJ}a_I a_J = \frac{1}{2}\sum_{I\neq J} w_{IJ}a_I a_J = a(f)^T {\bf W} a(f)
\end{equation}
where 
\begin{equation}
    W_{IJ} = \frac{1}{2}\frac{T_{IJ}}{T_\mathrm{obs}} \chi^2_{IJ}\;\;\mathrm{for}\; I\neq J
\end{equation}
and
\begin{equation}
    W_{II}=0.
\end{equation}
The approximation for $\mathcal{N}^{-1}(f)$ is exact for a white noise-only PSD, so the pulsar weights are
\begin{align}
    a_I(f) = \mathcal{N}_I^{-1}(f)\mathcal{R}(f) &= \frac{\mathcal{T}(f)\mathcal{R}(f)}{P_I(f)}\\
    	&= \frac{\mathcal{T}(f)\mathcal{R}(f)}{2\Delta t}\sigma_I^{-2}.
\end{align}
Then the GWB S/N is exactly in the quadratic form
\begin{equation}
    \rho^2 = p^T{\bf Q}p,
\end{equation}
where 
\begin{equation}
    {\bf Q} = 2T_\mathrm{obs}
    \left[\int^{f_\mathrm{Nyq}} df S_h(f)^2 
    \left(\frac{\mathcal{T}(f)\mathcal{R}(f)}{2\Delta t}\right)^2 \right]{\bf W}
\end{equation}
and 
\begin{equation}
    p \equiv \sigma_I^{-2}.
\end{equation}
For the white noise-only case, the function to optimize is 
\begin{equation}
    F(t) \equiv \rho^2(t) = p(t)^T {\bf Q} p(t),
\end{equation}
where
\begin{equation}
	p_i(t_i) = \sigma_i^{-2}(t_i).
\end{equation}
and
\begin{equation}
    p(t) =
    \begin{bmatrix}
        p_1(t_1) \\
        \vdots \\
        p_N(t_N)
    \end{bmatrix}.
\end{equation}
The differential is 
\begin{equation}
    dF = d\left(p^T {\bf Q} p \right) = (dp)^T {\bf Q} p + p^T {\bf Q}(dp) = 2 p^T{\bf Q} (dp).
\end{equation}
Let $u(t) \equiv {\bf Q} p(t)$, so that 
\begin{equation}
    dF = 2 u(t)^T dp.
\end{equation}
Componentwise,
\begin{equation}
    \frac{\partial F}{\partial t_i} = 2 u_i(t) \frac{dp_i}{dt_i}.
\end{equation}
Subsituting,
\begin{equation}
    \frac{dp_i}{dt_i} = \frac{d}{dt_i}(\sigma_i^{-2}) = -2\sigma_i^{-3}\sigma_i'(t_i)
\end{equation}
gives the gradient as 
\begin{equation}
    (\nabla F(t))_i = \frac{\partial F}{\partial t_i} = -4\sigma_i(t_i)^{-3}\sigma_i'(t_i)({\bf Q}p(t)).
\end{equation}

\subsection{White Noise and Red Noise}

A white noise-only pulsar PSD is unrealistic in the intermediate-signal regime where
red noise induced from the GWB will be present in all pulsars and some pulsars may contain
intrinsic red spin noise:
\begin{equation}
	    P_I(f,\tint_i) = 2\sigmaTOA_{,I}(\tint_{,i})^2\Delta t + \
	    \frac{\AGWB^2}{12\pi^2}\left(\frac{f}{\fyr} \right)^{-\gammaGWB}\fyr^{-3} + \
		\frac{\Ared^2}{12\pi^2}\left(\frac{f}{\fyr} \right)^{-\gammared}\fyr^{-3},
\end{equation}
where $\gammaGWB = 3 - 2\alphaGWB$. 
In this case, $\rho^2$ is not 
quatratic in $\sigma^{-2}$. Instead it's quadratic in 
\begin{equation}
	 a_I(f) \approx \frac{\mathcal{T}(f)\mathcal{R}(f)}{P_I(f)}.
\end{equation}
We discretize the integral over GW frequency and write the expression in terms of $a_I(f_k)$. Then,
\begin{equation}
    \rho^2\simeq
    2T_{\rm obs}\sum_k 
    w_k
    S_h^2(f_k)
    \sum_{I<J}
    \frac{T_{IJ}}{T_{\rm obs}}
    \chi^2_{IJ}a_I(f_k)a_J(f_k).
\end{equation}
where, for trapezoidal integration,
\begin{equation}
    w_0 = w_{n-1} = \frac{\Delta f}{2}, \;\; w_k = \Delta f \; (1\leq k \leq n-2).
\end{equation}
Now define a stacked variable $\tilde{p}$ made up of the discretized inverse pulsar PSDs:
\begin{equation}
    \tilde{p} =
    \begin{bmatrix}
        a_1(f_1) \\
        \vdots \\
        a_N(f_1) \\
        a_1(f_2) \\
        \vdots \\
        a_N(f_2) \\
        \vdots
    \end{bmatrix}
    \in \mathbb{R}^{N \times N_f}.
\end{equation}
Then define a block-diagonal matrix 
\begin{equation}
    \tilde{Q} =
    \begin{bmatrix}
        Q(f_1) & 0      & \cdots \\
        0      & Q(f_2) & \cdots \\
        \vdots & \vdots & \ddots
        \end{bmatrix}
\end{equation}
where each block has entries
\begin{align}
    Q_{IJ}(f_k) &=
    \begin{cases}
    T_{\mathrm{obs}} \, w_kS_h^2(f_k)\, \frac{T_{IJ}}{T_\mathrm{obs}}\chi_{IJ}^2, & I \neq J, \\
    0, & I = J
    \end{cases}\\
    &=
    \begin{cases}
    2T_{\mathrm{obs}} \, w_k S_h^2(f_k)\, {\bf W}, & I \neq J, \\
    0, & I = J
    \end{cases}
\end{align}
so that
\begin{equation}
    \rho^2 = \sum_k a(f_k)^T{\bf Q}(f_k)a(f_k).
\end{equation}
Then recover the quadratic form in $a_I(f_k)$ as
\begin{equation}
    \rho^2 = \tilde{p}^{\,T} \, \tilde{Q} \, \tilde{p}.
\end{equation}
To derive the gradient, let
\begin{equation}
    c_k \equiv \mathcal{T}(f_k)\mathcal{R}(f_k).
\end{equation}
Then 
\begin{equation}
    a_I(f_k; t_I) = \frac{1}{S_I(f_k;t_I)} = \frac{c_k}{P_I(f_k; t_I)}.
\end{equation}
The function to optimize is 
\begin{align}
    F(t) \equiv \rho^2(t) &= \tilde{p}^T{\bf \tilde{Q}}\tilde{p}\\
    &= \sum_k a(f_k;t)^T Q(f_k)a(f_k;t).
\end{align}
The differential is 
\begin{equation}
    dF = 2\sum_k a(f_k;t)^TQ(f_k)da(f_k;t).
\end{equation}
Let $u(f_k;t) = Q(f_k)a(f_k;t)$. Then,
\begin{equation}
    dF = 2\sum_k u(f_k;t)^Tda(f_k;t).
\end{equation}
Since $a_I(f_k;t)$ depends only on $t_I$,
\begin{equation}
    \frac{\partial F}{\partial t_i} = 2\sum_k u_i(f_k;t) \frac{\partial a_i(f_k; t)}{\partial t_i}.
\end{equation}
Recall that $a_i = c_k / P_i$, so that
\begin{equation}
    \frac{\partial a_i(f_k;t)}{\partial t_i} = -c_k\frac{1}{\left(P_i(f_k;t)\right)^2}
                                                  \frac{\partial P_i(f_k;t)}{\partial t_i}.
\end{equation}
Only the white noise depends on $t_i$, so
\begin{equation}
    \frac{\partial P_i(f_k;t)}{\partial t_i} = 4\Delta \sigma_i(t_i)\sigma_i'(t_i).
\end{equation}
Plugging back in,
\begin{equation}
    \frac{\partial a_i(f_k;t)}{\partial t_i} = -c_k\frac{4\Delta t \sigma_i(t_i)\sigma_i'(t_i)}
                                                {\left(P_i(f_k;t)\right)^2},
\end{equation}
then
\begin{equation}
    \frac{\partial F}{\partial t_i} = -8\Delta t\sum_k u_i(f_k;t)\frac{c_k\sigma_i(t_i)\sigma_i'(t_i)}
                                                {\left(P_i(f_k;t)\right)^2}
\end{equation}
and finally
\begin{equation}
       (\nabla F(t))_i = \frac{\partial F}{\partial t_i} = -8\Delta t \sigma_i(t_i)\sigma_i'(t_i)
                                    \sum_{k=1}^{N_f} \left[(Q_k a_k)_i \right]\frac{a_i(f_k; t)}
                                                {P_i(f_k;t)}.
\end{equation}
$Q_k$ is independent of the noise and can therefore be pre-computed.
\section{A Stationarity and Feasibility Diagnostic for the Optimal Time}\label{appx:kktresid}
Consider the equality- and inequality-constrained optimization problem 
\begin{align}
     \min_{\tvec \in \mathbb{R}^n} \quad F(\tvec) &\equiv -\rho^2(\tvec)/\rho^2(\teq)\\
    \mathrm{s.t.}
     \sum_{i=1}^\Npsrs t_{\mathrm{int},i} &= B,  \\
     \tmini \le \tint_{,i} & \le \tmaxi. 
\end{align}
There is one equality constraint-- the budget, $B$. There are $2\Npsrs$ inequality constraints for available time for each pulsar.
At any feasible $\tvec$, the active set of constraints, $\mathcal{A}(\tint)$, is the equality constraint index $\mathcal{E}$ and the set of inequality constraint indices, $\mathcal{I}$, where $\tint$ for pulsar $i$ is at a boundary.
The Lagrangian for this problem is therefore 
\begin{equation}
    \mathcal{L}(\tvec,\lambda,\mu^+_i, \mu^-_i) = F(\tvec) -
    \lambda\!\left(\sum_i \tint_{,i} - B\right) -
    \sum_{i\in\mathcal{A}(\tint)}  \mu^-_i (\tint_{,i} - \tmini) -
    \sum_{i\in\mathcal{A}(\tint)} \mu^+_i (\tmaxi-\tint_{,i}),
\end{equation}
where $\lambda$ is the budget-equality Lagrange multiplier, $\mu^-_i$ are the minimum per-pulsar allocation Lagrange multipliers, and $\mu^+_i$ are the maximum per-pulsar allocation Lagrange multipliers. The Karush-Kuhn-Tucker (KKT) conditions are the first-order necessary conditions for $\topt$ to be a local optimum \citep{nocedalwright}.
For this problem the KKT conditions are:

\begin{align}
    \nabla_t \mathcal{L}(\topt,\lambda^*, \boldsymbol{\mu^{+}}^*, \boldsymbol{\mu^{-}}^*) &= 
               \nabla_t F(\topt) - \lambda^* - \boldsymbol{\mu^{-}}^* + \boldsymbol{\mu^{+}}^* = \bf{0} \label{eq:stationarity} \\
    \label{eq:kkt_budget_feas}
    \sum_i \tint_{,i}^* -B &= 0 \\
    \tmini \le \tint_{,i}^* & \le \tmaxi \\
    {\mu^-_i}^* \geq 0, &\quad {\mu^+_i}^* \geq 0 \quad \mathrm{for\ all}\ i \in \mathcal{I}\label{eq:kktfeas} \\
    {\mu^-_i}^* (\tint_{,i}^* - \tmini) = 0, &\quad  {\mu^+_i}^* (\tmaxi-\tint_{,i}^*) = 0
    									\label{eq:kktcomplement}
\end{align}
Equation \eqref{eq:kktcomplement} implies that ${\mu^-_i}^* = {\mu^+_i}^* = 0$ for inactive inequality constraints, i.e. where a pulsar is optimally between its maximum and minimum time. Together with the stationarity condition, Equation \eqref{eq:stationarity},
this implies that when $\tmin < \tint_{,i}^* < \tmax$,
\begin{equation}
	\left.\frac{\partial F}{\partial \tint_{,i}} \right|_{\tint_{,i}^*} - \lambda^* = 0
			\quad \mathrm{for\ all}\ i \notin \mathcal{A}(\tint).
			\label{eq:kkt_gradeq}
\end{equation}
Equations \eqref{eq:stationarity} and \eqref{eq:kktfeas} imply that
\begin{align}
	\label{eq:kkt_boundgrad_min}
	\left.\frac{\partial F}{\partial \tint_{,i}} \right|_{\tint_{,i}^*} - \lambda^* \geq 0 
		\quad \mathrm{when}\ \tint_{,i}^* = \tmini,\\
    \label{eq:kkt_boundgrad_max}
	\left.\frac{\partial F}{\partial \tint_{,i}} \right|_{\tint_{,i}^*} - \lambda^* \leq 0 
		\quad \mathrm{when}\ \tint_{,i}^* = \tmaxi.
\end{align}
We compute a residual, $\epsilon_\mathrm{KKT}$, which measures the maximum normalized violation of 
the stationarity condition \eqref{eq:stationarity} and the budget feasibility \eqref{eq:kkt_budget_feas}.

\begin{equation}
    \epsilon_\mathrm{KKT} = \max
        \left[\frac{||r||_\infty}{\max \left(1,||\nabla F||_\infty,|\lambda^*|,||\vec{\mu}^*||_\infty \right)},
        \frac{|\sum_i t_{\mathrm{int},i}^* - B||}{\max \left(1, |B|\right)}\right],
\end{equation}
where
\begin{equation}
    r_i =   \left.\frac{\partial F}{\partial t_{\mathrm{int},i}}\right|_{\tint_{,i}^*} - \lambda^* -
    			{\mu^-_i}^* +  {\mu^+_i}^*.
\end{equation}
We consider the solution $\topt$ to be numerically consistent with KKT stationarity and budget feasibility if 
$\epsilon_\mathrm{KKT}$ is approximately zero to within machine precision.
\section{Implied Red Noise Amplitude Distribution}\label{appx:rednoise}
\citet{sc10_rednoise} express the red component of the RMS residuals after a 2nd-order polynomial fit as
\begin{equation}
	\hat{\sigma}_\mathrm{TN,2} = C_2 \nu^\alpha|\dot{\nu}_{-15}|^\beta\mathrm{\ \mu s\ } T^\gamma,
\end{equation}
where $\nu_{-15}$ is in units of $10^{-15}\;\mathrm{s^{-2}}$.
They define a diagnostic parameter $\zeta$ which represents how far the measured
timing noise $\sigma_\mathrm{TN,2}$ deviates from the global fit. The post-fit and
measured red noise are related by
\begin{equation}
	\sigma_\mathrm{TN,2} = \zeta \hat{\sigma}_\mathrm{TN,2} = \zeta \
							C_2 \nu^\alpha|\dot{\nu}_{-15}|^\beta\mathrm{\ \mu s\ }T^\gamma.
\end{equation}
Substituting the spin period and Shklovskii-corrected, intrinsic period derivative for the spin frequency
and frequency derivative,
\begin{equation}\label{eq:cssigred}
	\sigma_\mathrm{TN,2} = \zeta \hat{\sigma}_\mathrm{TN,2} = \zeta \
							C_2 P^{-\alpha}\left(\frac{|\Pdotint|}{10^{-15}P^2}\right)^\beta\mathrm{\ \mu s\ } T^\gamma.
\end{equation}
\citet{sc10_rednoise} present the PDF of measured RMS residuals as 
\begin{align}
	f_{\sigma_\mathrm{TN,2}}(\sigma_\mathrm{TN,2}) &= \frac{1}{\sqrt{2\pi\delta^2}}\
							\exp\left[-\left(\frac{\ln(\hat{\sigma}_\mathrm{TN,2}/\sigma_\mathrm{TN,2})^2}{2\delta^2}\right)\right].
\end{align}
While they say that $\sigma_\mathrm{TN,2}$ is distributed log-normally, what the above equation really implies is
that $\ln (\sigma_\mathrm{TN,2})$ is distributed normally, a subtle distinction implied by the lack of
$1 / \sigma_\mathrm{TN,2}$ in the prefactor. If 
\begin{equation}
	\ln (\sigma_\mathrm{TN,2}) \sim \mathcal{N}\left(\ln(\hat{\sigma}_\mathrm{TN,2}),\delta^2 \right),
\end{equation}
then it is implied that
\begin{align}
	\ln (\zeta) = \ln (\sigma_\mathrm{TN,2}) - \ln(\hat{\sigma}_\mathrm{TN,2})\\
	\ln (\zeta) \sim \mathcal{N}(0,\delta^2).
\end{align}
\cite{ng9_excessnoise} fit additional red noise measurements
and upper-limits from NANOGrav and PPTA MSPs to further constrain the red noise distribution.
We adopt their values for 
$\ln C_2$, $\alpha$, $\beta$, $\gamma$, and $\delta$ from
row ${\rm MSP_{10,PPTA} + NANO}$ of their Table 4.
They model the red noise as a stationary
Gaussian process that is parameterized by a timing residual power spectrum of the form
\begin{equation}\label{eq:strainpsd}
	P(f) = A_\mathrm{r}^2\left(\frac{f}{1\mathrm{yr}^{-1}}\right)^{-\gamma_\mathrm{red}}
\end{equation}
where $A_\mathrm{r}$ is the amplitude of the red noise in $\mathrm{\mu s\ yr^{1/2}}$ and
$\gamma_\mathrm{red}$ is the (positive) spectral index. From the noise model, $\sigma_\mathrm{TN,2}$
can be estimated as \citep{ng9_excessnoise}
\begin{align}\label{eq:ngsigred}
	\sigma_\mathrm{TN,2} &\approx \left(\int^\infty_{1/T} P(f)df\right)^{1/2}\nonumber\\[1em]
						&= 3\cdot 10^{-3}\mathrm{\ \mu s\ }(\gamma_\mathrm{red} - 1)^{-1/2}\
						\left(\frac{A_\mathrm{r}}{3\cdot 10^{-3}\mathrm{\mu s\ yr^{1/2}}}\right)\
						{T_\mathrm{yr}}^{(\gamma_\mathrm{red}-1)/2},
\end{align}
which is only valid for power-law spectra when $\gamma_\mathrm{red} > 1$.
To express the red noise amplitude in terms of $P$ and $|\Pdotint|$, equate Equations \eqref{eq:cssigred} and 
\eqref{eq:ngsigred}:
\begin{align}
	\zeta C_2 P^{-\alpha}\left(\frac{|\Pdotint|}{10^{-15}P^2}\right)^\beta\mathrm{\ \mu s\ } T^\gamma&=\
	3\cdot 10^{-3}\mathrm{\mu s\ }(\gamma_\mathrm{red}-1)^{-1/2}\
						\left(\frac{A_\mathrm{r}}{3\cdot 10^{-3}\mathrm{\mu s\ yr^{1/2}}}\right)
						T_{\rm yr}^{(\gamma_\mathrm{red}-1)/2}\\
	A_\mathrm{r} &= (\gamma_\mathrm{red}-1)^{1/2}\zeta C_2 P^{-\alpha}
					\left(\frac{|\Pdotint|}{10^{-15}P^2}\right)^\beta T^\gamma T_{\rm yr}^{-(\gamma_\mathrm{red}-1)/2}
					\mathrm{\ \mu s\ yr^{1/2}}.
\end{align}
Using the relationship between the spectral index and spin-noise scaling parameter \citep{ng9_excessnoise},
$\gamma_\mathrm{red}=2\gamma + 1$, cancels out the dependence on timespan:
\begin{equation}
		A_\mathrm{r} = (\gamma_\mathrm{red}-1)^{1/2}\zeta C_2 P^{-\alpha}
					\left(\frac{|\Pdotint|}{10^{-15}P^2}\right)^\beta
					\mathrm{\ \mu s\ yr^{1/2}}.
\end{equation}
Convert to a dimensionless strain amplitude (by equating the second term of 
Equation \ref{eq:psr_psd} and Equation \ref{eq:strainpsd}):
\begin{equation}
	\Ared = A_\mathrm{r}\frac{\sqrt{12}\pi}{1\ \mathrm{yr} \times 10^{6}}.
\end{equation}
In our single-realization of the PTA (Equation \ref{eq:sc10RN}), we assume the mean value of the distribution, i.e. $\zeta = 1$.
In this model, $\gammared$ is implicitly fixed at a 5.6 for all pulsars.
\bibliography{ms-main}{}

@ARTICLE{ne2025,
       author = {{Ocker}, Stella Koch and {Cordes}, James M.},
        title = "{NE2025: An Updated Electron Density Model for the Galactic Interstellar Medium}",
      journal = {\apj},
         year = 2026,
        month = may,
       volume = {1002},
       number = {1},
          eid = {3},
        pages = {3},
          doi = {10.3847/1538-4357/ae5825},
archivePrefix = {arXiv},
       eprint = {2602.11838},
 primaryClass = {astro-ph.GA},
       adsurl = {https://ui.adsabs.harvard.edu/abs/2026ApJ..1002....3O}
}

@ARTICLE{larsen24,
       author = {{Larsen}, Bjorn and {Mingarelli}, Chiara M.~F. and {Hazboun}, Jeffrey S. and {Chalumeau}, Aur{\'e}lien and {Good}, Deborah C. and {Simon}, Joseph and {Agazie}, Gabriella and {Anumarlapudi}, Akash and {Archibald}, Anne M. and {Arzoumanian}, Zaven and {Baker}, Paul T. and {Brook}, Paul R. and {Cromartie}, H. Thankful and {Crowter}, Kathryn and {DeCesar}, Megan E. and {Demorest}, Paul B. and {Dolch}, Timothy and {Ferrara}, Elizabeth C. and {Fiore}, William and {Fonseca}, Emmanuel and {Freedman}, Gabriel E. and {Garver-Daniels}, Nate and {Gentile}, Peter A. and {Glaser}, Joseph and {Jennings}, Ross J. and {Jones}, Megan L. and {Kaplan}, David L. and {Kerr}, Matthew and {Lam}, Michael T. and {Lorimer}, Duncan R. and {Luo}, Jing and {Lynch}, Ryan S. and {McEwen}, Alexander and {McLaughlin}, Maura A. and {McMann}, Natasha and {Meyers}, Bradley W. and {Ng}, Cherry and {Nice}, David J. and {Pennucci}, Timothy T. and {Perera}, Benetge B.~P. and {Pol}, Nihan S. and {Radovan}, Henri A. and {Ransom}, Scott M. and {Ray}, Paul S. and {Schmiedekamp}, Ann and {Schmiedekamp}, Carl and {Shapiro-Albert}, Brent J. and {Stairs}, Ingrid H. and {Stovall}, Kevin and {Susobhanan}, Abhimanyu and {Swiggum}, Joseph K. and {Wahl}, Haley M. and {Champion}, David J. and {Cognard}, Isma{\"e}l and {Guillemot}, Lucas and {Hu}, Huanchen and {Keith}, Michael J. and {Liu}, Kuo and {McKee}, James W. and {Parthasarathy}, Aditya and {Perrodin}, Delphine and {Possenti}, Andrea and {Shaifullah}, Golam M. and {Theureau}, Gilles},
        title = "{The NANOGrav 15 yr Data Set: Chromatic Gaussian Process Noise Models for Six Pulsars}",
      journal = {\apj},
         year = 2024,
        month = sep,
       volume = {972},
       number = {1},
          eid = {49},
        pages = {49},
          doi = {10.3847/1538-4357/ad5291},
archivePrefix = {arXiv},
       eprint = {2405.14941},
 primaryClass = {astro-ph.HE},
       adsurl = {https://ui.adsabs.harvard.edu/abs/2024ApJ...972...49L}
}

@ARTICLE{meerkat_thousand_xv,
       author = {{Karastergiou}, A. and {Johnston}, S. and {Posselt}, B. and {Oswald}, L.~S. and {Kramer}, M. and {Weltevrede}, P.},
        title = "{The Thousand-Pulsar-Array programme on MeerKAT ─ XV. A comparison of the radio emission properties of slow and millisecond pulsars}",
      journal = {\mnras},
         year = 2024,
        month = aug,
       volume = {532},
       number = {3},
        pages = {3558-3566},
          doi = {10.1093/mnras/stae1694},
archivePrefix = {arXiv},
       eprint = {2407.06836},
 primaryClass = {astro-ph.HE},
       adsurl = {https://ui.adsabs.harvard.edu/abs/2024MNRAS.532.3558K}
}

@manual{gbtobsguide,
    author       = {{GBT Support Staff}},
    title        = {Proposer's Guide for the Green Bank Telescope},
    organization = {Green Bank Observatory},
    year         = {2017},
    month        = jun
}

@ARTICLE{ppta_evidence,
       author = {{Reardon}, Daniel J. and {Zic}, Andrew and {Shannon}, Ryan M. and {Hobbs}, George B. and {Bailes}, Matthew and {Di Marco}, Valentina and {Kapur}, Agastya and {Rogers}, Axl F. and {Thrane}, Eric and {Askew}, Jacob and {Bhat}, N.~D. Ramesh and {Cameron}, Andrew and {Cury{\l}o}, Ma{\l}gorzata and {Coles}, William A. and {Dai}, Shi and {Goncharov}, Boris and {Kerr}, Matthew and {Kulkarni}, Atharva and {Levin}, Yuri and {Lower}, Marcus E. and {Manchester}, Richard N. and {Mandow}, Rami and {Miles}, Matthew T. and {Nathan}, Rowina S. and {Os{\l}owski}, Stefan and {Russell}, Christopher J. and {Spiewak}, Ren{\'e}e and {Zhang}, Songbo and {Zhu}, Xing-Jiang},
        title = "{Search for an Isotropic Gravitational-wave Background with the Parkes Pulsar Timing Array}",
      journal = {\apjl},
         year = 2023,
        month = jul,
       volume = {951},
       number = {1},
          eid = {L6},
        pages = {L6},
          doi = {10.3847/2041-8213/acdd02},
archivePrefix = {arXiv},
       eprint = {2306.16215},
 primaryClass = {astro-ph.HE},
       adsurl = {https://ui.adsabs.harvard.edu/abs/2023ApJ...951L...6R}
}

@article{epta_inpta_evidence,
	author = {{EPTA Collaboration and InPTA Collaboration} and {Antoniadis, J.} and {Arumugam, P.} and {Arumugam, S.} and {Babak, S.} and {Bagchi, M.} and {Bak Nielsen, A.-S.} and {Bassa, C. G.} and {Bathula, A.} and {Berthereau, A.} and {Bonetti, M.} and {Bortolas, E.} and {Brook, P. R.} and {Burgay, M.} and {Caballero, R. N.} and {Chalumeau, A.} and {Champion, D. J.} and {Chanlaridis, S.} and {Chen, S.} and {Cognard, I.} and {Dandapat, S.} and {Deb, D.} and {Desai, S.} and {Desvignes, G.} and {Dhanda-Batra, N.} and {Dwivedi, C.} and {Falxa, M.} and {Ferdman, R. D.} and {Franchini, A.} and {Gair, J. R.} and {Goncharov, B.} and {Gopakumar, A.} and {Graikou, E.} and {Grießmeier, J.-M.} and {Guillemot, L.} and {Guo, Y. J.} and {Gupta, Y.} and {Hisano, S.} and {Hu, H.} and {Iraci, F.} and {Izquierdo-Villalba, D.} and {Jang, J.} and {Jawor, J.} and {Janssen, G. H.} and {Jessner, A.} and {Joshi, B. C.} and {Kareem, F.} and {Karuppusamy, R.} and {Keane, E. F.} and {Keith, M. J.} and {Kharbanda, D.} and {Kikunaga, T.} and {Kolhe, N.} and {Kramer, M.} and {Krishnakumar, M. A.} and {Lackeos, K.} and {Lee, K. J.} and {Liu, K.} and {Liu, Y.} and {Lyne, A. G.} and {McKee, J. W.} and {Maan, Y.} and {Main, R. A.} and {Mickaliger, M. B.} and {Niţu, I. C.} and {Nobleson, K.} and {Paladi, A. K.} and {Parthasarathy, A.} and {Perera, B. B. P.} and {Perrodin, D.} and {Petiteau, A.} and {Porayko, N. K.} and {Possenti, A.} and {Prabu, T.} and {Quelquejay Leclere, H.} and {Rana, P.} and {Samajdar, A.} and {Sanidas, S. A.} and {Sesana, A.} and {Shaifullah, G.} and {Singha, J.} and {Speri, L.} and {Spiewak, R.} and {Srivastava, A.} and {Stappers, B. W.} and {Surnis, M.} and {Susarla, S. C.} and {Susobhanan, A.} and {Takahashi, K.} and {Tarafdar, P.} and {Theureau, G.} and {Tiburzi, C.} and {van der Wateren, E.} and {Vecchio, A.} and {Venkatraman Krishnan, V.} and {Verbiest, J. P. W.} and {Wang, J.} and {Wang, L.} and {Wu, Z.}},
	title = {The second data release from the European Pulsar Timing Array - III. Search for gravitational wave signals},
	DOI= "10.1051/0004-6361/202346844",
	url= "https://doi.org/10.1051/0004-6361/202346844",
	journal = {\aap},
	year = 2023,
	volume = 678,
	pages = "A50",
}

@ARTICLE{jaffe03,
       author = {{Jaffe}, A.~H. and {Backer}, D.~C.},
        title = "{Gravitational Waves Probe the Coalescence Rate of Massive Black Hole Binaries}",
      journal = {\apj},
         year = 2003,
        month = feb,
       volume = {583},
       number = {2},
        pages = {616-631},
          doi = {10.1086/345443},
archivePrefix = {arXiv},
       eprint = {astro-ph/0210148},
 primaryClass = {astro-ph},
       adsurl = {https://ui.adsabs.harvard.edu/abs/2003ApJ...583..616J}
}

@ARTICLE{sesana04,
       author = {{Sesana}, Alberto and {Haardt}, Francesco and {Madau}, Piero and {Volonteri}, Marta},
        title = "{Low-Frequency Gravitational Radiation from Coalescing Massive Black Hole Binaries in Hierarchical Cosmologies}",
      journal = {\apj},
         year = 2004,
        month = aug,
       volume = {611},
       number = {2},
        pages = {623-632},
          doi = {10.1086/422185},
archivePrefix = {arXiv},
       eprint = {astro-ph/0401543},
 primaryClass = {astro-ph},
       adsurl = {https://ui.adsabs.harvard.edu/abs/2004ApJ...611..623S}
}

@misc{kuhn_tucker51,
 author = {Kuhn, H. W. and Tucker, A. W.},
 title = {Nonlinear programming},
 year = {1951},
 language = {English},
 howpublished = {Proc. {Berkeley} {Sympos}. math. {Statist}. {Probability}, {California} {July} 31 - {August} 12, 1950, 481-492 (1951).},
 zbMATH = {3067835},
 Zbl = {0044.05903}
}

@phdthesis{karush39,
title = {Minima of functions of several variables with inequalities as side conditions.},
author = {Karush, William},
year = {1939},
school = {Thesis (S.M.)--University of Chicago, Department of Mathematics, December 1939.},
}

@book{dsa_science_doc,
  editor    = {McLaughlin, Maura and Walter, Fabian and Hallinan, Gregg and Ravi, Vikram},
  title     = {{DSA} Community Science Document},
  publisher = {California Institute of Technology},
  year      = {2026},
  month     = mar,
  note      = {Version 1}
}

@ARTICLE{larsen26,
       author = {{Larsen}, Bjorn and {Baier}, Jeremy G. and {Oliver}, Daniel J. and {Wayt}, Kalista and {Chang}, Yu-Ting and {Hazboun}, Jeffrey S. and {Mingarelli}, Chiara M.~F. and {Simon}, Joseph and {Miles}, Matthew T. and {Agazie}, Gabriella and {Anumarlapudi}, Akash and {Archibald}, Anne M. and {Arzoumanian}, Zaven and {Baker}, Paul T. and {Brook}, Paul R. and {Cromartie}, H. Thankful and {Crowter}, Kathryn and {DeCesar}, Megan E. and {Demorest}, Paul B. and {Dolch}, Timothy and {Ferrara}, Elizabeth C. and {Fiore}, William and {Fonseca}, Emmanuel and {Freedman}, Gabriel E. and {Garver-Daniels}, Nate and {Gentile}, Peter A. and {Glaser}, Joseph and {Good}, Deborah C. and {Jennings}, Ross J. and {Jones}, Megan L. and {Kaplan}, David L. and {Kerr}, Matthew and {Lam}, Michael T. and {Lorimer}, Duncan R. and {Luo}, Jing and {Lynch}, Ryan S. and {McEwen}, Alexander and {McLaughlin}, Maura A. and {McMann}, Natasha and {Meyers}, Bradley W. and {Ng}, Cherry and {Nice}, David J. and {Pennucci}, Timothy T. and {Perera}, Benetge B.~P. and {Pol}, Nihan S. and {Radovan}, Henri A. and {Ransom}, Scott M. and {Ray}, Paul S. and {Schmiedekamp}, Ann and {Schmiedekamp}, Carl and {Shapiro-Albert}, Brent J. and {Stairs}, Ingrid H. and {Stovall}, Kevin and {Susobhanan}, Abhimanyu and {Swiggum}, Joseph K. and {Wahl}, Haley M.},
        title = "{The NANOGrav 15 yr Dataset: Customized Chromatic Noise Models}",
      journal = {\apj},
         year = 2026,
        month = jul,
       volume = {1005},
       number = {1},
          eid = {29},
        pages = {29},
          doi = {10.3847/1538-4357/ae677c},
archivePrefix = {arXiv},
       eprint = {2606.28571},
 primaryClass = {astro-ph.HE},
       adsurl = {https://ui.adsabs.harvard.edu/abs/2026ApJ..1005...29L}
}

@ARTICLE{ng15_custnoise,
       author = {{Agarwal}, Nikita and {Agazie}, Gabriella and {Amosso}, Alessandra and {Anumarlapudi}, Akash and {Archibald}, Anne M. and {Arzoumanian}, Zaven and {Ashok}, Anjana and {Baier}, Jeremy G. and {Baker}, Paul T. and {Becsy}, Bence and {Blecha}, Laura and {Brazier}, Adam and {Brook}, Paul R. and {Burke-Spolaor}, Sarah and {Burnette}, Rand and {Case}, Robin and {Casey-Clyde}, J. Andrew and {Chang}, Yu-Ting and {Charisi}, Maria and {Chatterjee}, Shami and {Cohen}, Tyler and {Cordes}, James M. and {Cornish}, Neil J. and {Crawford}, Fronefield and {Cromartie}, H. Thankful and {Crowter}, Kathryn and {DeCesar}, Megan E. and {Demorest}, Paul B. and {Deng}, Heling and {Dey}, Lankeswar and {Dolch}, Timothy and {Doskoch}, Graham M. and {Ferrara}, Elizabeth C. and {Fiore}, William and {Fonseca}, Emmanuel and {Freedman}, Gabriel E. and {Gardiner}, Emiko C. and {Garver-Daniels}, Nate and {Gentile}, Peter A. and {Gersbach}, Kyle A. and {Glaser}, Joseph and {Good}, Deborah C. and {Gultekin}, Kayhan and {Gundersen}, Aiden and {Harris}, C.~J. and {Hashemi Asl}, Doa and {Hazboun}, Jeffrey S. and {Jennings}, Ross J. and {Johnson}, Aaron D. and {Jones}, Megan L. and {Kaplan}, David L. and {Sreekumar}, Anala K. and {Kelley}, Luke Zoltan and {Kerr}, Matthew and {Key}, Joey S. and {Laal}, Nima and {Lam}, Michael T. and {Lamb}, William G. and {Larsen}, Bjorn and {Lazio}, T. Joseph W. and {Lewandowska}, Natalia and {Liu}, Tingting and {Lorimer}, Duncan R. and {Luo}, Jing and {Lynch}, Ryan S. and {Ma}, Chung-Pei and {Madison}, Dustin R. and {Martsen}, Ashley and {Matt}, Cayenne and {McEwen}, Alexander and {McKee}, James W. and {McLaughlin}, Maura A. and {McMann}, Natasha and {Meyers}, Bradley W. and {Meyers}, Patrick M. and {Miles}, Matthew T. and {Mingarelli}, Chiara M.~F. and {Mitridate}, Andrea and {Ng}, Cherry and {Nice}, David J. and {Nichols}, Shania and {Ocker}, Stella K. and {Oliver}, Daniel J. and {Olum}, Ken D. and {Pennucci}, Timothy T. and {Perera}, Benetge B.~P. and {Petrov}, Polina and {Pol}, Nihan S. and {Radovan}, Henri A. and {Ransom}, Scott M. and {Ray}, Paul S. and {Romano}, Joseph D. and {Runnoe}, Jessie C. and {Saffer}, Alexander and {Sardesai}, Shashwat C. and {Schmiedekamp}, Ann and {Schmiedekamp}, Carl and {Schmitz}, Kai and {Schult}, Levi and {Shapiro-Albert}, Brent J. and {Siemens}, Xavier and {Simon}, Joseph and {Sosa Fiscella}, Sophia V. and {Stairs}, Ingrid H. and {Stinebring}, Daniel R. and {Stovall}, Kevin and {Strahler}, Robin and {Susobhanan}, Abhimanyu and {Swiggum}, Joseph K. and {Taylor}, Jacob and {Taylor}, Stephen R. and {Thompson}, Mercedes S. and {Turner}, Jacob E. and {Vallisneri}, Michele and {van Haasteren}, Rutger and {Verbiest}, Joris P.~W. and {Vigeland}, Sarah J. and {Wahl}, Haley M. and {Wayt}, Kalista and {Wilson}, Kevin P. and {Witt}, Caitlin A. and {Wright}, David and {Young}, Olivia},
        title = "{The NANOGrav 15 yr Data Set: Impacts of Customized Chromatic Noise Models on Gravitational Wave Analyses}",
      journal = {arXiv e-prints},
         year = 2026,
        month = jun,
          eid = {arXiv:2606.28554},
        pages = {arXiv:2606.28554},
          doi = {10.48550/arXiv.2606.28554},
archivePrefix = {arXiv},
       eprint = {2606.28554},
 primaryClass = {astro-ph.CO},
       adsurl = {https://ui.adsabs.harvard.edu/abs/2026arXiv260628554A}
}

@ARTICLE{ng15_noisebudget,
       author = {{Agazie}, Gabriella and {Anumarlapudi}, Akash and {Archibald}, Anne M. and {Arzoumanian}, Zaven and {Baker}, Paul T. and {B{\'e}csy}, Bence and {Blecha}, Laura and {Brazier}, Adam and {Brook}, Paul R. and {Burke-Spolaor}, Sarah and {Charisi}, Maria and {Chatterjee}, Shami and {Cohen}, Tyler and {Cordes}, James M. and {Cornish}, Neil J. and {Crawford}, Fronefield and {Cromartie}, H. Thankful and {Crowter}, Kathryn and {Decesar}, Megan E. and {Demorest}, Paul B. and {Dolch}, Timothy and {Drachler}, Brendan and {Ferrara}, Elizabeth C. and {Fiore}, William and {Fonseca}, Emmanuel and {Freedman}, Gabriel E. and {Garver-Daniels}, Nate and {Gentile}, Peter A. and {Glaser}, Joseph and {Good}, Deborah C. and {Guertin}, Lydia and {G{\"u}ltekin}, Kayhan and {Hazboun}, Jeffrey S. and {Jennings}, Ross J. and {Johnson}, Aaron D. and {Jones}, Megan L. and {Kaiser}, Andrew R. and {Kaplan}, David L. and {Kelley}, Luke Zoltan and {Kerr}, Matthew and {Key}, Joey S. and {Laal}, Nima and {Lam}, Michael T. and {Lamb}, William G. and {Lazio}, T. Joseph W. and {Lewandowska}, Natalia and {Liu}, Tingting and {Lorimer}, Duncan R. and {Luo}, Jing and {Lynch}, Ryan S. and {Ma}, Chung-Pei and {Madison}, Dustin R. and {McEwen}, Alexander and {McKee}, James W. and {McLaughlin}, Maura A. and {McMann}, Natasha and {Meyers}, Bradley W. and {Mingarelli}, Chiara M.~F. and {Mitridate}, Andrea and {Ng}, Cherry and {Nice}, David J. and {Ocker}, Stella Koch and {Olum}, Ken D. and {Pennucci}, Timothy T. and {Perera}, Benetge B.~P. and {Pol}, Nihan S. and {Radovan}, Henri A. and {Ransom}, Scott M. and {Ray}, Paul S. and {Romano}, Joseph D. and {Sardesai}, Shashwat C. and {Schmiedekamp}, Ann and {Schmiedekamp}, Carl and {Schmitz}, Kai and {Shapiro-Albert}, Brent J. and {Siemens}, Xavier and {Simon}, Joseph and {Siwek}, Magdalena S. and {Stairs}, Ingrid H. and {Stinebring}, Daniel R. and {Stovall}, Kevin and {Susobhanan}, Abhimanyu and {Swiggum}, Joseph K. and {Taylor}, Stephen R. and {Turner}, Jacob E. and {Unal}, Caner and {Vallisneri}, Michele and {Vigeland}, Sarah J. and {Wahl}, Haley M. and {Witt}, Caitlin A. and {Young}, Olivia and {Nanograv Collaboration}},
        title = "{The NANOGrav 15 yr Data Set: Detector Characterization and Noise Budget}",
      journal = {\apjl},
         year = 2023,
        month = jul,
       volume = {951},
       number = {1},
          eid = {L10},
        pages = {L10},
          doi = {10.3847/2041-8213/acda88},
archivePrefix = {arXiv},
       eprint = {2306.16218},
 primaryClass = {astro-ph.HE},
       adsurl = {https://ui.adsabs.harvard.edu/abs/2023ApJ...951L..10A}
}

@ARTICLE{cordesshannon12,
       author = {{Cordes}, J.~M. and {Shannon}, R.~M.},
        title = "{Minimum Requirements for Detecting a Stochastic Gravitational Wave Background Using Pulsars}",
      journal = {\apj},
         year = 2012,
        month = may,
       volume = {750},
       number = {2},
          eid = {89},
        pages = {89},
          doi = {10.1088/0004-637X/750/2/89},
archivePrefix = {arXiv},
       eprint = {1106.4047},
 primaryClass = {astro-ph.GA},
       adsurl = {https://ui.adsabs.harvard.edu/abs/2012ApJ...750...89C}
}

@ARTICLE{baier25,
       author = {{Baier}, Jeremy G. and {Hazboun}, Jeffrey S. and {Romano}, Joseph D.},
        title = "{A sensitivity curve approach to tuning a pulsar timing array in the detection era}",
      journal = {Classical and Quantum Gravity},
         year = 2025,
        month = apr,
       volume = {42},
       number = {7},
          eid = {075008},
        pages = {075008},
          doi = {10.1088/1361-6382/adbbab},
archivePrefix = {arXiv},
       eprint = {2409.00336},
 primaryClass = {astro-ph.HE},
       adsurl = {https://ui.adsabs.harvard.edu/abs/2025CQGra..42g5008B}
}

@ARTICLE{pennucci19,
       author = {{Pennucci}, Timothy T.},
        title = "{Frequency-dependent Template Profiles for High-precision Pulsar Timing}",
      journal = {\apj},
         year = 2019,
        month = jan,
       volume = {871},
       number = {1},
          eid = {34},
        pages = {34},
          doi = {10.3847/1538-4357/aaf6ef},
archivePrefix = {arXiv},
       eprint = {1812.02006},
 primaryClass = {astro-ph.HE},
       adsurl = {https://ui.adsabs.harvard.edu/abs/2019ApJ...871...34P}
}

@ARTICLE{sc10_rednoise,
       author = {{Shannon}, Ryan M. and {Cordes}, James M.},
        title = "{Assessing the Role of Spin Noise in the Precision Timing of Millisecond Pulsars}",
      journal = {\apj},
         year = 2010,
        month = dec,
       volume = {725},
       number = {2},
        pages = {1607-1619},
          doi = {10.1088/0004-637X/725/2/1607},
archivePrefix = {arXiv},
       eprint = {1010.4794},
 primaryClass = {astro-ph.SR},
       adsurl = {https://ui.adsabs.harvard.edu/abs/2010ApJ...725.1607S}
}

@ARTICLE{al22,
       author = {{Aggarwal}, Kshitij and {Lorimer}, Duncan. R.},
        title = "{On the radio spectra of Galactic millisecond pulsars}",
      journal = {arXiv e-prints},
         year = 2022,
        month = mar,
          eid = {arXiv:2203.05560},
        pages = {arXiv:2203.05560},
          doi = {10.48550/arXiv.2203.05560},
archivePrefix = {arXiv},
       eprint = {2203.05560},
 primaryClass = {astro-ph.HE},
       adsurl = {https://ui.adsabs.harvard.edu/abs/2022arXiv220305560A}
}

@ARTICLE{nicetaylor95,
       author = {{Nice}, D.~J. and {Taylor}, J.~H.},
        title = "{PSR J2019+2425 and PSR J2322+2057 and the Proper Motions of Millisecond Pulsars}",
      journal = {\apj},
         year = 1995,
        month = mar,
       volume = {441},
        pages = {429},
          doi = {10.1086/175367},
       adsurl = {https://ui.adsabs.harvard.edu/abs/1995ApJ...441..429N}
}

@ARTICLE{shklovskii,
       author = {{Shklovskii}, I.~S.},
        title = "{Possible Causes of the Secular Increase in Pulsar Periods.}",
      journal = {\sovast},
         year = 1970,
        month = feb,
       volume = {13},
        pages = {562},
       adsurl = {https://ui.adsabs.harvard.edu/abs/1970SvA....13..562S}
}

@ARTICLE{ridley_lorimer_10,
       author = {{Ridley}, J.~P. and {Lorimer}, D.~R.},
        title = "{Isolated pulsar spin evolution on the diagram}",
      journal = {\mnras},
         year = 2010,
        month = may,
       volume = {404},
       number = {2},
        pages = {1081-1088},
          doi = {10.1111/j.1365-2966.2010.16342.x},
archivePrefix = {arXiv},
       eprint = {1001.2483},
 primaryClass = {astro-ph.GA},
       adsurl = {https://ui.adsabs.harvard.edu/abs/2010MNRAS.404.1081R}
}

@article{crl08,
  author = {{Champion}, D. J. and {Ransom}, S. M. and {Lazarus}, P. and {Camilo}, F. and {Bassa}, C. and {Kaspi}, V. M. and {Nice}, D. J. and {Freire}, P. C. C. and {Stairs}, I. H. and {van Leeuwen}, J. and {Stappers}, B. W. and {Cordes}, J. M. and {Hessels}, J. W. T. and {Lorimer}, D. R. and {Arzoumanian}, Z. and {Backer}, D. C. and {Bhat}, N. D. R. and {Chatterjee}, S. and {Cognard}, I. and {Deneva}, J. S. and {Faucher-Gigu{\`e}re}, C.-A. and {Gaensler}, B. M. and {Han}, J. and {Jenet}, F. A. and {Kasian}, L. and {Kondratiev}, V. I. and {Kramer}, M. and {Lazio}, J. and {McLaughlin}, M. A. and {Venkataraman}, A. and {Vlemmings}, W.},
  title = {An Eccentric Binary Millisecond Pulsar in the Galactic Plane},
  journal = {sci},
  year = {2008},
  volume = {320},
  pages = {1309},
  doi = {10.1126/science.1157580}
}

@ARTICLE{psrcat,
       author = {{Manchester}, R.~N. and {Hobbs}, G.~B. and {Teoh}, A. and {Hobbs}, M.},
        title = "{The Australia Telescope National Facility Pulsar Catalogue}",
      journal = {\aj},
         year = 2005,
        month = apr,
       volume = {129},
       number = {4},
        pages = {1993-2006},
          doi = {10.1086/428488},
archivePrefix = {arXiv},
       eprint = {astro-ph/0412641},
 primaryClass = {astro-ph},
       adsurl = {https://ui.adsabs.harvard.edu/abs/2005AJ....129.1993M}
}

@ARTICLE{chime_frb1,
       author = {{CHIME/FRB Collaboration} and {Amiri}, Mandana and {Andersen}, Bridget C. and {Bandura}, Kevin and {Berger}, Sabrina and {Bhardwaj}, Mohit and {Boyce}, Michelle M. and {Boyle}, P.~J. and {Brar}, Charanjot and {Breitman}, Daniela and {Cassanelli}, Tomas and {Chawla}, Pragya and {Chen}, Tianyue and {Cliche}, J.-F. and {Cook}, Amanda and {Cubranic}, Davor and {Curtin}, Alice P. and {Deng}, Meiling and {Dobbs}, Matt and {Dong}, Fengqiu Adam and {Eadie}, Gwendolyn and {Fandino}, Mateus and {Fonseca}, Emmanuel and {Gaensler}, B.~M. and {Giri}, Utkarsh and {Good}, Deborah C. and {Halpern}, Mark and {Hill}, Alex S. and {Hinshaw}, Gary and {Josephy}, Alexander and {Kaczmarek}, Jane F. and {Kader}, Zarif and {Kania}, Joseph W. and {Kaspi}, Victoria M. and {Landecker}, T.~L. and {Lang}, Dustin and {Leung}, Calvin and {Li}, Dongzi and {Lin}, Hsiu-Hsien and {Masui}, Kiyoshi W. and {McKinven}, Ryan and {Mena-Parra}, Juan and {Merryfield}, Marcus and {Meyers}, Bradley W. and {Michilli}, Daniele and {Milutinovic}, Nikola and {Mirhosseini}, Arash and {M{\"u}nchmeyer}, Moritz and {Naidu}, Arun and {Newburgh}, Laura and {Ng}, Cherry and {Patel}, Chitrang and {Pen}, Ue-Li and {Petroff}, Emily and {Pinsonneault-Marotte}, Tristan and {Pleunis}, Ziggy and {Rafiei-Ravandi}, Masoud and {Rahman}, Mubdi and {Ransom}, Scott M. and {Renard}, Andre and {Sanghavi}, Pranav and {Scholz}, Paul and {Shaw}, J. Richard and {Shin}, Kaitlyn and {Siegel}, Seth R. and {Sikora}, Andrew E. and {Singh}, Saurabh and {Smith}, Kendrick M. and {Stairs}, Ingrid and {Tan}, Chia Min and {Tendulkar}, S.~P. and {Vanderlinde}, Keith and {Wang}, Haochen and {Wulf}, Dallas and {Zwaniga}, A.~V.},
        title = "{The First CHIME/FRB Fast Radio Burst Catalog}",
      journal = {\apjs},
         year = 2021,
        month = dec,
       volume = {257},
       number = {2},
          eid = {59},
        pages = {59},
          doi = {10.3847/1538-4365/ac33ab},
archivePrefix = {arXiv},
       eprint = {2106.04352},
 primaryClass = {astro-ph.HE},
       adsurl = {https://ui.adsabs.harvard.edu/abs/2021ApJS..257...59C}
}

@ARTICLE{bangale24,
       author = {{Bangale}, P. and {Bhattacharyya}, B. and {Camilo}, F. and {Clark}, C.~J. and {Cognard}, I. and {DeCesar}, M.~E. and {Ferrara}, E.~C. and {Gentile}, P. and {Guillemot}, L. and {Hessels}, J.~W.~T. and {Johnson}, T.~J. and {Kerr}, M. and {McLaughlin}, M.~A. and {Nieder}, L. and {Ransom}, S.~M. and {Ray}, P.~S. and {Roberts}, M.~S.~E. and {Roy}, J. and {Sanpa-arsa}, S. and {Theureau}, G. and {Wolff}, M.~T.},
        title = "{A 350 MHz Green Bank Telescope Survey of Unassociated Fermi LAT Sources: Discovery and Timing of 10 Millisecond Pulsars}",
      journal = {\apj},
         year = 2024,
        month = may,
       volume = {966},
       number = {2},
          eid = {161},
        pages = {161},
          doi = {10.3847/1538-4357/ad2994},
archivePrefix = {arXiv},
       eprint = {2402.09366},
 primaryClass = {astro-ph.HE},
       adsurl = {https://ui.adsabs.harvard.edu/abs/2024ApJ...966..161B}
}

@ARTICLE{spiewak22_meertime,
       author = {{Spiewak}, R. and {Bailes}, M. and {Miles}, M.~T. and {Parthasarathy}, A. and {Reardon}, D.~J. and {Shamohammadi}, M. and {Shannon}, R.~M. and {Bhat}, N.~D.~R. and {Buchner}, S. and {Cameron}, A.~D. and {Camilo}, F. and {Geyer}, M. and {Johnston}, S. and {Karastergiou}, A. and {Keith}, M. and {Kramer}, M. and {Serylak}, M. and {van Straten}, W. and {Theureau}, G. and {Venkatraman Krishnan}, V.},
        title = "{The MeerTime Pulsar Timing Array: A census of emission properties and timing potential}",
      journal = {\pasa},
         year = 2022,
        month = jul,
       volume = {39},
          eid = {e027},
        pages = {e027},
          doi = {10.1017/pasa.2022.19},
archivePrefix = {arXiv},
       eprint = {2204.04115},
 primaryClass = {astro-ph.HE},
       adsurl = {https://ui.adsabs.harvard.edu/abs/2022PASA...39...27S}
}

@PHDTHESIS{sanpaarsa16,
       author = {{Sanpa-Arsa}, Siraprapa},
        title = "{Searching for new millisecond pulsars with the GBT in Fermi unassociated sources}",
       school = {University of Virginia},
         year = 2016,
        month = jan,
       adsurl = {https://ui.adsabs.harvard.edu/abs/2016PhDT.......539S}
}

@ARTICLE{kramer98,
       author = {{Kramer}, Michael and {Xilouris}, Kiriaki M. and {Lorimer}, Duncan R. and {Doroshenko}, Oleg and {Jessner}, Axel and {Wielebinski}, Richard and {Wolszczan}, Alexander and {Camilo}, Fernando},
        title = "{The Characteristics of Millisecond Pulsar Emission. I. Spectra, Pulse Shapes, and the Beaming Fraction}",
      journal = {\apj},
         year = 1998,
        month = jul,
       volume = {501},
       number = {1},
        pages = {270-285},
          doi = {10.1086/305790},
archivePrefix = {arXiv},
       eprint = {astro-ph/9801177},
 primaryClass = {astro-ph},
       adsurl = {https://ui.adsabs.harvard.edu/abs/1998ApJ...501..270K}
}

@ARTICLE{kuniyoshi15,
       author = {{Kuniyoshi}, M. and {Verbiest}, J.~P.~W. and {Lee}, K.~J. and {Adebahr}, B. and {Kramer}, M. and {Noutsos}, A.},
        title = "{Low-frequency spectral turn-overs in millisecond pulsars studied from imaging observations}",
      journal = {\mnras},
         year = 2015,
        month = oct,
       volume = {453},
       number = {1},
        pages = {828-836},
          doi = {10.1093/mnras/stv1604},
archivePrefix = {arXiv},
       eprint = {1507.03732},
 primaryClass = {astro-ph.GA},
       adsurl = {https://ui.adsabs.harvard.edu/abs/2015MNRAS.453..828K}
}

@article{andersen23,
doi = {10.3847/1538-3881/acec78},
url = {https://doi.org/10.3847/1538-3881/acec78},
year = {2023},
month = {sep},
publisher = {The American Astronomical Society},
volume = {166},
number = {4},
pages = {138},
author = {Andersen, Bridget C. and Patel, Chitrang and Brar, Charanjot and Boyle, P. J. and Fonseca, Emmanuel and Kaspi, Victoria M. and Masui, Kiyoshi W. and Mena-Parra, Juan and Merryfield, Marcus and Meyers, Bradley W. and Sand, Ketan R. and Scholz, Paul and Siegel, Seth R. and Singh, Saurabh},
title = {Flux Calibration of CHIME/FRB Intensity Data},
journal = {The Astronomical Journal}
}

@ARTICLE{chime_overview,
       author = {{CHIME/Pulsar Collaboration} and {Amiri}, M. and {Bandura}, K.~M. and {Boyle}, P.~J. and {Brar}, C. and {Cliche}, J.-F. and {Crowter}, K. and {Cubranic}, D. and {Demorest}, P.~B. and {Denman}, N.~T. and {Dobbs}, M. and {Dong}, F.~Q. and {Fandino}, M. and {Fonseca}, E. and {Good}, D.~C. and {Halpern}, M. and {Hill}, A.~S. and {H{\"o}fer}, C. and {Kaspi}, V.~M. and {Landecker}, T.~L. and {Leung}, C. and {Lin}, H.-H. and {Luo}, J. and {Masui}, K.~W. and {McKee}, J.~W. and {Mena-Parra}, J. and {Meyers}, B.~W. and {Michilli}, D. and {Naidu}, A. and {Newburgh}, L. and {Ng}, C. and {Patel}, C. and {Pinsonneault-Marotte}, T. and {Ransom}, S.~M. and {Renard}, A. and {Scholz}, P. and {Shaw}, J.~R. and {Sikora}, A.~E. and {Stairs}, I.~H. and {Tan}, C.~M. and {Tendulkar}, S.~P. and {Tretyakov}, I. and {Vanderlinde}, K. and {Wang}, H. and {Wang}, X.},
        title = "{The CHIME Pulsar Project: System Overview}",
      journal = {\apjs},
         year = 2021,
        month = jul,
       volume = {255},
       number = {1},
          eid = {5},
        pages = {5},
          doi = {10.3847/1538-4365/abfdcb},
archivePrefix = {arXiv},
       eprint = {2008.05681},
 primaryClass = {astro-ph.IM},
       adsurl = {https://ui.adsabs.harvard.edu/abs/2021ApJS..255....5C}
}

@INPROCEEDINGS{ng18,
       author = {{Ng}, Cherry},
        title = "{Pulsar science with the CHIME telescope}",
    booktitle = {Pulsar Astrophysics the Next Fifty Years},
         year = 2018,
       editor = {{Weltevrede}, P. and {Perera}, B.~B.~P. and {Preston}, L.~L. and {Sanidas}, S.},
       series = {IAU Symposium},
       volume = {337},
        month = aug,
        pages = {179-182},
          doi = {10.1017/S1743921317010638},
archivePrefix = {arXiv},
       eprint = {1711.02104},
 primaryClass = {astro-ph.IM},
       adsurl = {https://ui.adsabs.harvard.edu/abs/2018IAUS..337..179N}
}

@INPROCEEDINGS{flygare_dsa,
  author={Flygare, Jonas},
  booktitle={2024 IEEE International Symposium on Antennas and Propagation and INC/USNC‐URSI Radio Science Meeting (AP-S/INC-USNC-URSI)}, 
  title={Wideband Low-Loss Feed Design for the DSA-2000 Ambient Temperature Array in Radio Astronomy}, 
  year={2024},
  volume={},
  number={},
  pages={5-6},
  doi={10.1109/AP-S/INC-USNC-URSI52054.2024.10686435}}

@ARTICLE{nancay22_J2205,
       author = {{Desvignes}, G. and {Cognard}, I. and {Smith}, D.~A. and {Champion}, D. and {Guillemot}, L. and {Kramer}, M. and {Lespagnol}, P. and {Octau}, F. and {Theureau}, G.},
        title = "{The SPAN512 mid-latitude pulsar survey at the Nan{\c{c}}ay Radio Telescope}",
      journal = {\aap},
         year = 2022,
        month = nov,
       volume = {667},
          eid = {A79},
        pages = {A79},
          doi = {10.1051/0004-6361/202244171},
archivePrefix = {arXiv},
       eprint = {2209.01806},
 primaryClass = {astro-ph.HE},
       adsurl = {https://ui.adsabs.harvard.edu/abs/2022A&A...667A..79D}
}

@book{nocedalwright,
  title={Numerical Optimization},
  author={Nocedal, J. and Wright, S.},
  isbn={9780387400655},
  lccn={2006923897},
  series={Springer Series in Operations Research and Financial Engineering},
  url={https://books.google.com/books?id=VbHYoSyelFcC},
  year={2006},
  publisher={Springer New York}
}

@book{conn00_trustregion,
author = {Conn, Andrew R. and Gould, Nicholas I. M. and Toint, Philippe L.},
title = {Trust Region Methods},
publisher = {Society for Industrial and Applied Mathematics},
year = {2000},
doi = {10.1137/1.9780898719857},
address = {},
edition   = {},
URL = {https://epubs.siam.org/doi/abs/10.1137/1.9780898719857},
eprint = {https://epubs.siam.org/doi/pdf/10.1137/1.9780898719857}
}

@ARTICLE{hazboun19,
       author = {{Hazboun}, Jeffrey S. and {Romano}, Joseph D. and {Smith}, Tristan L.},
        title = "{Realistic sensitivity curves for pulsar timing arrays}",
      journal = {\prd},
         year = 2019,
        month = nov,
       volume = {100},
       number = {10},
          eid = {104028},
        pages = {104028},
          doi = {10.1103/PhysRevD.100.104028},
archivePrefix = {arXiv},
       eprint = {1907.04341},
 primaryClass = {gr-qc},
       adsurl = {https://ui.adsabs.harvard.edu/abs/2019PhRvD.100j4028H}
}

@software{FrequencyOptimizer,
	author = {{Lam}, M.~T. and {Cohen}, Tyler and {Lynch}, R.~S.},
	title = {FrequencyOptimizer: Calculator for optimal frequencies analysis of pulsars},
	month = may,
	year = 2026,
	version = {v0.1},
	publisher = {github},
	url = {https://github.com/mtlam/FrequencyOptimizer}
}

@ARTICLE{cordes10,
       author = {{Cordes}, J.~M. and {Shannon}, R.~M.},
        title = "{A Measurement Model for Precision Pulsar Timing}",
      journal = {arXiv e-prints},
         year = 2010,
        month = oct,
          eid = {arXiv:1010.3785},
        pages = {arXiv:1010.3785},
          doi = {10.48550/arXiv.1010.3785},
archivePrefix = {arXiv},
       eprint = {1010.3785},
 primaryClass = {astro-ph.IM},
       adsurl = {https://ui.adsabs.harvard.edu/abs/2010arXiv1010.3785C}
}

@ARTICLE{gitika25,
       author = {{Gitika}, Pratyasha and {Shannon}, Ryan M. and {Bailes}, Matthew and {Reardon}, Daniel J. and {Miles}, Matthew T. and {Champion}, David J. and {Grunthal}, Kathrin},
        title = "{Optimising the MeerKAT pulsar timing array and towards precision pulsar timing with SKA-mid}",
      journal = {\pasa},
         year = 2025,
        month = oct,
       volume = {42},
          eid = {e146},
        pages = {e146},
          doi = {10.1017/pasa.2025.10107},
archivePrefix = {arXiv},
       eprint = {2510.03139},
 primaryClass = {astro-ph.HE},
       adsurl = {https://ui.adsabs.harvard.edu/abs/2025PASA...42..146G}
}

@ARTICLE{lam19_jitter,
       author = {{Lam}, M.~T. and {McLaughlin}, M.~A. and {Arzoumanian}, Z. and {Blumer}, H. and {Brook}, P.~R. and {Cromartie}, H.~T. and {Demorest}, P.~B. and {DeCesar}, M.~E. and {Dolch}, T. and {Ellis}, J.~A. and {Ferdman}, R.~D. and {Ferrara}, E.~C. and {Fonseca}, E. and {Garver-Daniels}, N. and {Gentile}, P.~A. and {Jones}, M.~L. and {Lorimer}, D.~R. and {Lynch}, R.~S. and {Ng}, C. and {Nice}, D.~J. and {Pennucci}, T.~T. and {Ransom}, S.~M. and {Spiewak}, R. and {Stairs}, I.~H. and {Stovall}, K. and {Swiggum}, J.~K. and {Vigeland}, S.~J. and {Zhu}, W.~W.},
        title = "{The NANOGrav 12.5 yr Data Set: The Frequency Dependence of Pulse Jitter in Precision Millisecond Pulsars}",
      journal = {\apj},
         year = 2019,
        month = feb,
       volume = {872},
       number = {2},
          eid = {193},
        pages = {193},
          doi = {10.3847/1538-4357/ab01cd},
archivePrefix = {arXiv},
       eprint = {1809.03058},
 primaryClass = {astro-ph.HE},
       adsurl = {https://ui.adsabs.harvard.edu/abs/2019ApJ...872..193L}
}

@article{cordes16_freqDM,
doi = {10.3847/0004-637X/817/1/16},
url = {https://doi.org/10.3847/0004-637X/817/1/16},
year = {2016},
month = {jan},
publisher = {The American Astronomical Society},
volume = {817},
number = {1},
pages = {16},
author = {Cordes, J. M. and Shannon, R. M. and Stinebring, D. R.},
title = {FREQUENCY-DEPENDENT DISPERSION MEASURES AND IMPLICATIONS FOR PULSAR TIMING},
journal = {The Astrophysical Journal}
}

@ARTICLE{lawson87,
       author = {{Lawson}, K.~D. and {Mayer}, C.~J. and {Osborne}, J.~L. and {Parkinson}, M.~L.},
        title = "{Variations in the spectral index of the galactic radio continuum emission in the northern hemisphere.}",
      journal = {\mnras},
         year = 1987,
        month = mar,
       volume = {225},
        pages = {307-327},
          doi = {10.1093/mnras/225.2.307},
       adsurl = {https://ui.adsabs.harvard.edu/abs/1987MNRAS.225..307L}
}

@ARTICLE{haslam82,
       author = {{Haslam}, C.~G.~T. and {Salter}, C.~J. and {Stoffel}, H. and {Wilson}, W.~E.},
        title = "{A 408-MHZ All-Sky Continuum Survey. II. The Atlas of Contour Maps}",
      journal = {\aaps},
         year = 1982,
        month = jan,
       volume = {47},
        pages = {1},
       adsurl = {https://ui.adsabs.harvard.edu/abs/1982A&AS...47....1H}
}

@ARTICLE{psrpoppy,
       author = {{Bates}, S.~D. and {Lorimer}, D.~R. and {Rane}, A. and {Swiggum}, J.},
        title = "{PSRPOPPy: an open-source package for pulsar population simulations}",
      journal = {\mnras},
         year = 2014,
        month = apr,
       volume = {439},
       number = {3},
        pages = {2893-2902},
          doi = {10.1093/mnras/stu157},
archivePrefix = {arXiv},
       eprint = {1311.3427},
 primaryClass = {astro-ph.IM},
       adsurl = {https://ui.adsabs.harvard.edu/abs/2014MNRAS.439.2893B}
}

@article{Lee2012,
   title={The optimal schedule for pulsar timing array observations: <i>The optimal schedule for pulsar timing array observations</i>},
   volume={423},
   ISSN={0035-8711},
   url={http://dx.doi.org/10.1111/j.1365-2966.2012.21070.x},
   DOI={10.1111/j.1365-2966.2012.21070.x},
   number={3},
   journal={Monthly Notices of the Royal Astronomical Society},
   publisher={Oxford University Press (OUP)},
   author={Lee, K. J. and Bassa, C. G. and Janssen, G. H. and Karuppusamy, R. and Kramer, M. and Smits, R. and Stappers, B. W.},
   year={2012},
   month=May, pages={2642–2655} }

@article{gbt350_fermi,
doi = {10.3847/1538-4357/ad2994},
url = {https://doi.org/10.3847/1538-4357/ad2994},
year = {2024},
month = {may},
publisher = {The American Astronomical Society},
volume = {966},
number = {2},
pages = {161},
author = {Bangale, P. and Bhattacharyya, B. and Camilo, F. and Clark, C. J. and Cognard, I. and DeCesar, M. E. and Ferrara, E. C. and Gentile, P. and Guillemot, L. and Hessels, J. W. T. and Johnson, T. J. and Kerr, M. and McLaughlin, M. A. and Nieder, L. and Ransom, S. M. and Ray, P. S. and Roberts, M. S. E. and Roy, J. and Sanpa-arsa, S. and Theureau, G. and Wolff, M. T.},
title = {A 350 MHz Green Bank Telescope Survey of Unassociated Fermi LAT Sources: Discovery and Timing of 10 Millisecond Pulsars},
journal = {The Astrophysical Journal}
}

@ARTICLE{christy14,
       author = {{Christy}, Brian and {Anella}, Ryan and {Lommen}, Andrea and {Finn}, Lee Samuel and {Camuccio}, Richard and {Handzo}, Emma},
        title = "{Optimization of NANOGrav's Time Allocation for Maximum Sensitivity to Single Sources}",
      journal = {\apj},
         year = 2014,
        month = oct,
       volume = {794},
       number = {2},
          eid = {163},
        pages = {163},
          doi = {10.1088/0004-637X/794/2/163},
archivePrefix = {arXiv},
       eprint = {1409.7722},
 primaryClass = {astro-ph.IM},
       adsurl = {https://ui.adsabs.harvard.edu/abs/2014ApJ...794..163C}
}

@article{middleton25,
    author = {Middleton, Hannah and Shannon, Ryan M and Bailes, Matthew and Cameron, Andrew D and Corongiu, Alessandro and Geyer, Marisa and Jones, Max and Kramer, Michael and Miles, Matthew T and Parthasarathy, Aditya and Possenti, Andrea and Reardon, Daniel J},
    title = {A simple optimization for the MeerKAT pulsar timing array},
    journal = {Monthly Notices of the Royal Astronomical Society},
    volume = {540},
    number = {1},
    pages = {603-611},
    year = {2025},
    month = {06},
    issn = {0035-8711},
    doi = {10.1093/mnras/staf748},
    url = {https://doi.org/10.1093/mnras/staf748},
    eprint = {https://academic.oup.com/mnras/article-pdf/540/1/603/63154552/staf748.pdf},
}

@article{Burke_Spolaor_2019,
   title={The astrophysics of nanohertz gravitational waves},
   volume={27},
   ISSN={1432-0754},
   url={http://dx.doi.org/10.1007/s00159-019-0115-7},
   DOI={10.1007/s00159-019-0115-7},
   number={1},
   journal={The Astronomy and Astrophysics Review},
   publisher={Springer Science and Business Media LLC},
   author={Burke-Spolaor, Sarah and Taylor, Stephen R. and Charisi, Maria and Dolch, Timothy and Hazboun, Jeffrey S. and Holgado, A. Miguel and Kelley, Luke Zoltan and Lazio, T. Joseph W. and Madison, Dustin R. and McMann, Natasha and Mingarelli, Chiara M. F. and Rasskazov, Alexander and Siemens, Xavier and Simon, Joseph J. and Smith, Tristan L.},
   year={2019},
   month=jun }

@INPROCEEDINGS{hallinan19_dsa2000,
       author = {{Hallinan}, Gregg and {Ravi}, V. and {Weinreb}, S. and {Kocz}, J. and {Huang}, Y. and {Woody}, D.~P. and {Lamb}, J. and {D'Addario}, L. and {Catha}, M. and {Law}, C. and {Kulkarni}, S.~R. and {Phinney}, E.~S. and {Eastwood}, M.~W. and {Bouman}, K. and {McLaughlin}, M. and {Ransom}, S. and {Siemens}, X. and {Cordes}, J. and {Lynch}, R. and {Kaplan}, D. and {Brazier}, A. and {Bhatnagar}, S. and {Myers}, S. and {Walter}, F. and {Gaensler}, B.},
        title = "{The DSA-2000 {\textemdash} A Radio Survey Camera}",
    booktitle = {Bulletin of the American Astronomical Society},
         year = 2019,
       volume = {51},
        month = sep,
          eid = {255},
        pages = {255},
          doi = {10.48550/arXiv.1907.07648},
archivePrefix = {arXiv},
       eprint = {1907.07648},
 primaryClass = {astro-ph.IM},
       adsurl = {https://ui.adsabs.harvard.edu/abs/2019BAAS...51g.255H}
}

@ARTICLE{hasasia,
       author = {{Hazboun}, Jeffrey and {Romano}, Joseph and {Smith}, Tristan},
        title = "{Hasasia: A Python package for Pulsar Timing Array Sensitivity Curves}",
      journal = {The Journal of Open Source Software},
         year = 2019,
        month = oct,
       volume = {4},
       number = {42},
          eid = {1775},
        pages = {1775},
          doi = {10.21105/joss.01775},
       adsurl = {https://ui.adsabs.harvard.edu/abs/2019JOSS....4.1775H}
}

@ARTICLE{ng12p5_gwb,
       author = {{Arzoumanian}, Zaven and {Baker}, Paul T. and {Blumer}, Harsha and {B{\'e}csy}, Bence and {Brazier}, Adam and {Brook}, Paul R. and {Burke-Spolaor}, Sarah and {Chatterjee}, Shami and {Chen}, Siyuan and {Cordes}, James M. and {Cornish}, Neil J. and {Crawford}, Fronefield and {Cromartie}, H. Thankful and {Decesar}, Megan E. and {Demorest}, Paul B. and {Dolch}, Timothy and {Ellis}, Justin A. and {Ferrara}, Elizabeth C. and {Fiore}, William and {Fonseca}, Emmanuel and {Garver-Daniels}, Nathan and {Gentile}, Peter A. and {Good}, Deborah C. and {Hazboun}, Jeffrey S. and {Holgado}, A. Miguel and {Islo}, Kristina and {Jennings}, Ross J. and {Jones}, Megan L. and {Kaiser}, Andrew R. and {Kaplan}, David L. and {Kelley}, Luke Zoltan and {Key}, Joey Shapiro and {Laal}, Nima and {Lam}, Michael T. and {Lazio}, T. Joseph W. and {Lorimer}, Duncan R. and {Luo}, Jing and {Lynch}, Ryan S. and {Madison}, Dustin R. and {McLaughlin}, Maura A. and {Mingarelli}, Chiara M.~F. and {Ng}, Cherry and {Nice}, David J. and {Pennucci}, Timothy T. and {Pol}, Nihan S. and {Ransom}, Scott M. and {Ray}, Paul S. and {Shapiro-Albert}, Brent J. and {Siemens}, Xavier and {Simon}, Joseph and {Spiewak}, Ren{\'e}e and {Stairs}, Ingrid H. and {Stinebring}, Daniel R. and {Stovall}, Kevin and {Sun}, Jerry P. and {Swiggum}, Joseph K. and {Taylor}, Stephen R. and {Turner}, Jacob E. and {Vallisneri}, Michele and {Vigeland}, Sarah J. and {Witt}, Caitlin A. and {Nanograv Collaboration}},
        title = "{The NANOGrav 12.5 yr Data Set: Search for an Isotropic Stochastic Gravitational-wave Background}",
      journal = {ApJ Letters},
         year = 2020,
        month = dec,
       volume = {905},
       number = {2},
          eid = {L34},
        pages = {L34},
          doi = {10.3847/2041-8213/abd401},
archivePrefix = {arXiv},
       eprint = {2009.04496},
 primaryClass = {astro-ph.HE},
       adsurl = {https://ui.adsabs.harvard.edu/abs/2020ApJ...905L..34A}
}

@ARTICLE{ng9_excessnoise,
       author = {{Lam}, M.~T. and {Cordes}, J.~M. and {Chatterjee}, S. and {Arzoumanian}, Z. and {Crowter}, K. and {Demorest}, P.~B. and {Dolch}, T. and {Ellis}, J.~A. and {Ferdman}, R.~D. and {Fonseca}, E. and {Gonzalez}, M.~E. and {Jones}, G. and {Jones}, M.~L. and {Levin}, L. and {Madison}, D.~R. and {McLaughlin}, M.~A. and {Nice}, D.~J. and {Pennucci}, T.~T. and {Ransom}, S.~M. and {Shannon}, R.~M. and {Siemens}, X. and {Stairs}, I.~H. and {Stovall}, K. and {Swiggum}, J.~K. and {Zhu}, W.~W.},
        title = "{The NANOGrav Nine-year Data Set: Excess Noise in Millisecond Pulsar Arrival Times}",
      journal = {\apj},
         year = 2017,
        month = jan,
       volume = {834},
       number = {1},
          eid = {35},
        pages = {35},
          doi = {10.3847/1538-4357/834/1/35},
archivePrefix = {arXiv},
       eprint = {1610.01731},
 primaryClass = {astro-ph.HE},
       adsurl = {https://ui.adsabs.harvard.edu/abs/2017ApJ...834...35L}
}

@ARTICLE{ng15_timing,
       author = {{Agazie}, Gabriella and {Alam}, Md Faisal and {Anumarlapudi}, Akash and {Archibald}, Anne M. and {Arzoumanian}, Zaven and {Baker}, Paul T. and {Blecha}, Laura and {Bonidie}, Victoria and {Brazier}, Adam and {Brook}, Paul R. and {Burke-Spolaor}, Sarah and {B{\'e}csy}, Bence and {Chapman}, Christopher and {Charisi}, Maria and {Chatterjee}, Shami and {Cohen}, Tyler and {Cordes}, James M. and {Cornish}, Neil J. and {Crawford}, Fronefield and {Cromartie}, H. Thankful and {Crowter}, Kathryn and {Decesar}, Megan E. and {Demorest}, Paul B. and {Dolch}, Timothy and {Drachler}, Brendan and {Ferrara}, Elizabeth C. and {Fiore}, William and {Fonseca}, Emmanuel and {Freedman}, Gabriel E. and {Garver-Daniels}, Nate and {Gentile}, Peter A. and {Glaser}, Joseph and {Good}, Deborah C. and {G{\"u}ltekin}, Kayhan and {Hazboun}, Jeffrey S. and {Jennings}, Ross J. and {Jessup}, Cody and {Johnson}, Aaron D. and {Jones}, Megan L. and {Kaiser}, Andrew R. and {Kaplan}, David L. and {Kelley}, Luke Zoltan and {Kerr}, Matthew and {Key}, Joey S. and {Kuske}, Anastasia and {Laal}, Nima and {Lam}, Michael T. and {Lamb}, William G. and {Lazio}, T. Joseph W. and {Lewandowska}, Natalia and {Lin}, Ye and {Liu}, Tingting and {Lorimer}, Duncan R. and {Luo}, Jing and {Lynch}, Ryan S. and {Ma}, Chung-Pei and {Madison}, Dustin R. and {Maraccini}, Kaleb and {McEwen}, Alexander and {McKee}, James W. and {McLaughlin}, Maura A. and {McMann}, Natasha and {Meyers}, Bradley W. and {Mingarelli}, Chiara M.~F. and {Mitridate}, Andrea and {Ng}, Cherry and {Nice}, David J. and {Ocker}, Stella Koch and {Olum}, Ken D. and {Panciu}, Elisa and {Pennucci}, Timothy T. and {Perera}, Benetge B.~P. and {Pol}, Nihan S. and {Radovan}, Henri A. and {Ransom}, Scott M. and {Ray}, Paul S. and {Romano}, Joseph D. and {Salo}, Laura and {Sardesai}, Shashwat C. and {Schmiedekamp}, Carl and {Schmiedekamp}, Ann and {Schmitz}, Kai and {Shapiro-Albert}, Brent J. and {Siemens}, Xavier and {Simon}, Joseph and {Siwek}, Magdalena S. and {Stairs}, Ingrid H. and {Stinebring}, Daniel R. and {Stovall}, Kevin and {Susobhanan}, Abhimanyu and {Swiggum}, Joseph K. and {Taylor}, Stephen R. and {Turner}, Jacob E. and {Unal}, Caner and {Vallisneri}, Michele and {Vigeland}, Sarah J. and {Wahl}, Haley M. and {Wang}, Qiaohong and {Witt}, Caitlin A. and {Young}, Olivia and {Nanograv Collaboration}},
        title = "{The NANOGrav 15 yr Data Set: Observations and Timing of 68 Millisecond Pulsars}",
      journal = {ApJ Letters},
         year = 2023,
        month = jul,
       volume = {951},
       number = {1},
          eid = {L9},
        pages = {L9},
          doi = {10.3847/2041-8213/acda9a},
archivePrefix = {arXiv},
       eprint = {2306.16217},
 primaryClass = {astro-ph.HE},
       adsurl = {https://ui.adsabs.harvard.edu/abs/2023ApJ...951L...9A}
}

@ARTICLE{dolch21_cyspec,
       author = {{Dolch}, Timothy and {Stinebring}, Dan R. and {Jones}, Glenn and {Zhu}, Hengrui and {Lynch}, Ryan S. and {Cohen}, Tyler and {Demorest}, Paul B. and {Lam}, Michael T. and {Levin}, Lina and {McLaughlin}, Maura A. and {Palliyaguru}, Nipuni T.},
        title = "{Deconvolving Pulsar Signals with Cyclic Spectroscopy: A Systematic Evaluation}",
      journal = {ApJ},
         year = 2021,
        month = jun,
       volume = {913},
       number = {2},
          eid = {98},
        pages = {98},
          doi = {10.3847/1538-4357/abf48b},
archivePrefix = {arXiv},
       eprint = {2008.10562},
 primaryClass = {astro-ph.HE},
       adsurl = {https://ui.adsabs.harvard.edu/abs/2021ApJ...913...98D}
}

@ARTICLE{liu23,
       author = {{Liu}, Tingting and {Cohen}, Tyler and {McGrath}, Casey and {Demorest}, Paul B. and {Vigeland}, Sarah J.},
        title = "{Multi-messenger Approaches to Supermassive Black Hole Binary Detection and Parameter Estimation. II. Optimal Strategies for a Pulsar Timing Array}",
      journal = {ApJ},
         year = 2023,
        month = mar,
       volume = {945},
       number = {1},
          eid = {78},
        pages = {78},
          doi = {10.3847/1538-4357/acb492},
archivePrefix = {arXiv},
       eprint = {2301.07135},
 primaryClass = {astro-ph.HE},
       adsurl = {https://ui.adsabs.harvard.edu/abs/2023ApJ...945...78L}
}

@ARTICLE{lam18_freqopt,
       author = {{Lam}, M.~T. and {McLaughlin}, M.~A. and others},
      journal = {ApJ},
         year = 2018,
        month = jul,
       volume = {861},
       number = {1},
          eid = {12},
        pages = {12},
          doi = {10.3847/1538-4357/aac48d},
archivePrefix = {arXiv},
       eprint = {1710.02272},
 primaryClass = {astro-ph.HE},
       adsurl = {https://ui.adsabs.harvard.edu/abs/2018ApJ...861...12L}
}

@ARTICLE{detweiler79,
       author = {{Detweiler}, S.},
      journal = {ApJ},
         year = 1979,
        month = dec,
       volume = {234},
        pages = {1100-1104}
}

@article{siemens13,
  author={Siemens, X. and Ellis, J. and others},
  journal={Classical and Quantum Gravity},
  volume={30},
  number={22},
  pages={224015},
  year={2013}
}

@article{rajwade16,
author = {Rajwade, K. M. and Lorimer, D. R. and others},
journal = {MNRAS},
volume = {471},
number = {1},
pages = {730-739},
year = {2017},
doi = {10.1093/mnras/stx1661},
URL = {http://dx.doi.org/10.1093/mnras/stx1661},
eprint = {/oup/backfile/content_public/journal/mnras/471/1/10.1093_mnras_stx1661/1/stx1661.pdf}
}

@ARTICLE{ne2001,
       author = {{Cordes}, J.~M. and {Lazio}, T.~J.~W.},
      journal = {arXiv e-prints},
         year = "2002",
        month = "Jul",
          eid = {astro-ph/0207156},
        pages = {astro-ph/0207156},
archivePrefix = {arXiv},
       eprint = {astro-ph/0207156},
 primaryClass = {astro-ph},
       adsurl = {https://ui.adsabs.harvard.edu/abs/2002astro.ph..7156C}
}

@ARTICLE{kramer99,
       author = {{Kramer}, Michael and {Lange}, Christoph and others},
      journal = {ApJ},
         year = 1999,
        month = dec,
       volume = {526},
       number = {2},
        pages = {957-975},
          doi = {10.1086/308042},
archivePrefix = {arXiv},
       eprint = {astro-ph/9906442},
 primaryClass = {astro-ph},
       adsurl = {https://ui.adsabs.harvard.edu/abs/1999ApJ...526..957K}
}

@ARTICLE{vigeland16_scalinglaws,
       author = {{Vigeland}, S.~J. and {Siemens}, X.},
      journal = {Physical Review D},
         year = 2016,
        month = dec,
       volume = {94},
       number = {12},
          eid = {123003},
        pages = {123003},
          doi = {10.1103/PhysRevD.94.123003},
archivePrefix = {arXiv},
       eprint = {1609.03656},
 primaryClass = {astro-ph.IM},
       adsurl = {https://ui.adsabs.harvard.edu/abs/2016PhRvD..94l3003V}
}

@article{ng15_gwb_evidence,
doi = {10.3847/2041-8213/acdac6},
url = {https://dx.doi.org/10.3847/2041-8213/acdac6},
year = {2023},
month = {jun},
publisher = {The American Astronomical Society},
volume = {951},
number = {1},
pages = {L8},
author = {Gabriella Agazie and Akash Anumarlapudi and Anne M. Archibald and Zaven Arzoumanian and Paul T. Baker and Bence Bécsy and Laura Blecha and Adam Brazier and Paul R. Brook and Sarah Burke-Spolaor and Rand Burnette and Robin Case and Maria Charisi and Shami Chatterjee and Katerina Chatziioannou and Belinda D. Cheeseboro and Siyuan Chen and Tyler Cohen and James M. Cordes and Neil J. Cornish and Fronefield Crawford and H. Thankful Cromartie and Kathryn Crowter and Curt J. Cutler and Megan E. DeCesar and Dallas DeGan and Paul B. Demorest and Heling Deng and Timothy Dolch and Brendan Drachler and Justin A. Ellis and Elizabeth C. Ferrara and William Fiore and Emmanuel Fonseca and Gabriel E. Freedman and Nate Garver-Daniels and Peter A. Gentile and Kyle A. Gersbach and Joseph Glaser and Deborah C. Good and Kayhan Gültekin and Jeffrey S. Hazboun and Sophie Hourihane and Kristina Islo and Ross J. Jennings and Aaron D. Johnson and Megan L. Jones and Andrew R. Kaiser and David L. Kaplan and Luke Zoltan Kelley and Matthew Kerr and Joey S. Key and Tonia C. Klein and Nima Laal and Michael T. Lam and William G. Lamb and T. Joseph W. Lazio and Natalia Lewandowska and Tyson B. Littenberg and Tingting Liu and Andrea Lommen and Duncan R. Lorimer and Jing Luo and Ryan S. Lynch and Chung-Pei Ma and Dustin R. Madison and Margaret A. Mattson and Alexander McEwen and James W. McKee and Maura A. McLaughlin and Natasha McMann and Bradley W. Meyers and Patrick M. Meyers and Chiara M. F. Mingarelli and Andrea Mitridate and Priyamvada Natarajan and Cherry Ng and David J. Nice and Stella Koch Ocker and Ken D. Olum and Timothy T. Pennucci and Benetge B. P. Perera and Polina Petrov and Nihan S. Pol and Henri A. Radovan and Scott M. Ransom and Paul S. Ray and Joseph D. Romano and Shashwat C. Sardesai and Ann Schmiedekamp and Carl Schmiedekamp and Kai Schmitz and Levi Schult and Brent J. Shapiro-Albert and Xavier Siemens and Joseph Simon and Magdalena S. Siwek and Ingrid H. Stairs and Daniel R. Stinebring and Kevin Stovall and Jerry P. Sun and Abhimanyu Susobhanan and Joseph K. Swiggum and Jacob Taylor and Stephen R. Taylor and Jacob E. Turner and Caner Unal and Michele Vallisneri and Rutger van Haasteren and Sarah J. Vigeland and Haley M. Wahl and Qiaohong Wang and Caitlin A. Witt and Olivia Young and The NANOGrav Collaboration},
title = {The NANOGrav 15 yr Data Set: Evidence for a Gravitational-wave Background},
journal = {The Astrophysical Journal Letters}
}
\bibliographystyle{aasjournalv7}



\end{document}